\documentclass[aps,pre,twocolumn,superscriptaddress,longbibliography]{revtex4-1}
\usepackage{color}
\usepackage[usenames,dvipsnames]{xcolor}

\usepackage{graphicx}
\usepackage{epsfig}
\usepackage{dcolumn}
\usepackage{bm}
\usepackage{mathrsfs}
\usepackage{multirow}
\usepackage[all]{xy}
\usepackage{pbox}
\usepackage{amssymb}
\usepackage{bbold}

\usepackage{color}
\definecolor{myblue}{RGB}{65,105,225}
\definecolor{mygreen}{RGB}{34,139,34}
\definecolor{myorange}{RGB}{255,69,0}
\usepackage[colorlinks,linkcolor=myorange,urlcolor=myblue,citecolor=myblue]{hyperref}

\usepackage{natbib}
\usepackage{dsfont}
\usepackage[usenames,dvipsnames]{xcolor}
\usepackage{amsmath}

\def\(({\left(}
\def\)){\right)}
\def\[[{\left[}
\def\]]{\right]}

\newcommand{\beq}{\begin{equation}}
\newcommand{\eeq}{\end{equation}}

\newcommand{\ben}{\begin{eqnarray}}
\newcommand{\een}{\end{eqnarray}}

\newcommand{\lan}{\left\langle}
\newcommand{\ran}{\right\rangle}

\newcommand{\la}{\langle}
\newcommand{\ra}{\rangle}
\newcommand{\be}{\begin{equation}}
\newcommand{\ee}{\end{equation}}

\newcommand{\ket}[1]{\left| #1 \ran}
\newcommand{\bra}[1]{\lan #1 \right|}

\begin{document}  

\title{Dynamical criticality in open systems: non-perturbative physics, microscopic origin and direct observation}

\author{Carlos P\'erez-Espigares}
\affiliation{School of Physics and Astronomy, and Centre for the Mathematics and Theoretical Physics of Quantum Non-Equilibrium Systems,
University of Nottingham, Nottingham, NG7 2RD, United Kingdom}
\affiliation{Departamento de Electromagnetismo y F\'{\i}sica de la Materia, and Institute Carlos I for Theoretical and Computational Physics, Universidad de Granada, Granada 18071, Spain}

\author{Federico Carollo}
\affiliation{School of Physics and Astronomy, and Centre for the Mathematics and Theoretical Physics of Quantum Non-Equilibrium Systems,
University of Nottingham, Nottingham, NG7 2RD, United Kingdom}

\author{Juan P. Garrahan}
\affiliation{School of Physics and Astronomy, and Centre for the Mathematics and Theoretical Physics of Quantum Non-Equilibrium Systems,
University of Nottingham, Nottingham, NG7 2RD, United Kingdom}

\author{Pablo I. Hurtado}
\affiliation{Departamento de Electromagnetismo y F\'{\i}sica de la Materia, and Institute Carlos I for Theoretical and Computational Physics, Universidad de Granada, Granada 18071, Spain}

\date{\today}

\begin{abstract}
Driven diffusive systems may undergo phase transitions to sustain atypical values of the current. This leads in some cases to symmetry-broken space-time trajectories which enhance the probability of such fluctuations. Here we shed light on both the macroscopic large deviation properties and the microscopic origin of such spontaneous symmetry breaking in the open weakly asymmetric exclusion process. By studying the joint fluctuations of the current and a collective order parameter, we uncover the full dynamical phase diagram for arbitrary boundary driving, which is reminiscent of a $\mathbb{Z}_2$ symmetry-breaking transition. The associated joint large deviation function becomes non-convex below the critical point, where a Maxwell-like violation of the additivity principle is observed. At the microscopic level, the dynamical phase transition is linked to an emerging degeneracy of the ground state of the microscopic generator, from which the optimal trajectories in the symmetry-broken phase follow. In addition, we observe this new symmetry-breaking phenomenon in extensive rare-event simulations, confirming our macroscopic and microscopic results.
\end{abstract}


\maketitle 

\emph{Introduction.}-- The discovery of dynamical phase transitions (DPTs) in the \emph{fluctuations} of nonequilibrium systems has attracted much attention in recent years \cite{bertini05a,bodineau05a,harris05a,bertini06a,bodineau07a,Lecomte2007,garrahan07a,garrahan09a,hurtado11a,ates12a,perez-espigares13a,harris13a,vaikuntanathan14a,mey14a,jack15a,baek15a,tsobgni16a,harris17a,lazarescu17a,brandner17a,karevski17a,carollo17a,baek17a,tizon-escamilla17b,shpielberg17a,baek18a,shpielberg18a,perez2018glassy,chleboun2018,Klymko2018,Whitelam2018,vroylandt2018non}. In contrast with standard critical phenomena \cite{binney92a,zinn-justin02a}, which occur at the configurational level, 
DPTs appear in trajectory space when conditioning the system to sustain an unlikely value of dynamical observables such as the time-integrated current \cite{bertini05a,bertini06a,derrida07a,hurtado14a,lazarescu15a,shpielberg18a}. 
DPTs thus manifest as a peculiar change in the properties of trajectories responsible for such rare events, making these trajectories far more probable than anticipated due to the emergence of ordered structures such as traveling waves \cite{bodineau05a,hurtado11a,perez-espigares13a,karevski17a}, condensates \cite{harris05a,harris13a,chleboun2018} or hyperuniform states \cite{jack15a,carollo17a,carollo2018current}. In all these cases, the hallmark of the DPT is the appearance of a singularity in the so-called large deviation function (LDF), which controls the probability of fluctuations and plays the role of a thermodynamic potential for nonequilibrium systems \cite{derrida07a,touchette09a,bertini15a}. 
DPTs play a key role to understand the physics of different systems, from glass formers \cite{garrahan07a,garrahan09a,hedges09a,chandler10a,pitard11a,speck12a,pinchaipat17a,abou17a} to micromasers and superconducting transistors \cite{garrahan11a,genway12a}, and applications such as DPT-based quantum thermal switches \cite{manzano14a,manzano16a,manzano17a}. Moreover, by making rare events typical with the use of Doob's transform \cite{doob57a,chetrite15a,chetrite15b} or optimal fields \cite{bertini15a}, one may exploit DPTs to engineer and control nonequilibrium systems with a desired statistics \emph{on demand} \cite{carollo2018making}. 

In the context of diffusive systems, DPTs in current statistics have been throughly studied for periodic settings \cite{bodineau05a,bertini06a,hurtado11a,perez-espigares13a,tizon-escamilla17b}, in which the broken symmetry is time translational invariance, giving rise to a violation of the so-called additivity principle via traveling-wave profiles \cite{bodineau04a,hurtado11a}. Nevertheless, it has not been until very recently that other kind of symmetry-breaking scenarios (involving e.g. particle-hole symmetry) have been predicted for open systems \cite{baek17a}, i.e.~in contact with boundary reservoirs. 
In particular, a perturbative Landau theory restricted to zero or small boundary gradient has been recently put forward \cite{baek17a,baek18a} which predicts $1^{\text{st}}$- and $2^{\text{nd}}$-order DPTs in some diffusive media. 
Key questions remain unanswered, however,
such as the direct numerical observation of this DPT, its microscopic origin, the non-perturbative physics beyond the critical point, or its existence under 
strong boundary driving, the latter being one of the most challenging problems in nonequilibrium physics.

\begin{figure}[b!]
\vspace{-0.2cm}\includegraphics[scale=0.28]{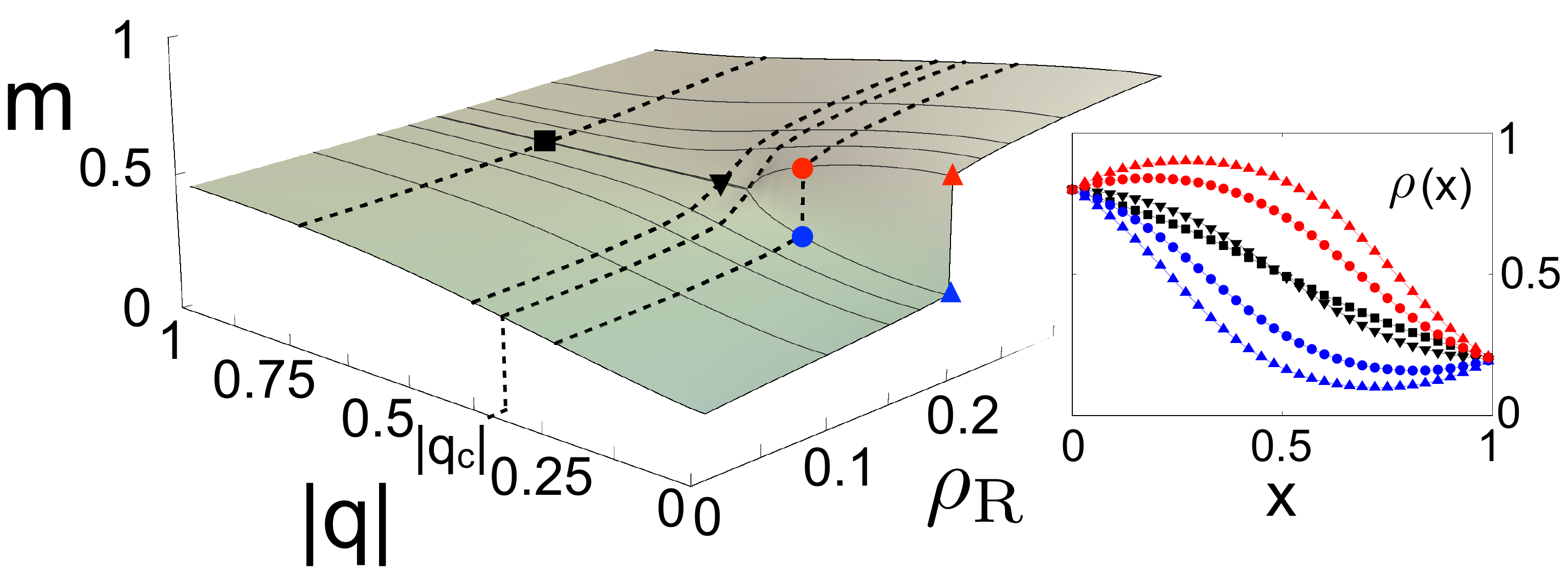}
\vspace{-0.4cm}
\caption{
Mass $m_q$ of the optimal trajectory responsible for a current fluctuation $q$ for different boundary drivings, 
with $\rho_{\text{L}}=0.8$, $\rho_{\text{R}}\in[0,0.4]$ and external field $E=4$. Inset: Optimal profiles for $\rho_{\text{R}}=0.2$ and $q$'s signaled in the main plot. 
}\label{fig1}
\end{figure}

\begin{figure}[t!]
\hspace{-0.2cm}\includegraphics[scale=0.355]{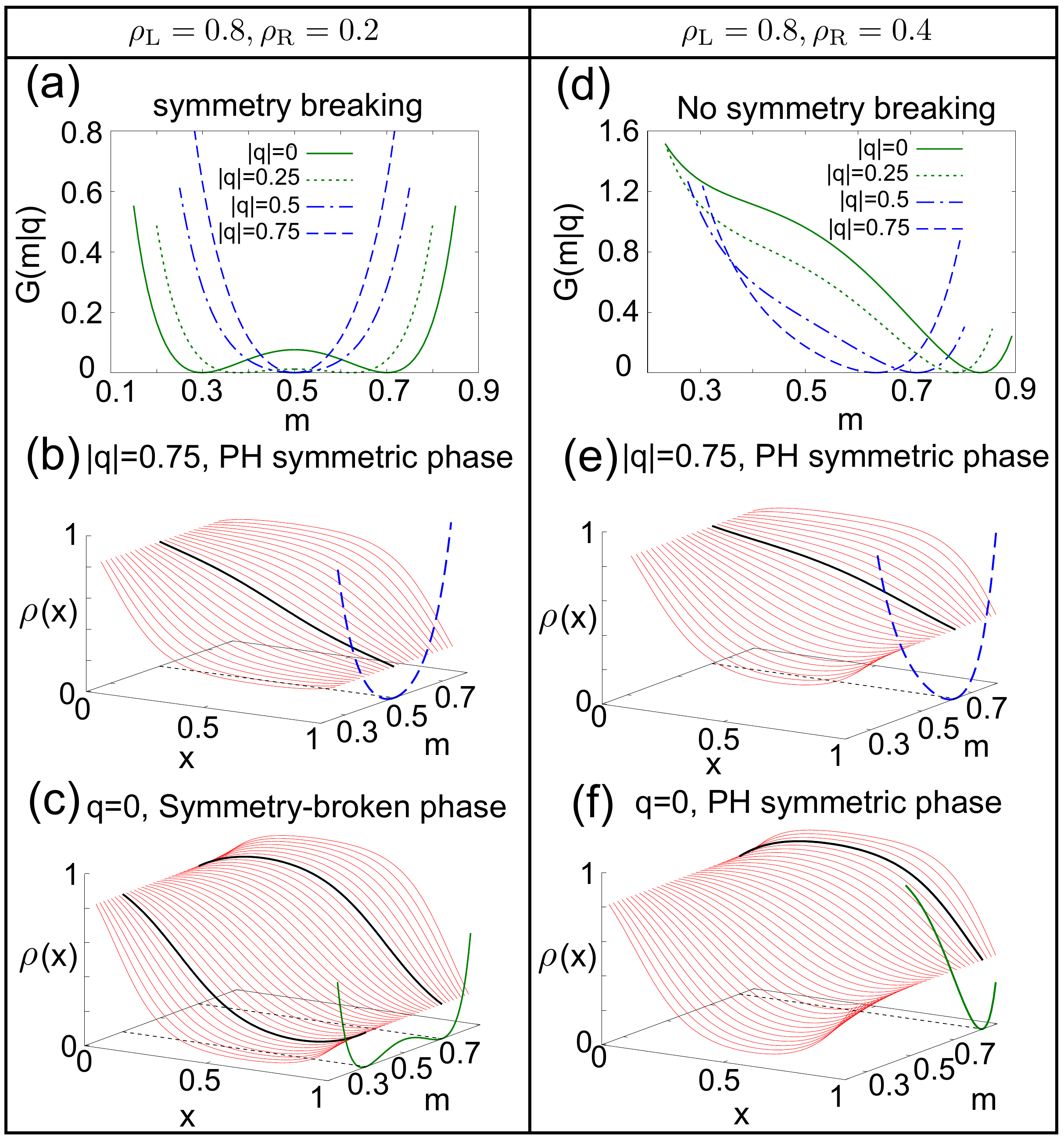}
\caption{
(a)  Conditional LDF $G(m|q)=G(m,q)-G(q)$ for $\rho_{\text{L}}=0.8$, $\rho_{\text{R}}=0.2$ and $E=4$ as a function of $m$ and different values of $q$. (b) $\rho_{m,q}(x)$ for $|q|=0.75$ and different $m$'s, together with the associated $G(m|q)$.
(c) Same results of panel (b) but for $q=0$. Two optimal profiles with high- and low-mass emerge (black solid lines). (d)-(f) Analogous results of panels (a)-(c) for $\rho_{\text{L}}=0.8$ and $\rho_{\text{R}}=0.4$.
}
\label{fig2}
\end{figure}

In this work we address these questions in a paradigmatic diffusive system, the open one-dimensional ($1d$) weakly asymmetric simple exclusion process (WASEP) \cite{wasep1masi89,wasep2gartner87}. 
In particular, by studying the joint fluctuations of the current $q$ and a novel collective order parameter defined by total mass ($m$),
we unveil analytically the full dynamical phase diagram for arbitrary boundary gradients, 
see Fig.~\ref{fig1}. A DPT is observed at a critical current $|q_c|$ for any boundary driving symmetric around the density $1/2$, i.e.~for $\rho_{\text{R}}=1-\rho_{\text{L}}$ (with $\rho_{\text{L}}$ and $\rho_{\text{R}}$ the left and right reservoir densities, respectively),
where the joint mass-current LDF $G(m,q)$ becomes non-convex (see Fig.~\ref{fig2}). This signals the breaking of the particle-hole (PH)
symmetry present in the governing action but no longer in the optimal trajectories associated to these atypical fluctuations:  for $|q|<|q_c|$ coexisting low- and high-mass trajectories appear with broken PH-symmetry. An asymmetric boundary gradient
favors one of the mass branches, deepening the associated minimum in $G(m,q)$. 
Interestingly, in the regime where $G(m,q)$ is non-convex, instanton-like time-dependent trajectories connecting the two local minima become optimal, demonstrating 
dynamical coexistence between the different symmetry-broken phases and signaling a violation of the additivity principle in open systems \cite{bodineau04a,hurtado09c,hurtado10a,bodineau05a,Gorissen2012,hurtado11a,perez-espigares13a,perez-espigares16a}. A spectral analysis of the microscopic dynamical generator of the WASEP shows that the DPT is triggered by an emerging degeneracy of the associated ground state, from which one can compute the density profiles of the symmetry-broken phase. We provide also the first direct observation of this phenomenon through extensive rare-event simulations \cite{giardina06a,lecomte07a,tailleur09a,giardina11a,nemoto16a,ray17a}. 
This work opens the door to studying DPTs in more complex scenarios, as e.g. open high-dimensional systems with multiple conservation laws, and represents a step forward in connecting current fluctuations with metastability and standard critical phenomena.

\emph{Model.}-- The WASEP belongs to a broad class of driven diffusive systems of fundamental interest \cite{wasep1masi89,wasep2gartner87,derrida07a}. Microscopically it consists of a $1d$ lattice of $L$ sites, each of which may be empty or occupied by one particle at most. Particles hop randomly to empty neighboring left (right) sites at a rate $\frac{1}{2}\text{e}^{-E/L}$ ($\frac{1}{2}\text{e}^{E/L}$), with $E$ an external field. In addition, particles are injected and removed at the leftmost (rightmost) site at rates $\alpha$ and $\gamma$ ($\delta$ and $\beta$), respectively, yielding in the diffusive limit boundary particle densities of $\rho_{\text{L}}=\alpha/(\alpha+\gamma)$ and $\rho_{\text{R}}=\delta/(\beta+\delta)$. At the mesoscopic level, driven diffusive systems like WASEP are characterized by a density field $\rho(x,t)$ which obeys a stochastic equation \cite{Spohn1991}
\be
\partial_t \rho = -\partial_x \Big(-D(\rho)\partial_x \rho + \sigma(\rho) E + \xi (x,t) \Big) \, ,
\label{langevin1}
\ee
with $D(\rho)$ and $\sigma(\rho)$ the diffusivity and mobility coefficients, which for WASEP are $D(\rho)=1/2$ and $\sigma (\rho)=\rho (1-\rho)$. The field $j(x,t)=-D(\rho)\partial_x \rho + \sigma(\rho) E + \xi (x,t)$ stands for the fluctuating current, and $\xi$ is a Gaussian white noise, with $\la\xi\ra=0$ and 
$\la \xi(x,t)\xi(x',t')\ra=L^{-1} \sigma(\rho) \delta(x-x') \delta(t-t')$, which accounts for microscopic fluctuations at the mesoscopic level. The density at the boundaries is fixed to $\rho(0,t)=\rho_{\text{L}}$ and $\rho(1,t)=\rho_{\text{R}}$ $\forall t$. 

\emph{DPT in the thermodynamics of currents.}--
When driven by $E\neq 0$ and/or $\rho_{\text{L}} \neq \rho_{\text{R}}$, the system relaxes to a nonequilibrium steady state characterized by an average current $\la q\ra$ and a non-trivial density profile $\rho_{\text{st}}(x)$ \cite{SM}. Moreover, we can associate to any trajectory $\{\rho(x,t),j(x,t)\}_0^{\tau}$ an empirical current $q=\tau^{-1} \int_0^{\tau} dt \int_0^1 dx~j(x,t)$.  In the following we show how from the structure of the probability of this current, $P(q)$, we can predict the existence of DPTs associated with spontaneous symmetry breaking.

The probability $P(\{\rho,j\}_0^{\tau})$ of any trajectory can be computed from Eq. (\ref{langevin1}) via a path integral formalism \cite{hurtado14a,bertini15a,derrida07a}, and scales in the large-size limit
as $P(\{\rho,j\}_0^{\tau})\sim \exp\{-L\, {\cal I}_{\tau}[\rho,j]\}$, with an action \cite{bertini15a}
\be
{\cal I}_{\tau}[\rho,j] = \int_0^{\tau} dt \int_0^1 dx \frac{\displaystyle \Big(j+D(\rho)\partial_x\rho - \sigma(\rho) E \Big)^2}{\displaystyle 2\sigma(\rho)} \, .
\label{MFT1}
\ee
The probability $P(\{\rho,j\}_0^{\tau})$ represents the ensemble of space-time trajectories, from which one can obtain the statistics of any observable depending on $\{\rho,j\}_0^{\tau}$.  In particular the probability of a given current $q$ can be obtained by minimizing the action functional (\ref{MFT1}) over all trajectories sustaining such current. This yields in the long-time limit $P(q)\sim \exp\{-\tau L G(q)\}$, with $G(q)=\lim_{\tau \rightarrow \infty} \frac{1}{\tau}\min^*_{\{\rho,j\}_0^{\tau}}{\cal I}_{\tau} (\rho,j)$ the current LDF, and $^*$ meaning that the minimization must be compatible with the prescribed constraints ($q,\rho_{\text{L,R}}$). The optimal trajectories $\rho_q(x,t)$ and $j_q(x,t)$ solution of this variational problem are then those adopted by the system in order to maintain the current $q$ over a long period of time, and turn out to be time-independent in many cases (a conjecture known as {\em additivity principle} \cite{bodineau04a}).

Just as in standard critical phenomena, the action (\ref{MFT1}) contains the symmetries which are eventually broken. For WASEP with $\rho_{\text{R}}=1-\rho_{\text{L}}$ it is easy to check that the action (\ref{MFT1}) is invariant under the transformation $\rho \to 1-\rho$, $x\to 1-x$, referred to as PH symmetry (resulting from the symmetry of $\sigma(\rho)$ around $\rho=1/2$). The optimal density profile $\rho_q(x)$ typically inherits this PH symmetry, mapping onto itself under the above transformation. However, as detailed in the Supp. Mat. \cite{SM}, for currents below a critical threshold ($|q|\le|q_c|$) and large enough $E$, two different (but equally) optimal profiles $\rho_q^\pm(x)$ appear such that $\rho_q^\pm(x) \to 1-\rho_q^\mp(1-x)$, see inset to Fig.~\ref{fig1}, giving rise to a second-order singularity in the current LDF. This spontaneous PH symmetry breaking can be easily understood \cite{baek17a,baek18a} by noting that, in order to sustain a low-current fluctuation, the system can react by either crowding with particles hence hindering motion, or rather emptying the lattice to minimize particle flow. Both tendencies break the action PH symmetry, eventually triggering the DPT. 

\emph{Order parameter fluctuations.}-- 
To better understand this DPT, we study the joint fluctuations of the current and an appropriate global order parameter for the transition, much in the spirit of the paradigmatic Ising model of standard critical behavior \cite{binney92a}. A natural choice for this order parameter is the total mass in the system, which clearly characterizes the DPT in this case but also in more complex scenarios.
Indeed, as shown in Fig.~\ref{fig1}, the typical mass during a current fluctuation, $m_q \equiv \int_0^1 dx \rho_q(x)$, exhibits a behavior strongly reminiscent of a standard $\mathbb{Z}_2$ phase transition, capturing the PH symmetry breaking. Defining the empirical mass for a trajectory as $m=\tau^{-1}\int_0^{\tau} dt \int_0^1 dx \rho(x,t)$, the probability of observing a joint mass-current fluctuation for long times and large system sizes scales as $P(m,q) \sim \exp\{-\tau L G(m,q)\}$, with $G(m,q)=\lim_{\tau \rightarrow \infty} \frac{1}{\tau}\min^{*}_{\{\rho,j\}_0^{\tau}}{\cal I}_{\tau} (\rho,j)$ being the mass-current LDF, such that $G(q)=\min_m G(m,q) = G(m_q,q)$. Within the additivity hypothesis \cite{bodineau04a,Gorissen2012,hurtado14a}
\be
G(m,q) = \min_{\rho(x)} \int_0^1 dx  \frac{\displaystyle \Big(q+D(\rho)\partial_x\rho - \sigma(\rho) E \Big)^2}{\displaystyle 2\sigma(\rho)}\, ,
\label{mainLDF1}
\ee
with the optimal profile $\rho_{m,q}(x)$ subject to the constraint $m = \int_0^1 dx \rho_{m,q}(x)$ as well as to fixed boundary conditions. The mass constraint can be implemented using a Lagrange multiplier, and we solve analitycally the resulting problem in terms of elliptic integrals and Jacobi elliptic functions, see \cite{SM}. We note that the $\rho_{m,q}(x)$ so obtained can be classified attending to their extrema.

Fig.~\ref{fig2} illustrates our results for strong boundary gradients, well beyond the linear nonequilibrium regime. In particular, for PH-symmetric boundaries ($\rho_{\text{R}}=1-\rho_{\text{L}}$), the conditional mass-current LDF $G(m|q)\equiv G(m,q)-G(q)$ exhibits a peculiar change of behavior at a critical current $|q_c|$, see panel \ref{fig2}.a: while for $|q|>|q_c|$ the LDF $G(m|q)$ displays a single minimum at $m_q=1/2$, with an associated PH-symmetric optimal profile (Fig. \ref{fig2}.b), for $|q|<|q_c|$ two equivalent minima $m_q^\pm$ appear in $G(m|q)$, each one associated with a PH-symmetry-broken optimal profile $\rho_q^\pm(x)$, see Fig. \ref{fig2}.c, such that $\rho_q^\pm(x) \to 1-\rho_q^\mp(1-x)$. The emergence of this non-convex regime in $G(m|q)$ signals a $2^{\text{nd}}$-order DPT to a PH-symmetry-broken dynamical phase. On the other hand, for PH-asymmetric  boundaries ($\rho_{\text{R}}\neq 1-\rho_{\text{L}}$), the governing action (\ref{MFT1}) is no longer PH-symmetric: the asymmetry favors one of the mass branches and the associated $G(m|q)$ displays a single \emph{global} minimum $\forall q$ and an unique optimal profile (see Fig. \ref{fig2}.d-f), explaining why no DPT is observed in this case \cite{Gorissen2012}. Still, $G(m|q)$ becomes non-convex for low enough currents, and for weak gradient asymmetry metastable-like local minima in $G(m|q)$ may appear \cite{SM}. 

\emph{Maxwell construction and additivity violation.}-- 
A natural question is whether time-dependent optimal trajectories exist which improve the additivity principle minimizers. 
The emergence of a non-convex regime in $G(m|q)$ for $|q|<|q_c|$ suggests a Maxwell-like instantonic solution in this region \cite{Bray89,baek18a,vroylandt2018non}. In particular, as we show in \cite{SM}, for PH-symmetric boundaries, fixed $|q|<|q_c|$ and $m\in (m_q^-,m_q^+)$, a trajectory which jumps smoothly (in a finite time) from $\rho_q^-(x)$ to $\rho_q^+(x)$ at time $t_0=\tau p$, with $p \equiv |m-m_q^+|/(m_q^+-m_q^-)$,  improves the additivity principle solution, yielding a straight Maxwell-like construction $G(m|q) = p G(m_q^-|q) + (1-p) G(m_q^+|q)$ for $m\in (m_q^-,m_q^+)$. This corresponds to a \emph{dynamical coexistence} of the different symmetry-broken phases for $|q|<|q_c|$, as expected for a $1^{\text{st}}$-order DPT, see Fig.~\ref{fig1}. Similar solutions exist for PH-asymmetric boundaries in regimes where $G(m|q)$ is non-convex, leading to \emph{metastable} dynamical coexistence, and we note that the role of the instanton around $|q|\approx |q_c|$ can be affected by how the $L\to\infty$ and $\tau\to \infty$ limits are taken \cite{baek18a}.

\emph{Microscopic results: Spectral analysis.}--  
Next we focus on the microscopic understanding of the symmetry-breaking DPT for current statistics. At the microscopic level, a configuration of the $1d$ WASEP is given by $C=\{n_k\}_{k=1,\ldots,L}$, where $n_k=0,1$ is the occupation number of the lattice's $k^{\text{th}}$ site. Within the quantum Hamiltonian formalism for the master equation \cite{schutz01a}, each configuration is represented as a vector in a Hilbert space, $\ket{C}=\bigotimes_{k=1}^L (n_k, 1-n_k)^T$, with $^T$ denoting transposition. The complete information about the system is contained in a vector $\ket{P}=(P(C_1),P(C_2),...)^T=\sum_i P(C_i)\ket{C_i}$, with $P(C_i)$ representing the probabilities of the different configurations $C_i$. This probability vector evolves according to the master equation $\partial_t\ket{P}={\mathbb W} \ket{P}$, where ${\mathbb W}$ defines the Markov generator of the dynamics. Such generator can be \emph{tilted} ${\mathbb W}^{\mu,\lambda}$ \cite{SM,touchette09a,hurtado14a} to bias the original stochastic dynamics in order to favor large (low) mass for  $\mu<0$ ($\mu>0$) and large (low) currents for $\lambda > 0$ ($\lambda < 0$), with $\mu$ and $\lambda$ the conjugate parameters to the microscopic mass and current observables, respectively. The connection between the biased dynamics and the large deviation properties of our system is established through the largest eigenvalue of ${\mathbb W}^{\mu,\lambda}$ \cite{touchette09a,garrahan18a}. Such eigenvalue, denoted by $\theta_0(\mu,\lambda)$, is nothing but the cumulant generating function of the observables $m$ and $q$, related to the LDF $G(m,q)$ via a Legendre transform, $\theta_0(\mu,\lambda)=L^{-1}\max_{m,q}[\lambda q - \mu L m - G(m,q)]$.

\begin{figure}[t!]
\vspace{-0.2cm}\includegraphics[scale=0.28]{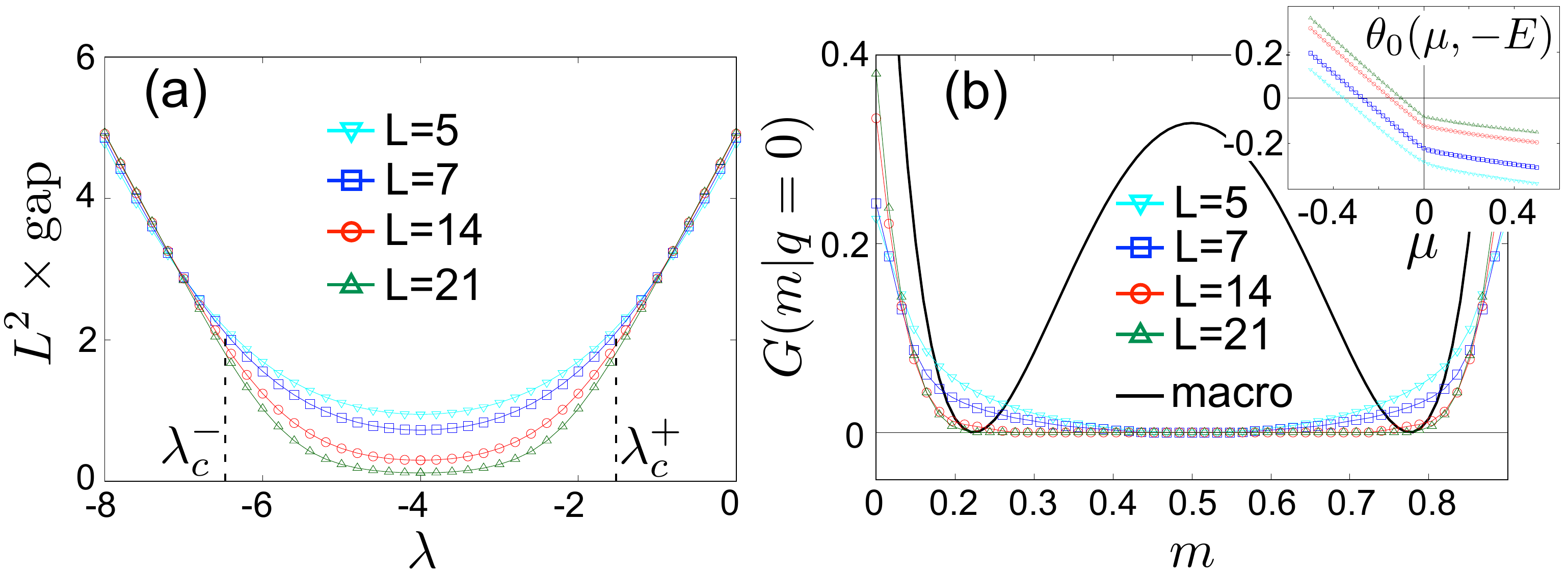}
\vspace{-0.4cm}
\caption{
(a) Scaled spectral gap of the tilted generator as a function of $\lambda$ for $\mu=0$ and different system sizes. (b) Main panel: LDF $G(m|q=0)$ obtained by Legendre transforming $L\theta_0(\mu,\lambda=-E)$ together with macroscopic predictions. For increasing system sizes the microscopic $G(m|q=0)$ converges to the convex envelope of the macroscopic prediction. Inset: $\theta_0(\mu,-E)$ for different system sizes. Notice the kink at $\mu=0$. In all cases $\rho_{\text{L}}=\rho_{\text{R}}=0.5$ and $E=4$.
}\label{fig3}
\end{figure}

We now consider exact numerical diagonalization of ${\mathbb W}^{\mu,\lambda}$ for a particular case of PH-symmetric boundaries and no mass bias ($\mu=0$). Fig.~\ref{fig3}.a shows that the diffusively-scaled spectral gap, $L^2[\theta_0(0,\lambda) - \theta_1(0,\lambda)]$, with $\theta_1(0,\lambda)$ the next-to-leading eigenvalue of ${\mathbb W}^{0,\lambda}$, tends to zero as $L$ increases in a region $\lambda_c^-<\lambda<\lambda_c^+$ (with $\lambda_c^{\pm}=-E\pm\sqrt{E^2-\pi^2}$) which corresponds to $|q|\le |q_c|=\sqrt{E^2-\pi^2}/4$ as predicted \cite{SM,baek17a}. This means that the $2^{\text{nd}}$-order DPT in current statistics unveiled above at the macroscopic level corresponds to an emerging degeneracy of the \emph{ground state} of ${\mathbb W}^{\mu,\lambda}$ (i.e. that corresponding to the leading eigenvalue), in which the sub-leading eigenvalue coalesces with the leading one. Moreover, by varying $\mu$ for $\lambda=-E$ (equiv. $q=0$) a remarkable $1^{\text{st}}$-order-like behavior associated with a kink of $\theta_0(\mu,\lambda=-E)$ at $\mu=0$ is found, see inset to Fig.~\ref{fig3}.b, consistent with the non-convex behavior of $G(m|q=0)$ found macroscopically and the associated dynamical coexistence of the two mass branches. Indeed, the numerical inverse Legendre transform of $\theta_0(\mu,\lambda=-E)$ converges to the convex envelope or Maxwell construction of the macroscopic prediction for $G(m|q=0)$, see Fig.~\ref{fig3}.b.

\begin{figure}[t!]
\vspace{0cm}\includegraphics[scale=0.245]{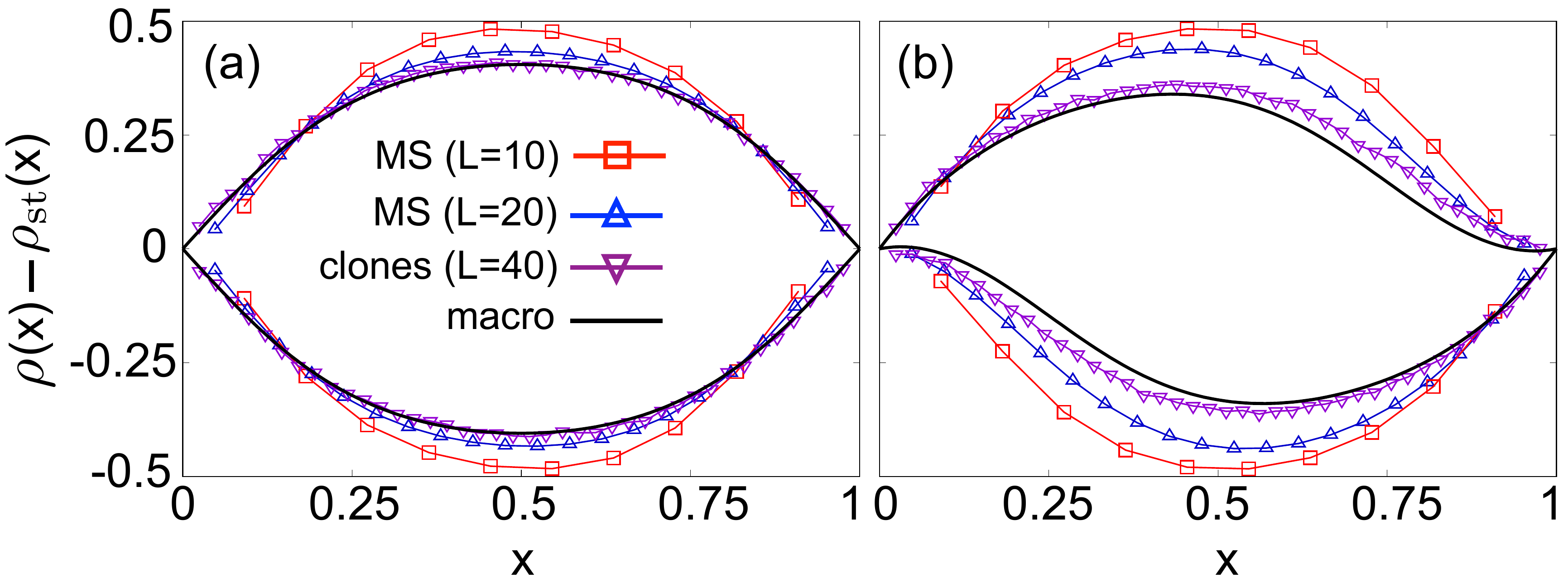}
\vspace{-0.4cm}
\caption{
(a) Optimal density profiles for the open WASEP with $\rho_{\text{L}}=\rho_{\text{R}}=0.5$ and $E=4$ conditioned to have a current $q=0$. Macroscopic predictions (black solid lines) and simulation results using the cloning algorithm for $L=40$ (purple down triangles).  Profiles associated with the extremal metastable states for $L=10$ (red squares) and $L=20$ (blue up triangles). (b) Same results for  $\rho_{\text{L}}=0.8$ and $\rho_{\text{R}}=0.2$. 
}\label{fig4}
\end{figure}

The eigenspace associated to $\theta_0(\mu,\lambda)$ contains the microscopic information about the typical trajectories responsible for a given fluctuation (as parametrized by $\lambda$ and $\mu$). In this way, the emergence of a degeneracy as $L$ increases points out to the appearance of two competing (symmetry-broken) states. For large but finite $L$, the spectral gap is small but non-zero and the eigenspace of $\theta_1(\mu,\lambda)$ defines a long-lived metastable state \cite{Gaveau06a,kurchan2009six,kasia16a,rose16a}. Using Doob's transform as a tool \cite{chetrite15a,carollo2018making}, one can show that any state in the degenerate (metastable) manifold is then given by a probability vector $\ket{P_{\text{MS}}^{c}} = \hat{L}_0 (\ket{R_0}+c \ket{R_1})$ \cite{SM}. Here $\ket{R_{i}}$ ($\ket{L_i}$) is the right (left) eigenvector associated with $\theta_i(\mu,\lambda)$ ($i=0,1$), and $\hat{L}_0$ is a diagonal matrix whose elements $(\hat{L}_0)_{ii}$ are the $i^{\text{th}}$ entries of $\ket{L_0}$. Moreover, $c\in[c_1,c_2]$ is a constant with $c_1$ ($c_2$) the smallest (largest) entry of the vector $\bra{L_1}\hat{L}_0^{-1}$. Interestingly, our microscopic approach shows that the high- and low-mass states in the symmetry-broken phase then correspond to the states $\ket{P_{\text{MS}}^{c_1,c_2}}$, from which the average density profile in each phase can be computed. Fig.~\ref{fig4} shows the profiles so obtained from the exact numerical diagonalization of ${\mathbb W}^{\mu,\lambda}$ for two different gradients, and the convergence to the macroscopic prediction as $L$ increases is clear.

\emph{Direct observation of the DPT.}--  
So far, we have obtained clear indications of a symmetry-breaking DPT both from a macroscopic approach and a microscopic (spectral) analysis.
The question remains as to whether this phenomenon is observable in simulations, which allow to reach larger system sizes. To address this we have performed extensive rare event simulations using the cloning Monte Carlo method \cite{giardina06a,lecomte07a,tailleur09a,giardina11a} to study current statistics in the open $1d$ WASEP. Starting from random initial configurations, we have measured the optimal density profiles adopted by the system to sustain a highly atypical current, namely $q=0$, using a population of $10^4$ clones and $L=40$. To capture the possible symmetry breaking, we average separately profiles with a total mass above and below $1/2$. Fig.~\ref{fig4}.a shows the result for $\rho_{\text{L}}=\rho_{\text{R}}=0.5$, while Fig.~\ref{fig4}.b displays data for $\rho_{\text{L}}=0.8$ and $\rho_{\text{R}}=0.2$ (in both cases $E=4$). The measured high- and low-mass optimal profiles again converge towards the macroscopic predictions, strongly supporting our results on the PH-symmetry-breaking scenario. 

\emph{Conclusions.}-- We have analyzed from a hydrodynamic, microscopic and computational point of view a $2^{\text{nd}}$-order DPT in the current statistis of a paradigmatic driven diffusive system, the open $1d$ WASEP, unveiling the full dynamical phase diagram for arbitrary current fluctuations and boundary driving. For that we have investigated the joint fluctuations of the current and a collective order parameter, the total mass in the system, finding that the associated LDF becomes non-convex for low enough currents. Microscopically, we link the observed DPT with an emerging degeneracy of the ground state of the tilted dynamical generator, from which the macroscopic optimal profiles can be computed. Our predictions are confirmed by the observation of this DPT phenomenon for the first time in rare event simulations.

We thank Vivien Lecomte for insightful discussions. The research leading to these results has received funding from the EPSRC Grant No. EP/M014266/1 and the Spanish Ministry MINECO project FIS2017-84256-P. C.P.E. acknowledges the funding received from the European Union's Horizon 2020 research and innovation programme under the Marie Sklodowska-Curie Cofund Programme Athenea3I Grant Agreement No. 754446. C.P.E, P.I.H and J.P.G acknowledge as well the hospitality and support of the International Centre for Theoretical Sciences (ICTS) in Bangalore (India), where part of this work was developed during the program \emph{Large deviation theory in statistical physics: Recent advances and future challenges} (Code: ICTS/Prog-ldt/2017/8). We are also grateful for access to the University of Nottingham High Performance Computing Facility, and for the computational resources and assistance provided by PROTEUS, the super-computing center of iC1 in Granada, Spain.

\let\oldaddcontentsline\addcontentsline
\renewcommand{\addcontentsline}[3]{}
\bibliography{referencias-BibDesk-v3}{}

\begin{thebibliography}{87}%
\makeatletter
\providecommand \@ifxundefined [1]{%
 \@ifx{#1\undefined}
}%
\providecommand \@ifnum [1]{%
 \ifnum #1\expandafter \@firstoftwo
 \else \expandafter \@secondoftwo
 \fi
}%
\providecommand \@ifx [1]{%
 \ifx #1\expandafter \@firstoftwo
 \else \expandafter \@secondoftwo
 \fi
}%
\providecommand \natexlab [1]{#1}%
\providecommand \enquote  [1]{``#1''}%
\providecommand \bibnamefont  [1]{#1}%
\providecommand \bibfnamefont [1]{#1}%
\providecommand \citenamefont [1]{#1}%
\providecommand \href@noop [0]{\@secondoftwo}%
\providecommand \href [0]{\begingroup \@sanitize@url \@href}%
\providecommand \@href[1]{\@@startlink{#1}\@@href}%
\providecommand \@@href[1]{\endgroup#1\@@endlink}%
\providecommand \@sanitize@url [0]{\catcode `\\12\catcode `\$12\catcode
  `\&12\catcode `\#12\catcode `\^12\catcode `\_12\catcode `\%12\relax}%
\providecommand \@@startlink[1]{}%
\providecommand \@@endlink[0]{}%
\providecommand \url  [0]{\begingroup\@sanitize@url \@url }%
\providecommand \@url [1]{\endgroup\@href {#1}{\urlprefix }}%
\providecommand \urlprefix  [0]{URL }%
\providecommand \Eprint [0]{\href }%
\providecommand \doibase [0]{http://dx.doi.org/}%
\providecommand \selectlanguage [0]{\@gobble}%
\providecommand \bibinfo  [0]{\@secondoftwo}%
\providecommand \bibfield  [0]{\@secondoftwo}%
\providecommand \translation [1]{[#1]}%
\providecommand \BibitemOpen [0]{}%
\providecommand \bibitemStop [0]{}%
\providecommand \bibitemNoStop [0]{.\EOS\space}%
\providecommand \EOS [0]{\spacefactor3000\relax}%
\providecommand \BibitemShut  [1]{\csname bibitem#1\endcsname}%
\let\auto@bib@innerbib\@empty
\bibitem [{\citenamefont {Bertini}\ \emph {et~al.}(2005)\citenamefont
  {Bertini}, \citenamefont {Sole}, \citenamefont {Gabrielli}, \citenamefont
  {Jona-Lasinio},\ and\ \citenamefont {Landim}}]{bertini05a}%
  \BibitemOpen
  \bibfield  {author} {\bibinfo {author} {\bibfnamefont {L.}~\bibnamefont
  {Bertini}}, \bibinfo {author} {\bibfnamefont {A.~De}\ \bibnamefont {Sole}},
  \bibinfo {author} {\bibfnamefont {D.}~\bibnamefont {Gabrielli}}, \bibinfo
  {author} {\bibfnamefont {G.}~\bibnamefont {Jona-Lasinio}}, \ and\ \bibinfo
  {author} {\bibfnamefont {C.}~\bibnamefont {Landim}},\ }\bibfield  {title}
  {\enquote {\bibinfo {title} {Current fluctuations in stochastic lattice
  gases},}\ }\href
  {http://journals.aps.org/prl/abstract/10.1103/PhysRevLett.94.030601}
  {\bibfield  {journal} {\bibinfo  {journal} {Phys. Rev. Lett.}\ }\textbf
  {\bibinfo {volume} {94}},\ \bibinfo {pages} {030601} (\bibinfo {year}
  {2005})}\BibitemShut {NoStop}%
\bibitem [{\citenamefont {Bodineau}\ and\ \citenamefont
  {Derrida}(2005)}]{bodineau05a}%
  \BibitemOpen
  \bibfield  {author} {\bibinfo {author} {\bibfnamefont {T.}~\bibnamefont
  {Bodineau}}\ and\ \bibinfo {author} {\bibfnamefont {B.}~\bibnamefont
  {Derrida}},\ }\bibfield  {title} {\enquote {\bibinfo {title} {Distribution of
  current in nonequilibrium diffusive systems and phase transitions},}\ }\href
  {http://journals.aps.org/pre/abstract/10.1103/PhysRevE.72.066110} {\bibfield
  {journal} {\bibinfo  {journal} {Phys. Rev. E}\ }\textbf {\bibinfo {volume}
  {72}},\ \bibinfo {pages} {066110} (\bibinfo {year} {2005})}\BibitemShut
  {NoStop}%
\bibitem [{\citenamefont {Harris}\ \emph {et~al.}(2005)\citenamefont {Harris},
  \citenamefont {Rakos},\ and\ \citenamefont {Schutz}}]{harris05a}%
  \BibitemOpen
  \bibfield  {author} {\bibinfo {author} {\bibfnamefont {R.~J.}\ \bibnamefont
  {Harris}}, \bibinfo {author} {\bibfnamefont {A.}~\bibnamefont {Rakos}}, \
  and\ \bibinfo {author} {\bibfnamefont {G.~M.}\ \bibnamefont {Schutz}},\
  }\bibfield  {title} {\enquote {\bibinfo {title} {Current fluctuations in the
  zero-range process with open boundaries},}\ }\href@noop {} {\bibfield
  {journal} {\bibinfo  {journal} {J. Stat. Mech.}\ ,\ \bibinfo {pages}
  {P08003}} (\bibinfo {year} {2005})}\BibitemShut {NoStop}%
\bibitem [{\citenamefont {Bertini}\ \emph {et~al.}(2006)\citenamefont
  {Bertini}, \citenamefont {Sole}, \citenamefont {Gabrielli}, \citenamefont
  {Jona-Lasinio},\ and\ \citenamefont {Landim}}]{bertini06a}%
  \BibitemOpen
  \bibfield  {author} {\bibinfo {author} {\bibfnamefont {L.}~\bibnamefont
  {Bertini}}, \bibinfo {author} {\bibfnamefont {A.~De}\ \bibnamefont {Sole}},
  \bibinfo {author} {\bibfnamefont {D.}~\bibnamefont {Gabrielli}}, \bibinfo
  {author} {\bibfnamefont {G.}~\bibnamefont {Jona-Lasinio}}, \ and\ \bibinfo
  {author} {\bibfnamefont {C.}~\bibnamefont {Landim}},\ }\bibfield  {title}
  {\enquote {\bibinfo {title} {Nonequilibrium current fluctuations in
  stochastic lattice gases},}\ }\href
  {http://link.springer.com/article/10.1007/s10955-006-9056-4} {\bibfield
  {journal} {\bibinfo  {journal} {J. Stat. Phys.}\ }\textbf {\bibinfo {volume}
  {123}},\ \bibinfo {pages} {237--276} (\bibinfo {year} {2006})}\BibitemShut
  {NoStop}%
\bibitem [{\citenamefont {Bodineau}\ and\ \citenamefont
  {Derrida}(2007)}]{bodineau07a}%
  \BibitemOpen
  \bibfield  {author} {\bibinfo {author} {\bibfnamefont {T.}~\bibnamefont
  {Bodineau}}\ and\ \bibinfo {author} {\bibfnamefont {B.}~\bibnamefont
  {Derrida}},\ }\bibfield  {title} {\enquote {\bibinfo {title} {Cumulants and
  large deviations of the current through non-equilibrium steady states},}\
  }\href {\doibase https://doi.org/10.1016/j.crhy.2007.04.014} {\bibfield
  {journal} {\bibinfo  {journal} {C.R. Phys.}\ }\textbf {\bibinfo {volume}
  {8}},\ \bibinfo {pages} {540 -- 555} (\bibinfo {year} {2007})}\BibitemShut
  {NoStop}%
\bibitem [{\citenamefont {Lecomte}\ \emph {et~al.}(2007)\citenamefont
  {Lecomte}, \citenamefont {Appert-Rolland},\ and\ \citenamefont {van
  Wijland}}]{Lecomte2007}%
  \BibitemOpen
  \bibfield  {author} {\bibinfo {author} {\bibfnamefont {V.}~\bibnamefont
  {Lecomte}}, \bibinfo {author} {\bibfnamefont {C.}~\bibnamefont
  {Appert-Rolland}}, \ and\ \bibinfo {author} {\bibfnamefont {F.}~\bibnamefont
  {van Wijland}},\ }\bibfield  {title} {\enquote {\bibinfo {title}
  {Thermodynamic formalism for systems with {M}arkov dynamics},}\ }\href
  {\doibase 10.1007/s10955-006-9254-0} {\bibfield  {journal} {\bibinfo
  {journal} {J. Stat. Phys.}\ }\textbf {\bibinfo {volume} {127}},\ \bibinfo
  {pages} {51} (\bibinfo {year} {2007})}\BibitemShut {NoStop}%
\bibitem [{\citenamefont {Garrahan}\ \emph {et~al.}(2007)\citenamefont
  {Garrahan}, \citenamefont {Jack}, \citenamefont {Lecomte}, \citenamefont
  {Pitard}, \citenamefont {van Duijvendijk},\ and\ \citenamefont {van
  Wijland}}]{garrahan07a}%
  \BibitemOpen
  \bibfield  {author} {\bibinfo {author} {\bibfnamefont {J.~P.}\ \bibnamefont
  {Garrahan}}, \bibinfo {author} {\bibfnamefont {R.~L.}\ \bibnamefont {Jack}},
  \bibinfo {author} {\bibfnamefont {V.}~\bibnamefont {Lecomte}}, \bibinfo
  {author} {\bibfnamefont {E.}~\bibnamefont {Pitard}}, \bibinfo {author}
  {\bibfnamefont {K.}~\bibnamefont {van Duijvendijk}}, \ and\ \bibinfo {author}
  {\bibfnamefont {F.}~\bibnamefont {van Wijland}},\ }\bibfield  {title}
  {\enquote {\bibinfo {title} {Dynamical first-order phase transition in
  kinetically constrained models of glasses},}\ }\href
  {http://journals.aps.org/prl/abstract/10.1103/PhysRevLett.98.195702}
  {\bibfield  {journal} {\bibinfo  {journal} {Phys. Rev. Lett.}\ }\textbf
  {\bibinfo {volume} {98}},\ \bibinfo {pages} {195702} (\bibinfo {year}
  {2007})}\BibitemShut {NoStop}%
\bibitem [{\citenamefont {Garrahan}\ \emph {et~al.}(2009)\citenamefont
  {Garrahan}, \citenamefont {Jack}, \citenamefont {Lecomte}, \citenamefont
  {Pitard}, \citenamefont {van Duijvendijk},\ and\ \citenamefont {van
  Wijland}}]{garrahan09a}%
  \BibitemOpen
  \bibfield  {author} {\bibinfo {author} {\bibfnamefont {J.~P.}\ \bibnamefont
  {Garrahan}}, \bibinfo {author} {\bibfnamefont {R.~L.}\ \bibnamefont {Jack}},
  \bibinfo {author} {\bibfnamefont {V.}~\bibnamefont {Lecomte}}, \bibinfo
  {author} {\bibfnamefont {E.}~\bibnamefont {Pitard}}, \bibinfo {author}
  {\bibfnamefont {K.}~\bibnamefont {van Duijvendijk}}, \ and\ \bibinfo {author}
  {\bibfnamefont {F.}~\bibnamefont {van Wijland}},\ }\bibfield  {title}
  {\enquote {\bibinfo {title} {First-order dynamical phase transition in models
  of glasses: an approach based on ensembles of histories},}\ }\href
  {http://iopscience.iop.org/article/10.1088/1751-8113/42/7/075007} {\bibfield
  {journal} {\bibinfo  {journal} {J. Phys. A}\ }\textbf {\bibinfo {volume}
  {42}},\ \bibinfo {pages} {075007} (\bibinfo {year} {2009})}\BibitemShut
  {NoStop}%
\bibitem [{\citenamefont {Hurtado}\ and\ \citenamefont
  {Garrido}(2011)}]{hurtado11a}%
  \BibitemOpen
  \bibfield  {author} {\bibinfo {author} {\bibfnamefont {P.~I.}\ \bibnamefont
  {Hurtado}}\ and\ \bibinfo {author} {\bibfnamefont {P.~L.}\ \bibnamefont
  {Garrido}},\ }\bibfield  {title} {\enquote {\bibinfo {title} {Spontaneous
  symmetry breaking at the fluctuating level},}\ }\href
  {http://journals.aps.org/prl/abstract/10.1103/PhysRevLett.107.180601}
  {\bibfield  {journal} {\bibinfo  {journal} {Phys. Rev. Lett.}\ }\textbf
  {\bibinfo {volume} {107}},\ \bibinfo {pages} {180601} (\bibinfo {year}
  {2011})}\BibitemShut {NoStop}%
\bibitem [{\citenamefont {Ates}\ \emph {et~al.}(2012)\citenamefont {Ates},
  \citenamefont {Olmos}, \citenamefont {Garrahan},\ and\ \citenamefont
  {Lesanovsky}}]{ates12a}%
  \BibitemOpen
  \bibfield  {author} {\bibinfo {author} {\bibfnamefont {C.}~\bibnamefont
  {Ates}}, \bibinfo {author} {\bibfnamefont {B.}~\bibnamefont {Olmos}},
  \bibinfo {author} {\bibfnamefont {J.~P.}\ \bibnamefont {Garrahan}}, \ and\
  \bibinfo {author} {\bibfnamefont {I.}~\bibnamefont {Lesanovsky}},\ }\bibfield
   {title} {\enquote {\bibinfo {title} {Dynamical phases and intermittency of
  the dissipative quantum {Ising} model},}\ }\href
  {http://journals.aps.org/pra/abstract/10.1103/PhysRevA.85.043620} {\bibfield
  {journal} {\bibinfo  {journal} {Phys. Rev. A}\ }\textbf {\bibinfo {volume}
  {85}},\ \bibinfo {pages} {043620} (\bibinfo {year} {2012})}\BibitemShut
  {NoStop}%
\bibitem [{\citenamefont {P\'erez-Espigares}\ \emph {et~al.}(2013)\citenamefont
  {P\'erez-Espigares}, \citenamefont {Garrido},\ and\ \citenamefont
  {Hurtado}}]{perez-espigares13a}%
  \BibitemOpen
  \bibfield  {author} {\bibinfo {author} {\bibfnamefont {C.}~\bibnamefont
  {P\'erez-Espigares}}, \bibinfo {author} {\bibfnamefont {P.~L.}\ \bibnamefont
  {Garrido}}, \ and\ \bibinfo {author} {\bibfnamefont {P.~I.}\ \bibnamefont
  {Hurtado}},\ }\bibfield  {title} {\enquote {\bibinfo {title} {Dynamical phase
  transition for current statistics in a simple driven diffusive system},}\
  }\href {http://journals.aps.org/pre/abstract/10.1103/PhysRevE.87.032115}
  {\bibfield  {journal} {\bibinfo  {journal} {Phys. Rev. E}\ }\textbf {\bibinfo
  {volume} {87}},\ \bibinfo {pages} {032115} (\bibinfo {year}
  {2013})}\BibitemShut {NoStop}%
\bibitem [{\citenamefont {Harris}\ \emph {et~al.}(2013)\citenamefont {Harris},
  \citenamefont {Popkov},\ and\ \citenamefont {Sch{\"u}tz}}]{harris13a}%
  \BibitemOpen
  \bibfield  {author} {\bibinfo {author} {\bibfnamefont {R.~J.}\ \bibnamefont
  {Harris}}, \bibinfo {author} {\bibfnamefont {V.}~\bibnamefont {Popkov}}, \
  and\ \bibinfo {author} {\bibfnamefont {G.~M.}\ \bibnamefont {Sch{\"u}tz}},\
  }\bibfield  {title} {\enquote {\bibinfo {title} {Dynamics of instantaneous
  condensation in the {ZRP} conditioned on an atypical current},}\ }\href
  {\doibase 10.3390/e15115065} {\bibfield  {journal} {\bibinfo  {journal}
  {Entropy}\ }\textbf {\bibinfo {volume} {15}},\ \bibinfo {pages} {5065}
  (\bibinfo {year} {2013})}\BibitemShut {NoStop}%
\bibitem [{\citenamefont {Vaikuntanathan}\ \emph {et~al.}(2014)\citenamefont
  {Vaikuntanathan}, \citenamefont {Gingrich},\ and\ \citenamefont
  {Geissler}}]{vaikuntanathan14a}%
  \BibitemOpen
  \bibfield  {author} {\bibinfo {author} {\bibfnamefont {S.}~\bibnamefont
  {Vaikuntanathan}}, \bibinfo {author} {\bibfnamefont {T.~R.}\ \bibnamefont
  {Gingrich}}, \ and\ \bibinfo {author} {\bibfnamefont {P.~L.}\ \bibnamefont
  {Geissler}},\ }\bibfield  {title} {\enquote {\bibinfo {title} {Dynamic phase
  transitions in simple driven kinetic networks},}\ }\href
  {http://journals.aps.org/pre/abstract/10.1103/PhysRevE.89.062108} {\bibfield
  {journal} {\bibinfo  {journal} {Phys. Rev. E}\ }\textbf {\bibinfo {volume}
  {89}},\ \bibinfo {pages} {062108} (\bibinfo {year} {2014})}\BibitemShut
  {NoStop}%
\bibitem [{\citenamefont {Mey}\ \emph {et~al.}(2014)\citenamefont {Mey},
  \citenamefont {Geissler},\ and\ \citenamefont {Garrahan}}]{mey14a}%
  \BibitemOpen
  \bibfield  {author} {\bibinfo {author} {\bibfnamefont {A.~S. J.~S.}\
  \bibnamefont {Mey}}, \bibinfo {author} {\bibfnamefont {P.~L.}\ \bibnamefont
  {Geissler}}, \ and\ \bibinfo {author} {\bibfnamefont {J.~P.}\ \bibnamefont
  {Garrahan}},\ }\bibfield  {title} {\enquote {\bibinfo {title} {Rare-event
  trajectory ensemble analysis reveals metastable dynamical phases in lattice
  proteins},}\ }\href@noop {} {\bibfield  {journal} {\bibinfo  {journal}
  {Physical Review E}\ }\textbf {\bibinfo {volume} {89}},\ \bibinfo {pages}
  {032109} (\bibinfo {year} {2014})}\BibitemShut {NoStop}%
\bibitem [{\citenamefont {Jack}\ \emph {et~al.}(2015)\citenamefont {Jack},
  \citenamefont {Thompson},\ and\ \citenamefont {Sollich}}]{jack15a}%
  \BibitemOpen
  \bibfield  {author} {\bibinfo {author} {\bibfnamefont {R.~L.}\ \bibnamefont
  {Jack}}, \bibinfo {author} {\bibfnamefont {I.~R.}\ \bibnamefont {Thompson}},
  \ and\ \bibinfo {author} {\bibfnamefont {P.}~\bibnamefont {Sollich}},\
  }\bibfield  {title} {\enquote {\bibinfo {title} {Hyperuniformity and phase
  separation in biased ensembles of trajectories for diffusive systems},}\
  }\href {http://journals.aps.org/prl/abstract/10.1103/PhysRevLett.114.060601}
  {\bibfield  {journal} {\bibinfo  {journal} {Phys. Rev. Lett.}\ }\textbf
  {\bibinfo {volume} {114}},\ \bibinfo {pages} {060601} (\bibinfo {year}
  {2015})}\BibitemShut {NoStop}%
\bibitem [{\citenamefont {Baek}\ and\ \citenamefont {Kafri}(2015)}]{baek15a}%
  \BibitemOpen
  \bibfield  {author} {\bibinfo {author} {\bibfnamefont {Y.}~\bibnamefont
  {Baek}}\ and\ \bibinfo {author} {\bibfnamefont {Y.}~\bibnamefont {Kafri}},\
  }\bibfield  {title} {\enquote {\bibinfo {title} {Singularities in large
  deviation functions},}\ }\href
  {http://stacks.iop.org/1742-5468/2015/i=8/a=P08026} {\bibfield  {journal}
  {\bibinfo  {journal} {J. Stat. Mech.}\ }\textbf {\bibinfo {volume} {2015}},\
  \bibinfo {pages} {P08026} (\bibinfo {year} {2015})}\BibitemShut {NoStop}%
\bibitem [{\citenamefont {Nyawo}\ and\ \citenamefont
  {Touchette}(2016)}]{tsobgni16a}%
  \BibitemOpen
  \bibfield  {author} {\bibinfo {author} {\bibfnamefont {O.~Tsobgni}\
  \bibnamefont {Nyawo}}\ and\ \bibinfo {author} {\bibfnamefont
  {H.}~\bibnamefont {Touchette}},\ }\bibfield  {title} {\enquote {\bibinfo
  {title} {A minimal model of dynamical phase transition},}\ }\href
  {http://stacks.iop.org/0295-5075/116/i=5/a=50009} {\bibfield  {journal}
  {\bibinfo  {journal} {Europhys. Lett.}\ }\textbf {\bibinfo {volume} {116}},\
  \bibinfo {pages} {50009} (\bibinfo {year} {2016})}\BibitemShut {NoStop}%
\bibitem [{\citenamefont {Harris}\ and\ \citenamefont
  {Touchette}(2017)}]{harris17a}%
  \BibitemOpen
  \bibfield  {author} {\bibinfo {author} {\bibfnamefont {R.~J.}\ \bibnamefont
  {Harris}}\ and\ \bibinfo {author} {\bibfnamefont {H.}~\bibnamefont
  {Touchette}},\ }\bibfield  {title} {\enquote {\bibinfo {title} {Phase
  transitions in large deviations of reset processes},}\ }\href
  {http://stacks.iop.org/1751-8121/50/i=10/a=10LT01} {\bibfield  {journal}
  {\bibinfo  {journal} {J. Phys. A}\ }\textbf {\bibinfo {volume} {50}},\
  \bibinfo {pages} {10LT01} (\bibinfo {year} {2017})}\BibitemShut {NoStop}%
\bibitem [{\citenamefont {Lazarescu}(2017)}]{lazarescu17a}%
  \BibitemOpen
  \bibfield  {author} {\bibinfo {author} {\bibfnamefont {A.}~\bibnamefont
  {Lazarescu}},\ }\bibfield  {title} {\enquote {\bibinfo {title} {Generic
  dynamical phase transition in one-dimensional bulk-driven lattice gases with
  exclusion},}\ }\href {http://stacks.iop.org/1751-8121/50/i=25/a=254004}
  {\bibfield  {journal} {\bibinfo  {journal} {J. Phys. A}\ }\textbf {\bibinfo
  {volume} {50}},\ \bibinfo {pages} {254004} (\bibinfo {year}
  {2017})}\BibitemShut {NoStop}%
\bibitem [{\citenamefont {Brandner}\ \emph {et~al.}(2017)\citenamefont
  {Brandner}, \citenamefont {Maisi}, \citenamefont {Pekola}, \citenamefont
  {Garrahan},\ and\ \citenamefont {Flindt}}]{brandner17a}%
  \BibitemOpen
  \bibfield  {author} {\bibinfo {author} {\bibfnamefont {K.}~\bibnamefont
  {Brandner}}, \bibinfo {author} {\bibfnamefont {V.F.}\ \bibnamefont {Maisi}},
  \bibinfo {author} {\bibfnamefont {J.P.}\ \bibnamefont {Pekola}}, \bibinfo
  {author} {\bibfnamefont {J.P.}\ \bibnamefont {Garrahan}}, \ and\ \bibinfo
  {author} {\bibfnamefont {C.}~\bibnamefont {Flindt}},\ }\bibfield  {title}
  {\enquote {\bibinfo {title} {Experimental determination of dynamical
  {L}ee-{Y}ang zeros},}\ }\href
  {https://journals.aps.org/prl/abstract/10.1103/PhysRevLett.118.180601}
  {\bibfield  {journal} {\bibinfo  {journal} {Phys. Rev. Lett.}\ }\textbf
  {\bibinfo {volume} {118}} (\bibinfo {year} {2017})}\BibitemShut {NoStop}%
\bibitem [{\citenamefont {Karevski}\ and\ \citenamefont
  {Sch\"utz}(2017)}]{karevski17a}%
  \BibitemOpen
  \bibfield  {author} {\bibinfo {author} {\bibfnamefont {D.}~\bibnamefont
  {Karevski}}\ and\ \bibinfo {author} {\bibfnamefont {G.M.}\ \bibnamefont
  {Sch\"utz}},\ }\bibfield  {title} {\enquote {\bibinfo {title} {Conformal
  invariance in driven diffusive systems at high currents},}\ }\href
  {http://journals.aps.org/prl/abstract/10.1103/PhysRevLett.118.030601}
  {\bibfield  {journal} {\bibinfo  {journal} {Phys. Rev. Lett.}\ }\textbf
  {\bibinfo {volume} {118}} (\bibinfo {year} {2017})}\BibitemShut {NoStop}%
\bibitem [{\citenamefont {Carollo}\ \emph {et~al.}(2017)\citenamefont
  {Carollo}, \citenamefont {Garrahan}, \citenamefont {Lesanovsky},\ and\
  \citenamefont {P\'erez-Espigares}}]{carollo17a}%
  \BibitemOpen
  \bibfield  {author} {\bibinfo {author} {\bibfnamefont {F.}~\bibnamefont
  {Carollo}}, \bibinfo {author} {\bibfnamefont {J.~P.}\ \bibnamefont
  {Garrahan}}, \bibinfo {author} {\bibfnamefont {I.}~\bibnamefont
  {Lesanovsky}}, \ and\ \bibinfo {author} {\bibfnamefont {C.}~\bibnamefont
  {P\'erez-Espigares}},\ }\bibfield  {title} {\enquote {\bibinfo {title}
  {Fluctuating hydrodynamics, current fluctuations, and hyperuniformity in
  boundary-driven open quantum chains},}\ }\href {\doibase
  10.1103/PhysRevE.96.052118} {\bibfield  {journal} {\bibinfo  {journal} {Phys.
  Rev. E}\ }\textbf {\bibinfo {volume} {96}},\ \bibinfo {pages} {052118}
  (\bibinfo {year} {2017})}\BibitemShut {NoStop}%
\bibitem [{\citenamefont {Baek}\ \emph {et~al.}(2017)\citenamefont {Baek},
  \citenamefont {Kafri},\ and\ \citenamefont {Lecomte}}]{baek17a}%
  \BibitemOpen
  \bibfield  {author} {\bibinfo {author} {\bibfnamefont {Y.}~\bibnamefont
  {Baek}}, \bibinfo {author} {\bibfnamefont {Y.}~\bibnamefont {Kafri}}, \ and\
  \bibinfo {author} {\bibfnamefont {V.}~\bibnamefont {Lecomte}},\ }\bibfield
  {title} {\enquote {\bibinfo {title} {Dynamical symmetry breaking and phase
  transitions in driven diffusive systems},}\ }\href
  {http://journals.aps.org/prl/abstract/10.1103/PhysRevLett.118.030604}
  {\bibfield  {journal} {\bibinfo  {journal} {Phys. Rev. Lett.}\ }\textbf
  {\bibinfo {volume} {118}},\ \bibinfo {pages} {030604} (\bibinfo {year}
  {2017})}\BibitemShut {NoStop}%
\bibitem [{\citenamefont {Tiz\'on-Escamilla}\ \emph {et~al.}(2017)\citenamefont
  {Tiz\'on-Escamilla}, \citenamefont {P\'erez-Espigares}, \citenamefont
  {Garrido},\ and\ \citenamefont {Hurtado}}]{tizon-escamilla17b}%
  \BibitemOpen
  \bibfield  {author} {\bibinfo {author} {\bibfnamefont {N.}~\bibnamefont
  {Tiz\'on-Escamilla}}, \bibinfo {author} {\bibfnamefont {C.}~\bibnamefont
  {P\'erez-Espigares}}, \bibinfo {author} {\bibfnamefont {P.~L.}\ \bibnamefont
  {Garrido}}, \ and\ \bibinfo {author} {\bibfnamefont {P.~I.}\ \bibnamefont
  {Hurtado}},\ }\bibfield  {title} {\enquote {\bibinfo {title} {Order and
  symmetry breaking in the fluctuations of driven systems},}\ }\href {\doibase
  10.1103/PhysRevLett.119.090602} {\bibfield  {journal} {\bibinfo  {journal}
  {Phys. Rev. Lett.}\ }\textbf {\bibinfo {volume} {119}},\ \bibinfo {pages}
  {090602} (\bibinfo {year} {2017})}\BibitemShut {NoStop}%
\bibitem [{\citenamefont {Shpielberg}(2017)}]{shpielberg17a}%
  \BibitemOpen
  \bibfield  {author} {\bibinfo {author} {\bibfnamefont {O.}~\bibnamefont
  {Shpielberg}},\ }\bibfield  {title} {\enquote {\bibinfo {title} {Geometrical
  interpretation of dynamical phase transitions in boundary-driven systems},}\
  }\href {\doibase 10.1103/PhysRevE.96.062108} {\bibfield  {journal} {\bibinfo
  {journal} {Phys. Rev. E}\ }\textbf {\bibinfo {volume} {96}},\ \bibinfo
  {pages} {062108} (\bibinfo {year} {2017})}\BibitemShut {NoStop}%
\bibitem [{\citenamefont {Baek}\ \emph {et~al.}(2018)\citenamefont {Baek},
  \citenamefont {Kafri},\ and\ \citenamefont {Lecomte}}]{baek18a}%
  \BibitemOpen
  \bibfield  {author} {\bibinfo {author} {\bibfnamefont {Y.}~\bibnamefont
  {Baek}}, \bibinfo {author} {\bibfnamefont {Y.}~\bibnamefont {Kafri}}, \ and\
  \bibinfo {author} {\bibfnamefont {V.}~\bibnamefont {Lecomte}},\ }\bibfield
  {title} {\enquote {\bibinfo {title} {Dynamical phase transitions in the
  current distribution of driven diffusive channels},}\ }\href
  {http://stacks.iop.org/1751-8121/51/i=10/a=105001} {\bibfield  {journal}
  {\bibinfo  {journal} {J. Phys. A}\ }\textbf {\bibinfo {volume} {51}},\
  \bibinfo {pages} {105001} (\bibinfo {year} {2018})}\BibitemShut {NoStop}%
\bibitem [{\citenamefont {Shpielberg}\ \emph {et~al.}(2018)\citenamefont
  {Shpielberg}, \citenamefont {Nemoto},\ and\ \citenamefont
  {Caetano}}]{shpielberg18a}%
  \BibitemOpen
  \bibfield  {author} {\bibinfo {author} {\bibfnamefont {O.}~\bibnamefont
  {Shpielberg}}, \bibinfo {author} {\bibfnamefont {T.}~\bibnamefont {Nemoto}},
  \ and\ \bibinfo {author} {\bibfnamefont {J.}~\bibnamefont {Caetano}},\
  }\bibfield  {title} {\enquote {\bibinfo {title} {Universality in dynamical
  phase transitions of diffusive systems},}\ }\href {\doibase
  10.1103/PhysRevE.98.052116} {\bibfield  {journal} {\bibinfo  {journal} {Phys.
  Rev. E}\ }\textbf {\bibinfo {volume} {98}},\ \bibinfo {pages} {052116}
  (\bibinfo {year} {2018})}\BibitemShut {NoStop}%
\bibitem [{\citenamefont {P\'erez-Espigares}\ \emph {et~al.}(2018)\citenamefont
  {P\'erez-Espigares}, \citenamefont {Lesanovsky}, \citenamefont {Garrahan},\
  and\ \citenamefont {Guti\'errez}}]{perez2018glassy}%
  \BibitemOpen
  \bibfield  {author} {\bibinfo {author} {\bibfnamefont {C.}~\bibnamefont
  {P\'erez-Espigares}}, \bibinfo {author} {\bibfnamefont {I.}~\bibnamefont
  {Lesanovsky}}, \bibinfo {author} {\bibfnamefont {J.~P.}\ \bibnamefont
  {Garrahan}}, \ and\ \bibinfo {author} {\bibfnamefont {R.}~\bibnamefont
  {Guti\'errez}},\ }\bibfield  {title} {\enquote {\bibinfo {title} {Glassy
  dynamics due to a trajectory phase transition in dissipative rydberg
  gases},}\ }\href {\doibase 10.1103/PhysRevA.98.021804} {\bibfield  {journal}
  {\bibinfo  {journal} {Phys. Rev. A}\ }\textbf {\bibinfo {volume} {98}},\
  \bibinfo {pages} {021804} (\bibinfo {year} {2018})}\BibitemShut {NoStop}%
\bibitem [{\citenamefont {Chleboun}\ \emph {et~al.}(2018)\citenamefont
  {Chleboun}, \citenamefont {Grosskinsky},\ and\ \citenamefont
  {Pizzoferrato}}]{chleboun2018}%
  \BibitemOpen
  \bibfield  {author} {\bibinfo {author} {\bibfnamefont {P.}~\bibnamefont
  {Chleboun}}, \bibinfo {author} {\bibfnamefont {S.}~\bibnamefont
  {Grosskinsky}}, \ and\ \bibinfo {author} {\bibfnamefont {A.}~\bibnamefont
  {Pizzoferrato}},\ }\bibfield  {title} {\enquote {\bibinfo {title} {Current
  large deviations for partially asymmetric particle systems on a ring},}\
  }\href {http://stacks.iop.org/1751-8121/51/i=40/a=405001} {\bibfield
  {journal} {\bibinfo  {journal} {J. Phys. A}\ }\textbf {\bibinfo {volume}
  {51}},\ \bibinfo {pages} {405001} (\bibinfo {year} {2018})}\BibitemShut
  {NoStop}%
\bibitem [{\citenamefont {Klymko}\ \emph {et~al.}(2018)\citenamefont {Klymko},
  \citenamefont {Geissler}, \citenamefont {Garrahan},\ and\ \citenamefont
  {Whitelam}}]{Klymko2018}%
  \BibitemOpen
  \bibfield  {author} {\bibinfo {author} {\bibfnamefont {K.}~\bibnamefont
  {Klymko}}, \bibinfo {author} {\bibfnamefont {P.~L.}\ \bibnamefont
  {Geissler}}, \bibinfo {author} {\bibfnamefont {J.~P.}\ \bibnamefont
  {Garrahan}}, \ and\ \bibinfo {author} {\bibfnamefont {S.}~\bibnamefont
  {Whitelam}},\ }\bibfield  {title} {\enquote {\bibinfo {title} {Rare behavior
  of growth processes via umbrella sampling of trajectories},}\ }\href
  {\doibase 10.1103/PhysRevE.97.032123} {\bibfield  {journal} {\bibinfo
  {journal} {Phys. Rev. E}\ }\textbf {\bibinfo {volume} {97}},\ \bibinfo
  {pages} {032123} (\bibinfo {year} {2018})}\BibitemShut {NoStop}%
\bibitem [{\citenamefont {Whitelam}(2018)}]{Whitelam2018}%
  \BibitemOpen
  \bibfield  {author} {\bibinfo {author} {\bibfnamefont {S.}~\bibnamefont
  {Whitelam}},\ }\bibfield  {title} {\enquote {\bibinfo {title} {Large
  deviations in the presence of cooperativity and slow dynamics},}\ }\href
  {\doibase 10.1103/PhysRevE.97.062109} {\bibfield  {journal} {\bibinfo
  {journal} {Phys. Rev. E}\ }\textbf {\bibinfo {volume} {97}},\ \bibinfo
  {pages} {062109} (\bibinfo {year} {2018})}\BibitemShut {NoStop}%
\bibitem [{\citenamefont {Vroylandt}\ and\ \citenamefont
  {Verley}(2018)}]{vroylandt2018non}%
  \BibitemOpen
  \bibfield  {author} {\bibinfo {author} {\bibfnamefont {H.}~\bibnamefont
  {Vroylandt}}\ and\ \bibinfo {author} {\bibfnamefont {G.}~\bibnamefont
  {Verley}},\ }\bibfield  {title} {\enquote {\bibinfo {title} {Non equivalence
  of dynamical ensembles and emergent non ergodicity},}\ }\href
  {https://arxiv.org/abs/1806.11470} {\bibfield  {journal} {\bibinfo  {journal}
  {arXiv:1806.11470}\ } (\bibinfo {year} {2018})}\BibitemShut {NoStop}%
\bibitem [{\citenamefont {Binney}\ \emph {et~al.}(1992)\citenamefont {Binney},
  \citenamefont {Dowrick}, \citenamefont {Fisher},\ and\ \citenamefont
  {Newman}}]{binney92a}%
  \BibitemOpen
  \bibfield  {author} {\bibinfo {author} {\bibfnamefont {J.~J.}\ \bibnamefont
  {Binney}}, \bibinfo {author} {\bibfnamefont {N.~J.}\ \bibnamefont {Dowrick}},
  \bibinfo {author} {\bibfnamefont {A.~J.}\ \bibnamefont {Fisher}}, \ and\
  \bibinfo {author} {\bibfnamefont {M.}~\bibnamefont {Newman}},\ }\href@noop {}
  {\emph {\bibinfo {title} {The Theory of Critical Phenomena: An Introduction
  to the Renormalization Group}}}\ (\bibinfo  {publisher} {Oxford University
  Press, Inc.},\ \bibinfo {address} {New York, NY, USA},\ \bibinfo {year}
  {1992})\BibitemShut {NoStop}%
\bibitem [{\citenamefont {Zinn-Justin}(2002)}]{zinn-justin02a}%
  \BibitemOpen
  \bibfield  {author} {\bibinfo {author} {\bibfnamefont {J.}~\bibnamefont
  {Zinn-Justin}},\ }\href@noop {} {\emph {\bibinfo {title} {Quantum Field
  Theory and Critical Phenomena; 4th ed.}}},\ Internat. Ser. Mono. Phys.\
  (\bibinfo  {publisher} {Clarendon Press},\ \bibinfo {address} {Oxford},\
  \bibinfo {year} {2002})\BibitemShut {NoStop}%
\bibitem [{\citenamefont {Derrida}()}]{derrida07a}%
  \BibitemOpen
  \bibfield  {author} {\bibinfo {author} {\bibfnamefont {B.}~\bibnamefont
  {Derrida}},\ }\bibfield  {title} {\enquote {\bibinfo {title} {Non-equilibrium
  steady states: fluctuations and large deviations of the density and of the
  current},}\ }\href
  {http://iopscience.iop.org/article/10.1088/1742-5468/2007/07/P07023}
  {\bibinfo  {journal} {J. Stat. Mech. P07023 (2007)}\ }\BibitemShut {NoStop}%
\bibitem [{\citenamefont {Hurtado}\ \emph {et~al.}(2014)\citenamefont
  {Hurtado}, \citenamefont {Espigares}, \citenamefont {del Pozo},\ and\
  \citenamefont {Garrido}}]{hurtado14a}%
  \BibitemOpen
\bibfield  {journal} {  }\bibfield  {author} {\bibinfo {author} {\bibfnamefont
  {P.~I.}\ \bibnamefont {Hurtado}}, \bibinfo {author} {\bibfnamefont {C.~P.}\
  \bibnamefont {Espigares}}, \bibinfo {author} {\bibfnamefont {J.~J.}\
  \bibnamefont {del Pozo}}, \ and\ \bibinfo {author} {\bibfnamefont {P.~L.}\
  \bibnamefont {Garrido}},\ }\bibfield  {title} {\enquote {\bibinfo {title}
  {Thermodynamics of currents in nonequilibrium diffusive systems: theory and
  simulation},}\ }\href
  {http://link.springer.com/article/10.1007/s10955-013-0894-6} {\bibfield
  {journal} {\bibinfo  {journal} {J. Stat. Phys.}\ }\textbf {\bibinfo {volume}
  {154}},\ \bibinfo {pages} {214--264} (\bibinfo {year} {2014})}\BibitemShut
  {NoStop}%
\bibitem [{\citenamefont {Lazarescu}(2015)}]{lazarescu15a}%
  \BibitemOpen
  \bibfield  {author} {\bibinfo {author} {\bibfnamefont {A.}~\bibnamefont
  {Lazarescu}},\ }\bibfield  {title} {\enquote {\bibinfo {title} {The
  physicist's companion to current fluctuations: one-dimensional bulk-driven
  lattice gases},}\ }\href
  {http://iopscience.iop.org/article/10.1088/1751-8113/48/50/503001/meta}
  {\bibfield  {journal} {\bibinfo  {journal} {J. Phys. A}\ }\textbf {\bibinfo
  {volume} {48}},\ \bibinfo {pages} {503001} (\bibinfo {year}
  {2015})}\BibitemShut {NoStop}%
\bibitem [{\citenamefont {Carollo}\ \emph
  {et~al.}(2018{\natexlab{a}})\citenamefont {Carollo}, \citenamefont
  {Garrahan},\ and\ \citenamefont {Lesanovsky}}]{carollo2018current}%
  \BibitemOpen
  \bibfield  {author} {\bibinfo {author} {\bibfnamefont {F.}~\bibnamefont
  {Carollo}}, \bibinfo {author} {\bibfnamefont {J.~P.}\ \bibnamefont
  {Garrahan}}, \ and\ \bibinfo {author} {\bibfnamefont {I.}~\bibnamefont
  {Lesanovsky}},\ }\bibfield  {title} {\enquote {\bibinfo {title} {Current
  fluctuations in boundary-driven quantum spin chains},}\ }\href {\doibase
  10.1103/PhysRevB.98.094301} {\bibfield  {journal} {\bibinfo  {journal} {Phys.
  Rev. B}\ }\textbf {\bibinfo {volume} {98}},\ \bibinfo {pages} {094301}
  (\bibinfo {year} {2018}{\natexlab{a}})}\BibitemShut {NoStop}%
\bibitem [{\citenamefont {Touchette}(2009)}]{touchette09a}%
  \BibitemOpen
  \bibfield  {author} {\bibinfo {author} {\bibfnamefont {H.}~\bibnamefont
  {Touchette}},\ }\bibfield  {title} {\enquote {\bibinfo {title} {The large
  deviation approach to statistical mechanics},}\ }\href
  {http://dx.doi.org/10.1016/j.physrep.2009.05.002} {\bibfield  {journal}
  {\bibinfo  {journal} {Phys. Rep.}\ }\textbf {\bibinfo {volume} {478}},\
  \bibinfo {pages} {1--69} (\bibinfo {year} {2009})}\BibitemShut {NoStop}%
\bibitem [{\citenamefont {Bertini}\ \emph {et~al.}(2015)\citenamefont
  {Bertini}, \citenamefont {Sole}, \citenamefont {Gabrielli}, \citenamefont
  {Jona-Lasinio},\ and\ \citenamefont {Landim}}]{bertini15a}%
  \BibitemOpen
  \bibfield  {author} {\bibinfo {author} {\bibfnamefont {L.}~\bibnamefont
  {Bertini}}, \bibinfo {author} {\bibfnamefont {A.~De}\ \bibnamefont {Sole}},
  \bibinfo {author} {\bibfnamefont {D.}~\bibnamefont {Gabrielli}}, \bibinfo
  {author} {\bibfnamefont {G.}~\bibnamefont {Jona-Lasinio}}, \ and\ \bibinfo
  {author} {\bibfnamefont {C.}~\bibnamefont {Landim}},\ }\bibfield  {title}
  {\enquote {\bibinfo {title} {Macroscopic fluctuation theory},}\ }\href
  {http://journals.aps.org/rmp/abstract/10.1103/RevModPhys.87.593} {\bibfield
  {journal} {\bibinfo  {journal} {Rev. Mod. Phys.}\ }\textbf {\bibinfo {volume}
  {87}},\ \bibinfo {pages} {593--636} (\bibinfo {year} {2015})}\BibitemShut
  {NoStop}%
\bibitem [{\citenamefont {Hedges}\ \emph {et~al.}(2009)\citenamefont {Hedges},
  \citenamefont {Jack}, \citenamefont {Garrahan},\ and\ \citenamefont
  {Chandler}}]{hedges09a}%
  \BibitemOpen
  \bibfield  {author} {\bibinfo {author} {\bibfnamefont {L.~O.}\ \bibnamefont
  {Hedges}}, \bibinfo {author} {\bibfnamefont {R.~L.}\ \bibnamefont {Jack}},
  \bibinfo {author} {\bibfnamefont {J.~P.}\ \bibnamefont {Garrahan}}, \ and\
  \bibinfo {author} {\bibfnamefont {D.}~\bibnamefont {Chandler}},\ }\bibfield
  {title} {\enquote {\bibinfo {title} {Dynamic order-disorder in atomistic
  models of structural glass formers},}\ }\href
  {http://science.sciencemag.org/content/323/5919/1309} {\bibfield  {journal}
  {\bibinfo  {journal} {Science}\ }\textbf {\bibinfo {volume} {323}},\ \bibinfo
  {pages} {1309} (\bibinfo {year} {2009})}\BibitemShut {NoStop}%
\bibitem [{\citenamefont {Chandler}\ and\ \citenamefont
  {Garrahan}(2010)}]{chandler10a}%
  \BibitemOpen
  \bibfield  {author} {\bibinfo {author} {\bibfnamefont {D.}~\bibnamefont
  {Chandler}}\ and\ \bibinfo {author} {\bibfnamefont {J.~P.}\ \bibnamefont
  {Garrahan}},\ }\bibfield  {title} {\enquote {\bibinfo {title} {Dynamics on
  the way to forming glass: bubbles in space-time.}}\ }\href
  {http://www.annualreviews.org/doi/abs/10.1146/annurev.physchem.040808.090405}
  {\bibfield  {journal} {\bibinfo  {journal} {Annu. Rev. Phys. Chem.}\ }\textbf
  {\bibinfo {volume} {61}},\ \bibinfo {pages} {191--217} (\bibinfo {year}
  {2010})}\BibitemShut {NoStop}%
\bibitem [{\citenamefont {Pitard}\ \emph {et~al.}(2011)\citenamefont {Pitard},
  \citenamefont {Lecomte},\ and\ \citenamefont {Van~Wijland}}]{pitard11a}%
  \BibitemOpen
  \bibfield  {author} {\bibinfo {author} {\bibfnamefont {E.}~\bibnamefont
  {Pitard}}, \bibinfo {author} {\bibfnamefont {V.}~\bibnamefont {Lecomte}}, \
  and\ \bibinfo {author} {\bibfnamefont {F.}~\bibnamefont {Van~Wijland}},\
  }\bibfield  {title} {\enquote {\bibinfo {title} {Dynamic transition in an
  atomic glass former: A molecular-dynamics evidence},}\ }\href
  {http://epljournal.edpsciences.org/articles/epl/abs/2011/23/epl14011/epl14011.html}
  {\bibfield  {journal} {\bibinfo  {journal} {Europhys. Lett.}\ }\textbf
  {\bibinfo {volume} {96}},\ \bibinfo {pages} {56002} (\bibinfo {year}
  {2011})}\BibitemShut {NoStop}%
\bibitem [{\citenamefont {Speck}\ \emph {et~al.}(2012)\citenamefont {Speck},
  \citenamefont {Malins},\ and\ \citenamefont {Royall}}]{speck12a}%
  \BibitemOpen
  \bibfield  {author} {\bibinfo {author} {\bibfnamefont {T.}~\bibnamefont
  {Speck}}, \bibinfo {author} {\bibfnamefont {A.}~\bibnamefont {Malins}}, \
  and\ \bibinfo {author} {\bibfnamefont {C.~P.}\ \bibnamefont {Royall}},\
  }\bibfield  {title} {\enquote {\bibinfo {title} {First-order phase transition
  in a model glass former: Coupling of local structure and dynamics},}\ }\href
  {http://journals.aps.org/prl/abstract/10.1103/PhysRevLett.109.195703}
  {\bibfield  {journal} {\bibinfo  {journal} {Phys. Rev. Lett.}\ }\textbf
  {\bibinfo {volume} {109}},\ \bibinfo {pages} {195703} (\bibinfo {year}
  {2012})}\BibitemShut {NoStop}%
\bibitem [{\citenamefont {Pinchaipat}\ \emph {et~al.}(2017)\citenamefont
  {Pinchaipat}, \citenamefont {Campo}, \citenamefont {Turci}, \citenamefont
  {Hallett}, \citenamefont {Speck},\ and\ \citenamefont
  {Royall}}]{pinchaipat17a}%
  \BibitemOpen
  \bibfield  {author} {\bibinfo {author} {\bibfnamefont {R.}~\bibnamefont
  {Pinchaipat}}, \bibinfo {author} {\bibfnamefont {M.}~\bibnamefont {Campo}},
  \bibinfo {author} {\bibfnamefont {F.}~\bibnamefont {Turci}}, \bibinfo
  {author} {\bibfnamefont {J.}~\bibnamefont {Hallett}}, \bibinfo {author}
  {\bibfnamefont {T.}~\bibnamefont {Speck}}, \ and\ \bibinfo {author}
  {\bibfnamefont {C.~P.}\ \bibnamefont {Royall}},\ }\bibfield  {title}
  {\enquote {\bibinfo {title} {Experimental evidence for a structural-dynamical
  transition in trajectory space},}\ }\href
  {https://journals.aps.org/prl/abstract/10.1103/PhysRevLett.119.028004}
  {\bibfield  {journal} {\bibinfo  {journal} {Phys. Rev. Lett.}\ }\textbf
  {\bibinfo {volume} {119}},\ \bibinfo {pages} {028004} (\bibinfo {year}
  {2017})}\BibitemShut {NoStop}%
\bibitem [{\citenamefont {Abou}\ \emph {et~al.}(2018)\citenamefont {Abou},
  \citenamefont {Colin}, \citenamefont {Lecomte}, \citenamefont {Pitard},\ and\
  \citenamefont {van Wijland}}]{abou17a}%
  \BibitemOpen
  \bibfield  {author} {\bibinfo {author} {\bibfnamefont {B.}~\bibnamefont
  {Abou}}, \bibinfo {author} {\bibfnamefont {R.}~\bibnamefont {Colin}},
  \bibinfo {author} {\bibfnamefont {V.}~\bibnamefont {Lecomte}}, \bibinfo
  {author} {\bibfnamefont {E.}~\bibnamefont {Pitard}}, \ and\ \bibinfo {author}
  {\bibfnamefont {F.}~\bibnamefont {van Wijland}},\ }\bibfield  {title}
  {\enquote {\bibinfo {title} {Activity statistics in a colloidal glass former:
  experimental evidence for a dynamical transition},}\ }\href
  {https://doi.org/10.1063/1.5006924} {\bibfield  {journal} {\bibinfo
  {journal} {J.Chem. Phys.}\ }\textbf {\bibinfo {volume} {148}},\ \bibinfo
  {pages} {164502} (\bibinfo {year} {2018})}\BibitemShut {NoStop}%
\bibitem [{\citenamefont {Garrahan}\ \emph {et~al.}(2011)\citenamefont
  {Garrahan}, \citenamefont {Armour},\ and\ \citenamefont
  {Lesanovsky}}]{garrahan11a}%
  \BibitemOpen
  \bibfield  {author} {\bibinfo {author} {\bibfnamefont {J.~P.}\ \bibnamefont
  {Garrahan}}, \bibinfo {author} {\bibfnamefont {A.~D.}\ \bibnamefont
  {Armour}}, \ and\ \bibinfo {author} {\bibfnamefont {I.}~\bibnamefont
  {Lesanovsky}},\ }\bibfield  {title} {\enquote {\bibinfo {title} {Quantum
  trajectory phase transitions in the micromaser},}\ }\href
  {http://journals.aps.org/pre/abstract/10.1103/PhysRevE.84.021115} {\bibfield
  {journal} {\bibinfo  {journal} {Phys. Rev. E}\ }\textbf {\bibinfo {volume}
  {84}},\ \bibinfo {pages} {021115} (\bibinfo {year} {2011})}\BibitemShut
  {NoStop}%
\bibitem [{\citenamefont {Genway}\ \emph {et~al.}(2012)\citenamefont {Genway},
  \citenamefont {Garrahan}, \citenamefont {Lesanovsky},\ and\ \citenamefont
  {Armour}}]{genway12a}%
  \BibitemOpen
  \bibfield  {author} {\bibinfo {author} {\bibfnamefont {S.}~\bibnamefont
  {Genway}}, \bibinfo {author} {\bibfnamefont {J.~P.}\ \bibnamefont
  {Garrahan}}, \bibinfo {author} {\bibfnamefont {I.}~\bibnamefont
  {Lesanovsky}}, \ and\ \bibinfo {author} {\bibfnamefont {A.~D.}\ \bibnamefont
  {Armour}},\ }\bibfield  {title} {\enquote {\bibinfo {title} {Phase
  transitions in trajectories of a superconducting single-electron transistor
  coupled to a resonator},}\ }\href
  {http://journals.aps.org/pre/abstract/10.1103/PhysRevE.85.051122} {\bibfield
  {journal} {\bibinfo  {journal} {Phys. Rev. E}\ }\textbf {\bibinfo {volume}
  {85}},\ \bibinfo {pages} {051122} (\bibinfo {year} {2012})}\BibitemShut
  {NoStop}%
\bibitem [{\citenamefont {Manzano}\ and\ \citenamefont
  {Hurtado}(2014)}]{manzano14a}%
  \BibitemOpen
  \bibfield  {author} {\bibinfo {author} {\bibfnamefont {D.}~\bibnamefont
  {Manzano}}\ and\ \bibinfo {author} {\bibfnamefont {P.~I.}\ \bibnamefont
  {Hurtado}},\ }\bibfield  {title} {\enquote {\bibinfo {title} {Symmetry and
  the thermodynamics of currents in open quantum systems},}\ }\href
  {https://journals.aps.org/prb/abstract/10.1103/PhysRevB.90.125138} {\bibfield
   {journal} {\bibinfo  {journal} {Phys. Rev. B}\ }\textbf {\bibinfo {volume}
  {90}},\ \bibinfo {pages} {125138} (\bibinfo {year} {2014})}\BibitemShut
  {NoStop}%
\bibitem [{\citenamefont {Manzano}\ and\ \citenamefont
  {Kyoseva}(2016)}]{manzano16a}%
  \BibitemOpen
  \bibfield  {author} {\bibinfo {author} {\bibfnamefont {D.}~\bibnamefont
  {Manzano}}\ and\ \bibinfo {author} {\bibfnamefont {E.}~\bibnamefont
  {Kyoseva}},\ }\bibfield  {title} {\enquote {\bibinfo {title} {An atomic
  symmetry-controlled thermal switch},}\ }\href
  {https://www.nature.com/articles/srep31161} {\bibfield  {journal} {\bibinfo
  {journal} {Sci. Rep.}\ }\textbf {\bibinfo {volume} {6}},\ \bibinfo {pages}
  {31161} (\bibinfo {year} {2016})}\BibitemShut {NoStop}%
\bibitem [{\citenamefont {Manzano}\ and\ \citenamefont
  {Hurtado}(2018)}]{manzano17a}%
  \BibitemOpen
  \bibfield  {author} {\bibinfo {author} {\bibfnamefont {D.}~\bibnamefont
  {Manzano}}\ and\ \bibinfo {author} {\bibfnamefont {P.I.}\ \bibnamefont
  {Hurtado}},\ }\bibfield  {title} {\enquote {\bibinfo {title} {Harnessing
  symmetry to control quantum transport},}\ }\href {\doibase
  10.1080/00018732.2018.1519981} {\bibfield  {journal} {\bibinfo  {journal}
  {Adv. in Phys.}\ }\textbf {\bibinfo {volume} {67}},\ \bibinfo {pages} {1}
  (\bibinfo {year} {2018})}\BibitemShut {NoStop}%
\bibitem [{\citenamefont {Doob}(1957)}]{doob57a}%
  \BibitemOpen
  \bibfield  {author} {\bibinfo {author} {\bibfnamefont {J.~L.}\ \bibnamefont
  {Doob}},\ }\bibfield  {title} {\enquote {\bibinfo {title} {Conditional
  {B}rownian motion and the boundary limits of harmonic functions},}\ }\href
  {https://eudml.org/doc/86928} {\bibfield  {journal} {\bibinfo  {journal}
  {Bull. Soc. Math. Fr.}\ }\textbf {\bibinfo {volume} {85}},\ \bibinfo {pages}
  {431} (\bibinfo {year} {1957})}\BibitemShut {NoStop}%
\bibitem [{\citenamefont {Chetrite}\ and\ \citenamefont
  {Touchette}(2015{\natexlab{a}})}]{chetrite15a}%
  \BibitemOpen
  \bibfield  {author} {\bibinfo {author} {\bibfnamefont {R.}~\bibnamefont
  {Chetrite}}\ and\ \bibinfo {author} {\bibfnamefont {H.}~\bibnamefont
  {Touchette}},\ }\bibfield  {title} {\enquote {\bibinfo {title} {Variational
  and optimal control representations of conditioned and driven processes},}\
  }\href
  {http://iopscience.iop.org/article/10.1088/1742-5468/2015/12/P12001/meta}
  {\bibfield  {journal} {\bibinfo  {journal} {J. Stat. Mech. P12001}\ }
  (\bibinfo {year} {2015}{\natexlab{a}})}\BibitemShut {NoStop}%
\bibitem [{\citenamefont {Chetrite}\ and\ \citenamefont
  {Touchette}(2015{\natexlab{b}})}]{chetrite15b}%
  \BibitemOpen
  \bibfield  {author} {\bibinfo {author} {\bibfnamefont {R.}~\bibnamefont
  {Chetrite}}\ and\ \bibinfo {author} {\bibfnamefont {H.}~\bibnamefont
  {Touchette}},\ }\bibfield  {title} {\enquote {\bibinfo {title}
  {Nonequilibrium {Markov} processes conditioned on large deviations},}\ }\href
  {https://link.springer.com/article/10.1007%2Fs00023-014-0375-8} {\bibfield
  {journal} {\bibinfo  {journal} {Ann. Henri Poincare}\ }\textbf {\bibinfo
  {volume} {16}},\ \bibinfo {pages} {2005} (\bibinfo {year}
  {2015}{\natexlab{b}})}\BibitemShut {NoStop}%
\bibitem [{\citenamefont {Carollo}\ \emph
  {et~al.}(2018{\natexlab{b}})\citenamefont {Carollo}, \citenamefont
  {Garrahan}, \citenamefont {Lesanovsky},\ and\ \citenamefont
  {P\'erez-Espigares}}]{carollo2018making}%
  \BibitemOpen
  \bibfield  {author} {\bibinfo {author} {\bibfnamefont {F.}~\bibnamefont
  {Carollo}}, \bibinfo {author} {\bibfnamefont {J.~P.}\ \bibnamefont
  {Garrahan}}, \bibinfo {author} {\bibfnamefont {I.}~\bibnamefont
  {Lesanovsky}}, \ and\ \bibinfo {author} {\bibfnamefont {C.}~\bibnamefont
  {P\'erez-Espigares}},\ }\bibfield  {title} {\enquote {\bibinfo {title}
  {Making rare events typical in {M}arkovian open quantum systems},}\ }\href
  {\doibase 10.1103/PhysRevA.98.010103} {\bibfield  {journal} {\bibinfo
  {journal} {Phys. Rev. A}\ }\textbf {\bibinfo {volume} {98}},\ \bibinfo
  {pages} {010103} (\bibinfo {year} {2018}{\natexlab{b}})}\BibitemShut
  {NoStop}%
\bibitem [{\citenamefont {Bodineau}\ and\ \citenamefont
  {Derrida}(2004)}]{bodineau04a}%
  \BibitemOpen
  \bibfield  {author} {\bibinfo {author} {\bibfnamefont {T.}~\bibnamefont
  {Bodineau}}\ and\ \bibinfo {author} {\bibfnamefont {B.}~\bibnamefont
  {Derrida}},\ }\bibfield  {title} {\enquote {\bibinfo {title} {Current
  fluctuations in nonequilibrium diffusive systems: {An} additivity
  principle},}\ }\href
  {http://journals.aps.org/prl/abstract/10.1103/PhysRevLett.92.180601}
  {\bibfield  {journal} {\bibinfo  {journal} {Phys. Rev. Lett.}\ }\textbf
  {\bibinfo {volume} {92}},\ \bibinfo {pages} {180601} (\bibinfo {year}
  {2004})}\BibitemShut {NoStop}%
\bibitem [{\citenamefont {De~Masi}\ \emph {et~al.}(1989)\citenamefont
  {De~Masi}, \citenamefont {Presutti},\ and\ \citenamefont
  {Scacciatelli}}]{wasep1masi89}%
  \BibitemOpen
  \bibfield  {author} {\bibinfo {author} {\bibfnamefont {A.}~\bibnamefont
  {De~Masi}}, \bibinfo {author} {\bibfnamefont {E.}~\bibnamefont {Presutti}}, \
  and\ \bibinfo {author} {\bibfnamefont {E.}~\bibnamefont {Scacciatelli}},\
  }\bibfield  {title} {\enquote {\bibinfo {title} {The weakly asymmetric simple
  exclusion process},}\ }\href
  {http://www.numdam.org/item/AIHPB_1989__25_1_1_0} {\bibfield  {journal}
  {\bibinfo  {journal} {Ann. Inst. Henri Poincar{\'e}}\ }\textbf {\bibinfo
  {volume} {25}},\ \bibinfo {pages} {1--38} (\bibinfo {year}
  {1989})}\BibitemShut {NoStop}%
\bibitem [{\citenamefont {G{\"a}rtner}(1987)}]{wasep2gartner87}%
  \BibitemOpen
  \bibfield  {author} {\bibinfo {author} {\bibfnamefont {J.}~\bibnamefont
  {G{\"a}rtner}},\ }\bibfield  {title} {\enquote {\bibinfo {title} {Convergence
  towards burger's equation and propagation of chaos for weakly asymmetric
  exclusion processes},}\ }\href
  {https://www.sciencedirect.com/science/article/pii/0304414987900408}
  {\bibfield  {journal} {\bibinfo  {journal} {Stoch. Proc. Appl.}\ }\textbf
  {\bibinfo {volume} {27}},\ \bibinfo {pages} {233--260} (\bibinfo {year}
  {1987})}\BibitemShut {NoStop}%
\bibitem [{\citenamefont {Hurtado}\ and\ \citenamefont
  {Garrido}(2009{\natexlab{a}})}]{hurtado09c}%
  \BibitemOpen
  \bibfield  {author} {\bibinfo {author} {\bibfnamefont {P.~I.}\ \bibnamefont
  {Hurtado}}\ and\ \bibinfo {author} {\bibfnamefont {P.~L.}\ \bibnamefont
  {Garrido}},\ }\bibfield  {title} {\enquote {\bibinfo {title} {Test of the
  {additivity} {principle} for {current} {fluctuations} in a {model} of {heat}
  conduction},}\ }\href
  {http://journals.aps.org/prl/abstract/10.1103/PhysRevLett.102.250601}
  {\bibfield  {journal} {\bibinfo  {journal} {Phys. Rev. Lett.}\ }\textbf
  {\bibinfo {volume} {102}},\ \bibinfo {pages} {250601} (\bibinfo {year}
  {2009}{\natexlab{a}})}\BibitemShut {NoStop}%
\bibitem [{\citenamefont {Hurtado}\ and\ \citenamefont
  {Garrido}(2010)}]{hurtado10a}%
  \BibitemOpen
  \bibfield  {author} {\bibinfo {author} {\bibfnamefont {P.~I.}\ \bibnamefont
  {Hurtado}}\ and\ \bibinfo {author} {\bibfnamefont {P.~L.}\ \bibnamefont
  {Garrido}},\ }\bibfield  {title} {\enquote {\bibinfo {title} {Large
  fluctuations of the macroscopic current in diffusive systems: {A} numerical
  test of the additivity principle},}\ }\href
  {http://journals.aps.org/pre/abstract/10.1103/PhysRevE.81.041102} {\bibfield
  {journal} {\bibinfo  {journal} {Phys. Rev. E}\ }\textbf {\bibinfo {volume}
  {81}},\ \bibinfo {pages} {041102} (\bibinfo {year} {2010})}\BibitemShut
  {NoStop}%
\bibitem [{\citenamefont {Gorissen}\ and\ \citenamefont
  {Vanderzande}(2012)}]{Gorissen2012}%
  \BibitemOpen
  \bibfield  {author} {\bibinfo {author} {\bibfnamefont {M.}~\bibnamefont
  {Gorissen}}\ and\ \bibinfo {author} {\bibfnamefont {C.}~\bibnamefont
  {Vanderzande}},\ }\bibfield  {title} {\enquote {\bibinfo {title} {Current
  fluctuations in the weakly asymmetric exclusion process with open
  boundaries},}\ }\href {\doibase 10.1103/PhysRevE.86.051114} {\bibfield
  {journal} {\bibinfo  {journal} {Phys. Rev. E}\ }\textbf {\bibinfo {volume}
  {86}},\ \bibinfo {pages} {051114} (\bibinfo {year} {2012})}\BibitemShut
  {NoStop}%
\bibitem [{\citenamefont {P\'erez-Espigares}\ \emph {et~al.}(2016)\citenamefont
  {P\'erez-Espigares}, \citenamefont {Garrido},\ and\ \citenamefont
  {Hurtado}}]{perez-espigares16a}%
  \BibitemOpen
  \bibfield  {author} {\bibinfo {author} {\bibfnamefont {C.}~\bibnamefont
  {P\'erez-Espigares}}, \bibinfo {author} {\bibfnamefont {P.~L.}\ \bibnamefont
  {Garrido}}, \ and\ \bibinfo {author} {\bibfnamefont {P.~I.}\ \bibnamefont
  {Hurtado}},\ }\bibfield  {title} {\enquote {\bibinfo {title} {Weak additivity
  principle for current statistics in $d$-dimensions},}\ }\href
  {http://journals.aps.org/pre/abstract/10.1103/PhysRevE.93.040103} {\bibfield
  {journal} {\bibinfo  {journal} {Phys. Rev. E}\ }\textbf {\bibinfo {volume}
  {93}},\ \bibinfo {pages} {040103(R)} (\bibinfo {year} {2016})}\BibitemShut
  {NoStop}%
\bibitem [{\citenamefont {Giardin\`a}\ \emph {et~al.}(2006)\citenamefont
  {Giardin\`a}, \citenamefont {Kurchan},\ and\ \citenamefont
  {Peliti}}]{giardina06a}%
  \BibitemOpen
  \bibfield  {author} {\bibinfo {author} {\bibfnamefont {C.}~\bibnamefont
  {Giardin\`a}}, \bibinfo {author} {\bibfnamefont {J.}~\bibnamefont {Kurchan}},
  \ and\ \bibinfo {author} {\bibfnamefont {L.}~\bibnamefont {Peliti}},\
  }\bibfield  {title} {\enquote {\bibinfo {title} {Direct evaluation of
  large-deviation functions},}\ }\href
  {http://journals.aps.org/prl/abstract/10.1103/PhysRevLett.96.120603}
  {\bibfield  {journal} {\bibinfo  {journal} {Phys. Rev. Lett.}\ }\textbf
  {\bibinfo {volume} {96}},\ \bibinfo {pages} {120603} (\bibinfo {year}
  {2006})}\BibitemShut {NoStop}%
\bibitem [{\citenamefont {Lecomte}\ and\ \citenamefont
  {Tailleur}()}]{lecomte07a}%
  \BibitemOpen
  \bibfield  {author} {\bibinfo {author} {\bibfnamefont {V.}~\bibnamefont
  {Lecomte}}\ and\ \bibinfo {author} {\bibfnamefont {J.}~\bibnamefont
  {Tailleur}},\ }\bibfield  {title} {\enquote {\bibinfo {title} {A numerical
  approach to large deviations in continuous time},}\ }\href
  {http://iopscience.iop.org/article/10.1088/1742-5468/2007/03/P03004/meta}
  {\bibinfo  {journal} {J. Stat. Mech. P03004 (2007)}\ }\BibitemShut {NoStop}%
\bibitem [{\citenamefont {Tailleur}\ and\ \citenamefont
  {Lecomte}(2009)}]{tailleur09a}%
  \BibitemOpen
\bibfield  {journal} {  }\bibfield  {author} {\bibinfo {author} {\bibfnamefont
  {J.}~\bibnamefont {Tailleur}}\ and\ \bibinfo {author} {\bibfnamefont
  {V.}~\bibnamefont {Lecomte}},\ }\bibfield  {title} {\enquote {\bibinfo
  {title} {Simulation of large deviation functions using population
  dynamics},}\ }\href {https://aip.scitation.org/doi/abs/10.1063/1.3082284}
  {\bibfield  {journal} {\bibinfo  {journal} {Modeling and Simulation of New
  Materials}\ }\textbf {\bibinfo {volume} {1091}},\ \bibinfo {pages} {212--219}
  (\bibinfo {year} {2009})}\BibitemShut {NoStop}%
\bibitem [{\citenamefont {Giardin\`a}\ \emph {et~al.}(2011)\citenamefont
  {Giardin\`a}, \citenamefont {Kurchan}, \citenamefont {Lecomte},\ and\
  \citenamefont {Tailleur}}]{giardina11a}%
  \BibitemOpen
  \bibfield  {author} {\bibinfo {author} {\bibfnamefont {C.}~\bibnamefont
  {Giardin\`a}}, \bibinfo {author} {\bibfnamefont {J.}~\bibnamefont {Kurchan}},
  \bibinfo {author} {\bibfnamefont {V.}~\bibnamefont {Lecomte}}, \ and\
  \bibinfo {author} {\bibfnamefont {J.}~\bibnamefont {Tailleur}},\ }\bibfield
  {title} {\enquote {\bibinfo {title} {Simulating rare events in dynamical
  processes},}\ }\href
  {http://link.springer.com/article/10.1007%2Fs10955-011-0350-4} {\bibfield
  {journal} {\bibinfo  {journal} {J. Stat. Phys.}\ }\textbf {\bibinfo {volume}
  {145}},\ \bibinfo {pages} {787--811} (\bibinfo {year} {2011})}\BibitemShut
  {NoStop}%
\bibitem [{\citenamefont {Nemoto}\ \emph {et~al.}(2016)\citenamefont {Nemoto},
  \citenamefont {Bouchet}, \citenamefont {Jack},\ and\ \citenamefont
  {Lecomte}}]{nemoto16a}%
  \BibitemOpen
  \bibfield  {author} {\bibinfo {author} {\bibfnamefont {T.}~\bibnamefont
  {Nemoto}}, \bibinfo {author} {\bibfnamefont {F.}~\bibnamefont {Bouchet}},
  \bibinfo {author} {\bibfnamefont {R.~L.}\ \bibnamefont {Jack}}, \ and\
  \bibinfo {author} {\bibfnamefont {V.}~\bibnamefont {Lecomte}},\ }\bibfield
  {title} {\enquote {\bibinfo {title} {Population dynamics method with a
  multi-canonical feedback control},}\ }\href
  {http://journals.aps.org/pre/abstract/10.1103/PhysRevE.93.062123} {\bibfield
  {journal} {\bibinfo  {journal} {Phys. Rev. E}\ }\textbf {\bibinfo {volume}
  {93}},\ \bibinfo {pages} {062123} (\bibinfo {year} {2016})}\BibitemShut
  {NoStop}%
\bibitem [{\citenamefont {Ray}\ \emph {et~al.}(2018)\citenamefont {Ray},
  \citenamefont {Chan},\ and\ \citenamefont {Limmer}}]{ray17a}%
  \BibitemOpen
  \bibfield  {author} {\bibinfo {author} {\bibfnamefont {U.}~\bibnamefont
  {Ray}}, \bibinfo {author} {\bibfnamefont {G.~Kin-Lic}\ \bibnamefont {Chan}},
  \ and\ \bibinfo {author} {\bibfnamefont {D.T.}\ \bibnamefont {Limmer}},\
  }\bibfield  {title} {\enquote {\bibinfo {title} {Exact fluctuations of
  nonequilibrium steady states from approximate auxiliary dynamics},}\ }\href
  {https://journals.aps.org/prl/abstract/10.1103/PhysRevLett.120.210602}
  {\bibfield  {journal} {\bibinfo  {journal} {Phys. Rev. Lett.}\ }\textbf
  {\bibinfo {volume} {120}},\ \bibinfo {pages} {210602} (\bibinfo {year}
  {2018})}\BibitemShut {NoStop}%
\bibitem [{\citenamefont {Spohn}(1991)}]{Spohn1991}%
  \BibitemOpen
  \bibfield  {author} {\bibinfo {author} {\bibfnamefont {H.}~\bibnamefont
  {Spohn}},\ }\href@noop {} {\emph {\bibinfo {title} {Large Scale Dynamics of
  Interacting Particles}}}\ (\bibinfo  {publisher} {Spinger Verlag},\ \bibinfo
  {year} {1991})\BibitemShut {NoStop}%
\bibitem [{SM()}]{SM}%
  \BibitemOpen
  \href@noop {} {\bibinfo  {journal} {See Supplemental Material for details}\
  }\BibitemShut {NoStop}%
\bibitem [{\citenamefont {Bray}\ and\ \citenamefont {McKane}(1989)}]{Bray89}%
  \BibitemOpen
\bibfield  {journal} {  }\bibfield  {author} {\bibinfo {author} {\bibfnamefont
  {A.~J.}\ \bibnamefont {Bray}}\ and\ \bibinfo {author} {\bibfnamefont {A.~J.}\
  \bibnamefont {McKane}},\ }\bibfield  {title} {\enquote {\bibinfo {title}
  {Instanton calculation of the escape rate for activation over a potential
  barrier driven by colored noise},}\ }\href {\doibase
  10.1103/PhysRevLett.62.493} {\bibfield  {journal} {\bibinfo  {journal} {Phys.
  Rev. Lett.}\ }\textbf {\bibinfo {volume} {62}},\ \bibinfo {pages} {493--496}
  (\bibinfo {year} {1989})}\BibitemShut {NoStop}%
\bibitem [{\citenamefont {Schutz}(2001)}]{schutz01a}%
  \BibitemOpen
  \bibfield  {author} {\bibinfo {author} {\bibfnamefont {G.~M.}\ \bibnamefont
  {Schutz}},\ }\href@noop {} {\emph {\bibinfo {title} {Exactly solvable models
  for many-body systems far from equilibrium}}},\ Vol.~\bibinfo {volume} {19}\
  (\bibinfo {year} {2001})\ pp.\ \bibinfo {pages} {1--251}\BibitemShut
  {NoStop}%
\bibitem [{\citenamefont {Garrahan}(2018)}]{garrahan18a}%
  \BibitemOpen
  \bibfield  {author} {\bibinfo {author} {\bibfnamefont {J.~P.}\ \bibnamefont
  {Garrahan}},\ }\bibfield  {title} {\enquote {\bibinfo {title} {Aspects of
  non-equilibrium in classical and quantum systems: Slow relaxation and
  glasses, dynamical large deviations, quantum non-ergodicity, and open quantum
  dynamics},}\ }\href
  {https://www.sciencedirect.com/science/article/pii/S0378437117313985?via%3Dihub}
  {\bibfield  {journal} {\bibinfo  {journal} {Physica A}\ }\textbf {\bibinfo
  {volume} {504}},\ \bibinfo {pages} {130} (\bibinfo {year}
  {2018})}\BibitemShut {NoStop}%
\bibitem [{\citenamefont {Gaveau}\ and\ \citenamefont
  {Schulman}(2006)}]{Gaveau06a}%
  \BibitemOpen
  \bibfield  {author} {\bibinfo {author} {\bibfnamefont {B.}~\bibnamefont
  {Gaveau}}\ and\ \bibinfo {author} {\bibfnamefont {L.~S.}\ \bibnamefont
  {Schulman}},\ }\bibfield  {title} {\enquote {\bibinfo {title} {Multiple
  phases in stochastic dynamics: Geometry and probabilities},}\ }\href
  {\doibase 10.1103/PhysRevE.73.036124} {\bibfield  {journal} {\bibinfo
  {journal} {Phys. Rev. E}\ }\textbf {\bibinfo {volume} {73}},\ \bibinfo
  {pages} {036124} (\bibinfo {year} {2006})}\BibitemShut {NoStop}%
\bibitem [{\citenamefont {a~pedagogical review~see
  J.~Kurchan}(2009)}]{kurchan2009six}%
  \BibitemOpen
  \bibfield  {author} {\bibinfo {author} {\bibfnamefont {For}\ \bibnamefont
  {a~pedagogical review~see J.~Kurchan}},\ }\bibfield  {title} {\enquote
  {\bibinfo {title} {Six out of equilibrium lectures},}\ }\href
  {https://arxiv.org/abs/0901.1271} {\bibfield  {journal} {\bibinfo  {journal}
  {arXiv:0901.1271}\ } (\bibinfo {year} {2009})}\BibitemShut {NoStop}%
\bibitem [{\citenamefont {Macieszczak}\ \emph {et~al.}(2016)\citenamefont
  {Macieszczak}, \citenamefont {Gu\ifmmode \mbox{\c{t}}\else
  \c{t}\fi{}\ifmmode~\u{a}\else \u{a}\fi{}}, \citenamefont {Lesanovsky},\ and\
  \citenamefont {Garrahan}}]{kasia16a}%
  \BibitemOpen
  \bibfield  {author} {\bibinfo {author} {\bibfnamefont {K.}~\bibnamefont
  {Macieszczak}}, \bibinfo {author} {\bibfnamefont {M.}~\bibnamefont
  {Gu\ifmmode \mbox{\c{t}}\else \c{t}\fi{}\ifmmode~\u{a}\else \u{a}\fi{}}},
  \bibinfo {author} {\bibfnamefont {I.}~\bibnamefont {Lesanovsky}}, \ and\
  \bibinfo {author} {\bibfnamefont {J.~P.}\ \bibnamefont {Garrahan}},\
  }\bibfield  {title} {\enquote {\bibinfo {title} {Towards a theory of
  metastability in open quantum dynamics},}\ }\href {\doibase
  10.1103/PhysRevLett.116.240404} {\bibfield  {journal} {\bibinfo  {journal}
  {Phys. Rev. Lett.}\ }\textbf {\bibinfo {volume} {116}},\ \bibinfo {pages}
  {240404} (\bibinfo {year} {2016})}\BibitemShut {NoStop}%
\bibitem [{\citenamefont {Rose}\ \emph {et~al.}(2016)\citenamefont {Rose},
  \citenamefont {Macieszczak}, \citenamefont {Lesanovsky},\ and\ \citenamefont
  {Garrahan}}]{rose16a}%
  \BibitemOpen
  \bibfield  {author} {\bibinfo {author} {\bibfnamefont {D.~C.}\ \bibnamefont
  {Rose}}, \bibinfo {author} {\bibfnamefont {K.}~\bibnamefont {Macieszczak}},
  \bibinfo {author} {\bibfnamefont {I.}~\bibnamefont {Lesanovsky}}, \ and\
  \bibinfo {author} {\bibfnamefont {J.~P.}\ \bibnamefont {Garrahan}},\
  }\bibfield  {title} {\enquote {\bibinfo {title} {Metastability in an open
  quantum ising model},}\ }\href {\doibase 10.1103/PhysRevE.94.052132}
  {\bibfield  {journal} {\bibinfo  {journal} {Phys. Rev. E}\ }\textbf {\bibinfo
  {volume} {94}},\ \bibinfo {pages} {052132} (\bibinfo {year}
  {2016})}\BibitemShut {NoStop}%
\bibitem [{\citenamefont {Derrida}(1998)}]{derrida98a}%
  \BibitemOpen
  \bibfield  {author} {\bibinfo {author} {\bibfnamefont {B.}~\bibnamefont
  {Derrida}},\ }\bibfield  {title} {\enquote {\bibinfo {title} {An exactly
  soluble non-equilibrium system: {The} asymmetric simple exclusion process},}\
  }\href {http://www.sciencedirect.com/science/article/pii/S0370157398000064}
  {\bibfield  {journal} {\bibinfo  {journal} {Phys. Rep.}\ }\textbf {\bibinfo
  {volume} {301}},\ \bibinfo {pages} {65--83} (\bibinfo {year}
  {1998})}\BibitemShut {NoStop}%
\bibitem [{\citenamefont {Chou}\ \emph {et~al.}(2011)\citenamefont {Chou},
  \citenamefont {Mallick},\ and\ \citenamefont {Zia}}]{chou11a}%
  \BibitemOpen
  \bibfield  {author} {\bibinfo {author} {\bibfnamefont {T.}~\bibnamefont
  {Chou}}, \bibinfo {author} {\bibfnamefont {K.}~\bibnamefont {Mallick}}, \
  and\ \bibinfo {author} {\bibfnamefont {R.~K.~P.}\ \bibnamefont {Zia}},\
  }\bibfield  {title} {\enquote {\bibinfo {title} {Non-equilibrium statistical
  mechanics: from a paradigmatic model to biological transport},}\ }\href@noop
  {} {\bibfield  {journal} {\bibinfo  {journal} {Reports On Progress In Phys.}\
  }\textbf {\bibinfo {volume} {74}},\ \bibinfo {pages} {116601} (\bibinfo
  {year} {2011})}\BibitemShut {NoStop}%
\bibitem [{\citenamefont {Hurtado}\ and\ \citenamefont
  {Garrido}(2009{\natexlab{b}})}]{hurtado09a}%
  \BibitemOpen
  \bibfield  {author} {\bibinfo {author} {\bibfnamefont {P.~I.}\ \bibnamefont
  {Hurtado}}\ and\ \bibinfo {author} {\bibfnamefont {P.~L.}\ \bibnamefont
  {Garrido}},\ }\bibfield  {title} {\enquote {\bibinfo {title} {Current
  fluctuations and statistics during a large deviation event in an exactly
  solvable transport model},}\ }\href
  {http://iopscience.iop.org/article/10.1088/1742-5468/2009/02/P02032/meta}
  {\bibfield  {journal} {\bibinfo  {journal} {J. Stat. Mech. P02032}\ }
  (\bibinfo {year} {2009}{\natexlab{b}})}\BibitemShut {NoStop}%
\bibitem [{\citenamefont {Saito}\ and\ \citenamefont {Dhar}(2011)}]{saito11a}%
  \BibitemOpen
  \bibfield  {author} {\bibinfo {author} {\bibfnamefont {K.}~\bibnamefont
  {Saito}}\ and\ \bibinfo {author} {\bibfnamefont {A.}~\bibnamefont {Dhar}},\
  }\bibfield  {title} {\enquote {\bibinfo {title} {Additivity {principle} in
  {high-dimensional} {deterministic} systems},}\ }\href
  {http://journals.aps.org/prl/abstract/10.1103/PhysRevLett.107.250601}
  {\bibfield  {journal} {\bibinfo  {journal} {Phys. Rev. Lett.}\ }\textbf
  {\bibinfo {volume} {107}},\ \bibinfo {pages} {250601} (\bibinfo {year}
  {2011})}\BibitemShut {NoStop}%
\bibitem [{\citenamefont {Hurtado}\ \emph {et~al.}(2011)\citenamefont
  {Hurtado}, \citenamefont {P\'erez-Espigares}, \citenamefont {del Pozo},\ and\
  \citenamefont {Garrido}}]{hurtado11b}%
  \BibitemOpen
  \bibfield  {author} {\bibinfo {author} {\bibfnamefont {P.~I.}\ \bibnamefont
  {Hurtado}}, \bibinfo {author} {\bibfnamefont {C.}~\bibnamefont
  {P\'erez-Espigares}}, \bibinfo {author} {\bibfnamefont {J.~J.}\ \bibnamefont
  {del Pozo}}, \ and\ \bibinfo {author} {\bibfnamefont {P.~L.}\ \bibnamefont
  {Garrido}},\ }\bibfield  {title} {\enquote {\bibinfo {title} {Symmetries in
  fluctuations far from equilibrium},}\ }\href
  {http://www.pnas.org/content/108/19/7704.short} {\bibfield  {journal}
  {\bibinfo  {journal} {Proc. Natl. Acad. Sci. USA}\ }\textbf {\bibinfo
  {volume} {108}},\ \bibinfo {pages} {7704--7709} (\bibinfo {year}
  {2011})}\BibitemShut {NoStop}%
\bibitem [{\citenamefont {Hurtado}\ \emph {et~al.}(2013)\citenamefont
  {Hurtado}, \citenamefont {Lasanta},\ and\ \citenamefont
  {Prados}}]{hurtado13a}%
  \BibitemOpen
  \bibfield  {author} {\bibinfo {author} {\bibfnamefont {P.~I.}\ \bibnamefont
  {Hurtado}}, \bibinfo {author} {\bibfnamefont {A.}~\bibnamefont {Lasanta}}, \
  and\ \bibinfo {author} {\bibfnamefont {A.}~\bibnamefont {Prados}},\
  }\bibfield  {title} {\enquote {\bibinfo {title} {Typical and rare
  fluctuations in nonlinear driven diffusive systems with dissipation},}\
  }\href {http://journals.aps.org/pre/abstract/10.1103/PhysRevE.88.022110}
  {\bibfield  {journal} {\bibinfo  {journal} {Phys. Rev. E}\ }\textbf {\bibinfo
  {volume} {88}},\ \bibinfo {pages} {022110} (\bibinfo {year}
  {2013})}\BibitemShut {NoStop}%
\bibitem [{\citenamefont {\v{Z}nidari\v{c}}(2014)}]{znidaric14a}%
  \BibitemOpen
  \bibfield  {author} {\bibinfo {author} {\bibfnamefont {M.}~\bibnamefont
  {\v{Z}nidari\v{c}}},\ }\bibfield  {title} {\enquote {\bibinfo {title} {Exact
  large-deviation statistics for a nonequilibrium quantum spin chain},}\ }\href
  {https://journals.aps.org/prl/abstract/10.1103/PhysRevLett.112.040602}
  {\bibfield  {journal} {\bibinfo  {journal} {Phys. Rev. Lett.}\ }\textbf
  {\bibinfo {volume} {112}},\ \bibinfo {pages} {040602} (\bibinfo {year}
  {2014})}\BibitemShut {NoStop}%
\bibitem [{\citenamefont {Shpielberg}\ and\ \citenamefont
  {Akkermans}(2016)}]{shpielberg16a}%
  \BibitemOpen
  \bibfield  {author} {\bibinfo {author} {\bibfnamefont {O.}~\bibnamefont
  {Shpielberg}}\ and\ \bibinfo {author} {\bibfnamefont {E.}~\bibnamefont
  {Akkermans}},\ }\bibfield  {title} {\enquote {\bibinfo {title} {Le
  {C}hatelier principle for out-of-equilibrium and boundary-driven systems:
  Application to dynamical phase transitions},}\ }\href
  {http://dx.doi.org/10.1103/PhysRevLett.116.240603} {\bibfield  {journal}
  {\bibinfo  {journal} {Phys. Rev. Lett.}\ }\textbf {\bibinfo {volume} {116}}
  (\bibinfo {year} {2016})}\BibitemShut {NoStop}%
\bibitem [{\citenamefont {Tiz{\'o}n-Escamilla}\ \emph
  {et~al.}(2017)\citenamefont {Tiz{\'o}n-Escamilla}, \citenamefont {Hurtado},\
  and\ \citenamefont {Garrido}}]{tizon-escamilla17a}%
  \BibitemOpen
  \bibfield  {author} {\bibinfo {author} {\bibfnamefont {N.}~\bibnamefont
  {Tiz{\'o}n-Escamilla}}, \bibinfo {author} {\bibfnamefont {P.~I.}\
  \bibnamefont {Hurtado}}, \ and\ \bibinfo {author} {\bibfnamefont {P.~L.}\
  \bibnamefont {Garrido}},\ }\bibfield  {title} {\enquote {\bibinfo {title}
  {Structure of the optimal path to a fluctuation},}\ }\href
  {http://journals.aps.org/pre/pdf/10.1103/PhysRevE.95.002100} {\bibfield
  {journal} {\bibinfo  {journal} {Phys. Rev. E}\ }\textbf {\bibinfo {volume}
  {95}},\ \bibinfo {pages} {002100} (\bibinfo {year} {2017})}\BibitemShut
  {NoStop}%
\bibitem [{\citenamefont {Byrd}\ and\ \citenamefont
  {Friedman}(1971)}]{byrd71a}%
  \BibitemOpen
  \bibfield  {author} {\bibinfo {author} {\bibfnamefont {P.~F.}\ \bibnamefont
  {Byrd}}\ and\ \bibinfo {author} {\bibfnamefont {M.~D.}\ \bibnamefont
  {Friedman}},\ }\href@noop {} {\emph {\bibinfo {title} {Handbook of Elliptic
  Integrals for Engineers and Scientists}}}\ (\bibinfo  {publisher}
  {Springer-Verlag},\ \bibinfo {year} {1971})\BibitemShut {NoStop}%
\end{thebibliography}%
\let\addcontentsline\oldaddcontentsline

\onecolumngrid
\newpage

\renewcommand\thesection{S\arabic{section}}
\renewcommand\theequation{S\arabic{equation}}
\renewcommand\thefigure{S\arabic{figure}}
\setcounter{equation}{0}

\begin{center}
{\Large SUPPLEMENTAL MATERIAL}
\end{center}
\begin{center}
{\Large Dynamical criticality in driven systems: Non-perturbative results, microscopic origin and direct observation}
\end{center}
\begin{center}
Carlos P\'erez-Espigares,$^{1,2}$ Federico Carollo,$^1$ Juan P. Garrahan,$^1$ and Pablo I. Hurtado$^2$
\end{center}
\begin{center}
$^1${\em School of Physics and Astronomy, and Centre for the Mathematics and Theoretical Physics of Quantum Non-Equilibrium Systems,
University of Nottingham, Nottingham, NG7 2RD, United Kingdom}\\
$^2${\em Departamento de Electromagnetismo y F\'{\i}sica 
de la Materia, and Institute Carlos I for Theoretical and Computational Physics, Universidad de Granada, Granada 18071, Spain}
\end{center}

%
%
%

\tableofcontents

\vspace{0.5cm}

In this Supplemental Material we solve the macroscopic fluctuation theory (MFT) equations for the joint current and mass fluctuations of the one-dimensional ($1d$) weakly asymmetric simple exclusion process (WASEP) coupled to boundary particle reservoirs at arbitrary densities or chemical potentials. This model belongs in a large class of driven diffusive systems of theoretical and technological interest. MFT \cite{bertini15a} provides a detailed description of dynamical fluctuations in general driven diffusive systems, starting from the hydrodynamic evolution equation for the system of interest and the sole knowledge of two transport coefficients, which can be measured experimentally. In particular, MFT offers explicit predictions for the large-deviation functions (LDFs) which characterize the fluctuations of different observables, as well as the associated trajectories in phase space responsible of these fluctuations. 

After a brief but self-consistent presentation of MFT in \S\ref{app1} and a characterization of the nonequilibrium steady state of the $1d$ open WASEP under arbitrary driving (see \S\ref{app2}), we proceed to solve analytically  in \S\ref{app3} the MFT equations for the joint mass-current statistics of this model, understanding along the way the symmetry-breaking dynamical phase transition described in the main text. Key to this calculation is the additivity conjecture \cite{bodineau04a}, which assumes that the optimal trajectories responsible of a trajectory are time-independent. We explore in \S\ref{app4} the possibility of additivity violations in the form of time-dependent, instantonic solutions to the MFT equations in regimes where the joint current-mass LDF becomes non-convex. Finally, we study in \S\ref{app5} from a microscopic point of view the DPT found at the macroscopic level, using in particular the quantum Hamiltonian formalism for the master equation and the tilted dynamical generator.

\newpage
\section{A crash course on MFT} 
\label{app1}

We hence consider systems described at the mesoscopic level by a continuity equation of the form
\be
\partial_t\rho + \partial_x j = 0 \, ,
\label{hydro}
\ee
where $\rho(x,t)$ and $j(x,t)$ are the density and current fields, respectively, and $x\in[0,1]$ and $t$ are the macroscopic space and time variables, obtained after a diffusive scaling limit such that $x=\tilde{x}/L$ and $t=\tilde{t}/L^2$, with $\tilde{x}$ and $\tilde{t}$ the equivalent microscopic variables and $L$ the system size in natural units. The system is coupled at the boundaries to particle reservoirs at densities $\rho_{\text{L,R}}$, so the boundary conditions for the density field are $\rho(0,t)=\rho_{\text{L}}$ and $\rho(1,t)=\rho_{\text{R}}$ $\forall t$. The current field in eq. (\ref{hydro}) is in general a fluctuating quantity, and can be written as
\begin{equation}
j(x,t)=-D(\rho)\partial_x \rho(x,t) +\sigma(\rho) E + \xi(x,t).
\label{current}
\end{equation}
The first two terms in the rhs are just Fick's law, which express the proportionality of the current to the density gradient and the external field $E$, with $D(\rho)$ and $\sigma(\rho)$ the diffusivity and mobility transport coefficients (which might be nonlinear functions of the local density).
The last term 
$\xi(x,t)$ is a \emph{weak} Gaussian white noise, 
such that
\begin{equation}
\langle \xi(x,t)\rangle=0, \qquad \langle \xi(x,t) \xi(x',t')\rangle = \frac{\sigma(\rho)}{L} \delta(x-x') \delta(t-t') \, .
\label{noise}
\end{equation}
This noise term accounts for all fast degrees of freedom which are integrated out in the coarse-graining proceduce which results in the mesoscopic hydrodynamic description (\ref{hydro})-(\ref{current}).
After some relaxation time, a system described by the above set of equations reaches a nonequilibrium steady state characterized by a (typically inhomogeneous) density profile $\rho_{\text{st}}(x)$ compatible with the above boundary conditions, and a nonzero average current $\la q\ra = -D(\rho_{\text{st}})\partial_x \rho_{\text{st}} +\sigma(\rho_{\text{st}}) E$ constant across space. Note that, for WASEP, the two key transport coefficients 
are $D(\rho)=\frac{1}{2}$ and $\sigma(\rho)=\rho(1-\rho)$ \cite{derrida98a,chou11a}, and Section \S\ref{app2} below describes the steady-state solution of the above hydrodynamic equations for the $1d$ open WASEP.

A simple path integral calculation starting from Eqs. (\ref{hydro})-(\ref{current}) then shows that the probability of a given field trajectory $\{\rho,j\}_0^{\tau}$  obeys a large deviation principle of the form $P(\{\rho,j\}_0^{\tau})\sim \exp(-L{\cal I}_{\tau}[\rho,j])$, with an action given by \cite{bertini15a,derrida07a,hurtado14a}
\be
{\cal I}_{\tau}[\rho,j] = \int_0^{\tau}dt\int_0^1dx \frac{[j+D(\rho)\partial_x\rho-E \sigma(\rho)]^2}{2\sigma(\rho)} \, ,
\label{Ipath}
\ee
with $\rho(x,t)$ and $j(x,t)$ coupled via the continuity equation $\partial_t\rho + \partial_x j = 0$ (in any other case ${\cal I}_{\tau}[\rho,j]\to \infty$). We are interested here in the joint statistics for fluctuations of the spacetime-integrated current $q$ and mass $m$. These two empirical observables are defined as
\ben
q &= & \frac{1}{\tau} \int_0^\tau dt\int_0^1 dx \, j(x,t) \, , \label{Acurrent}\\
m &= & \frac{1}{\tau} \int_0^\tau dt\int_0^1 dx \, \rho(x,t) \label{Amass} \, .
\een
The probability of observing a given $q$ and $m$ can now be written as a path integral over all possible trajectories $\{\rho,j\}_0^{\tau}$, weighted by its probability measure $P(\{ j,\rho\}_0^{\tau})$, and restricted to those trajectories compatible with the values of $q$ and $m$ in Eqs.~(\ref{Acurrent}) and (\ref{Amass}), respectively, the continuity equation (\ref{hydro}) at every point of space and time, and the fixed boundary conditions for the density field. For long times and large system sizes, this sum over trajectories is dominated by the associated saddle point and scales as $P(m,q)\sim \exp\{-\tau L G(m,q)\}$, where $G(m,q)$ is the mass-current large deviation function (LDF) given by
\beq
G(m,q)=\lim_{\tau \rightarrow \infty} \frac{1}{\tau}\min_{\{\rho,j\}_0^{\tau}}{\cal I}_{\tau} (\rho,j) \, .
\label{LDF0}
\eeq
The density and current fields solution of this variational problem, denoted here as 
$\rho_{m,q}(x,t)$ and $j_{m,q}(x,t)$, 
can be interpreted as the optimal trajectory the system follows in order to sustain a long-time mass and current joint fluctuation, and are in general time-dependent. 

However, in most applications of MFT to study fluctuations of time-integrated observables in open systems, such as (\ref{Acurrent})-(\ref{Amass}), it has been found that the optimal trajectory $\{\rho_{m,q},j_{m,q}\}_0^\tau$ is indeed \emph{time-independent}. Physically this means that, in order sustain a given mass-current long-time fluctuation, the system of interest settles after a negligible initial transient into a time-independent state (possibly followed by an equally negligible final transient). This property, known as Additivity Principle in literature \cite{bodineau04a,bertini05a,bertini06a,hurtado09a,hurtado11a,saito11a,hurtado11b,perez-espigares13a,hurtado13a,znidaric14a,bertini15a,shpielberg16a,perez-espigares16a,tizon-escamilla17a,tizon-escamilla17b}, strongly simplifies the variational problem at hand. In particular, the mass-current LDF now reads
\be
G(m,q) = \min_{\rho(x)} \int_0^1 dx \frac{\displaystyle \left[q+D(\rho)\rho'(x)-\sigma(\rho)E \right]^2}{\displaystyle 2\sigma(\rho)} \, .
\label{LDF1}
\ee
The optimal density profile solution of this simpler variational problem, $\rho_{m,q}(x)$, is subject to to the additional constraint
\be
m=\int_0^1 \rho_{m,q}(x)\,  dx \, , \label{Amass2}
\ee
and the optimal current field is simply $j_{m,q}(x)=q$ due to the continuity equation (\ref{hydro}) and the time-independence of the dominant trajectory. The integral constraint (\ref{Amass2}) can be implemented using a Lagrange multiplier $\lambda$ which will be fixed \emph{a posteriori} to enforce the constraint. We hence define a new function
\be
G(\lambda,q) =  \min_{\rho(x)} \int_0^1 dx \left\{ \frac{\displaystyle \left[q+D(\rho)\rho'(x)-\sigma(\rho)E \right]^2}{\displaystyle 2\sigma(\rho)} -\lambda\rho(x) \right\}\, .
\ee
The optimal density field for this variational problem is the solution of the following Euler-Lagrange equation
\be
q^2\left(\frac{1}{2\sigma(\rho)} \right)' + \frac{E^2}{2}\sigma'(\rho) - \rho''(x) \frac{D(\rho)^2}{\sigma(\rho)} - \rho'(x)^2 \left(\frac{D(\rho)^2}{2\sigma(\rho)} \right)' = \lambda, 
\ee
where the $'$ means derivative with respect to the argument, e.g. $\sigma'(\rho)=\frac{d\sigma(\rho)}{d\rho}$ and $\rho'(x)=\frac{d\rho(x)}{dx}$. Multiplying both sides of this equation by $\rho'(x)$, we arrive easily to
\be
\frac{d}{dx}\left[ \frac{q^2}{2\sigma(\rho)} - \lambda \rho(x) + \frac{E^2}{2}\sigma(\rho) - \rho'(x)^2 \frac{D(\rho)^2}{2\sigma(\rho)} \right] = 0 \, ,
\ee
which can be trivially integrated once to yield
\be
D(\rho)^2 \left( \frac{d\rho(x)}{dx}\right)^2 = q^2 + 2\left(K-\lambda\rho\right) \sigma(\rho) + E^2 \sigma(\rho)^2 \, , 
\label{ED1}
\ee
where $K$ is an integration constant which allows us to fix the correct boundary condition at one of the two ends, $\rho_{\lambda,q}(0)=\rho_{\text{L}}$ and $\rho_{\lambda,q}(1)=\rho_{\text{R}}$ (the other boundary value is given to solve the previous first-order differential equation).
Interestingly, the optimal density field solution of this differential equation does not depend on the sign of the current $q$ or the external field $E$, as they both appear squared in Eq. (\ref{ED1}). This fact is ultimately a macroscopic manifestation of the time-reversibility of microscopic dynamics. The value of the Lagrange multiplier $\lambda=\lambda(m,q)$ can be now fixed by imposing that the total mass associated to the solution $\rho_{\lambda,q}(x)$ of the above differential equation is just $m$, i.e.
\be
m = \int_0^1 \rho_{\lambda,q}(x)\,  dx \, .
\ee
Our aim in the following sections is to solve this variational problem for the $1d$ open WASEP, for which the key transport coefficients are $D(\rho)=1/2$ and $\sigma(\rho)=\rho(1-\rho)$. However, before proceeding with the analysis of fluctuations, we focus briefly on the steady state behavior.

\newpage
\section{Steady state for the $1d$ open WASEP}
\label{app2}

\begin{figure}
\includegraphics[width=18cm]{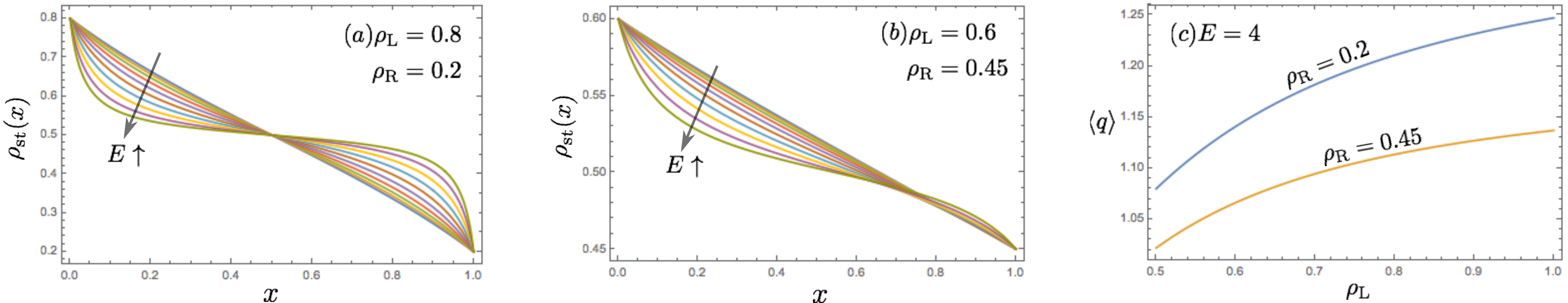}
\caption{(a) Steady-state density profile $\rho_{\text{st}}(x)$ for the $1d$ open WASEP for a symmetric gradient with boundary densities $\rho_{\text{L}}=0.8$ and $\rho_{\text{R}}=0.2$, and external fields $E\in[1,50]$ increasing as $E_k=50^{k/10}$ with $k\in[0,10]$. (b) Same results as in (a), but for an asymmetric gradient with boundary densities $\rho_{\text{L}}=0.6$ and $\rho_{\text{R}}=0.45$. (c) Steady state current $\la q\ra$ vs $\rho_{\text{L}}$ for external field $E=4$ and two right boundary densities, namely $\rho_{\text{R}}=0.2$ and $\rho_{\text{R}}=0.45$.}
\label{fig0SM}
\end{figure}

In this section we derive the steady state current $\la q \ra$ and density profile $\rho_{\text{st}}(x)$ for the $1d$ open WASEP driven by an arbitrary external density gradient and possibly by an additional external field $E$. These steady state properties are given by Fick's law, which for $D(\rho)=1/2$ and $\sigma(\rho)=\rho(1-\rho)$ simply reads
\be
\la q\ra = -\frac{1}{2}\frac{d\rho_{\text{st}}(x)}{dx} + E\rho_{\text{st}}(x)[1-\rho_{\text{st}}(x)] \, ,
\ee
with boundary conditions $\rho_{\text{st}}(0)=\rho_{\text{L}}$ and $\rho_{\text{st}}(1)=\rho_{\text{R}}$.
The previous equation can be easily solved
\be
x = \int_{\rho_{\text{L}}}^{\rho_{\text{st}}(x)} \frac{d\rho}{2[E\rho(1-\rho)-\la q \ra]} = -\frac{1}{\theta}\tan^{-1}\left(\frac{E}{\theta}(2\rho-1) \right)\Bigg |_{\rho_{\text{L}}}^{\rho_{\text{st}}(x)}  \equiv -\frac{1}{\theta} {\cal T}_\theta(\rho)\Bigg |_{\rho_{\text{L}}}^{\rho_{\text{st}}(x)} \, , \nonumber
\ee
with the definitions $\theta\equiv \sqrt{E(4\la q \ra - E)}$ and ${\cal T}_\theta(\rho)\equiv \tan^{-1}\left(\frac{E}{\theta}(2\rho-1) \right)$. Equivalently
\be
x = \frac{1}{\theta}\left [ {\cal T}_\theta(\rho_{\text{L}}) - {\cal T}_\theta(\rho_{\text{st}}(x)) \right] \, . \nonumber 
\ee
By imposing now that $\rho_{\text{st}}(1)=\rho_{\text{R}}$, we obtain an implicit equation for the constant $\theta$, i.e.
\be
\theta = {\cal T}_\theta(\rho_{\text{L}}) - {\cal T}_\theta(\rho_{\text{R}}) \, . \nonumber 
\ee 
This equation for $\theta(\rho_{\text{L}},\rho_{\text{R}},E)$ cannot be solved analytically in general. However, it might be solved numerically for every external parameter 3-tuple $(\rho_{\text{L}},\rho_{\text{R}},E)$. From this solution one can obtain the steady-state current
\be
\la q \ra = \frac{1}{4}\left(E + \frac{\theta^2}{E} \right)
\ee
and the stationary density profile
\be
\rho_{\text{st}}(x) = \frac{1}{2}\left \{1 + \frac{\theta}{E} \tan\left[ {\cal T}_\theta(\rho_{\text{L}}) - \theta x \right]  \right\} \, .
\ee
Fig. \ref{fig0SM} shows steady-state profiles and stationary currents for different values of $(\rho_{\text{L}},\rho_{\text{R}},E)$.

\newpage
\section{Joint mass-current fluctuations in the $1d$ open WASEP}
\label{app3}

We now return to our original problem of determining the optimal density field associated to a mass-current fluctuation in the $1d$ open WASEP. The governing differential equation (\ref{ED1}) reads in this case
\be
\frac{1}{4}\left(\frac{d\rho}{dx} \right)^2 = q^2 + 2(K-\lambda\rho)\rho(1-\rho) + E^2\rho^2(1-\rho)^2 \, , 
\label{ED2}
\ee
with boundary conditions $\rho(0)=\rho_{\text{L}}$ and $\rho(1)=\rho_{\text{R}}$. Without loss of generality, we assume from now on that $\rho_{\text{L}}\ge \rho_{\text{R}}$ and $E>0$; equivalent results to those described below hold in other situations. Note also that the case $E=0$ results in a simpler problem lacking any dynamical phase transition \cite{baek17a}, so it won't be studied here. The rhs of Eq. (\ref{ED2}) defines a fourth-order polynomial in $\rho$,
\be
\pi_0(\rho)\equiv q^2 + 2(K-\lambda\rho)\rho(1-\rho) + E^2\rho^2(1-\rho)^2 \equiv E^2 \pi(\rho) \, ,
\ee
whose roots will play a key role in the analysis of possible solutions. In particular, the real roots of $\pi(\rho)$ (equivalently $\pi_0(\rho)$) define the possible extrema of the optimal density field, though as we discuss next not all real roots correspond necessarily to extrema of the profile. 


A first observation is that, for $\rho_{\text{L}}>\rho_{\text{R}}$, no (local) extrema of the optimal profile $\rho_{\lambda,q}(x)$ can lie within the $\rho$-interval $(\rho_{\text{R}},\rho_{\text{L}})$. To see why, let's assume for a moment that there exists a local extremum $\rho_a\in (\rho_{\text{R}},\rho_{\text{L}})$, i.e. a real root $\rho_a\in \mathbb{R}$ such that $\pi(\rho_a)=0$ and $\pi'(\rho_a)\neq 0$. If $\rho_a=\rho(x_a)$ is a local maximum, it must be reached from below from both sides (as $x\to x_a^\pm$), and this is not possible since $\rho_{\text{L}} > \rho_a$. Equivalently, if $\rho_a$ is a local minimum it should be reached from above from both sides, and this is again not possible because $\rho_{\text{R}}<\rho_a$. Hence no local extrema of the density profile can lie in the interval $(\rho_{\text{R}},\rho_{\text{L}})$. Similarly, only one maximum can exists above $\rho_{\text{L}}$. Indeed, if two maxima $\rho_a>\rho_b>\rho_{\text{L}}$ exist (one local, the other global), they must be separated by a local minimum $\rho_c>\rho_{\text{L}}$. By definition, this local minimum must be reached from above from both sides, and this is again impossible since $\rho_{\text{L}}<\rho_c$. An equivalent argument shows that only one minimum can exists below $\rho_{\text{R}}$. Moreover, a numerical analysis of the differential equation (\ref{ED2}) shows that no inflection points, for which $\pi(\rho)=0=\pi'(\rho)$ simultaneously, are to be expected in the solutions, so we can safely assume that only maxima and minima are possible. These arguments therefore suggest that the optimal density profile solution of the Eq. (\ref{ED2}) can be either (a) monotonous, or contain (b) a single maximum, (c) a single minimum, or at most (d) one maximum and one minimum.



Before embarking on the general solution of the differential equation (\ref{ED2}), 
let us summarize the global solution strategy. As we will show below, the resulting density profile can be written as a rational function of Jacobi elliptic functions (either sn, cn or tn Jacobi functions \cite{byrd71a}, depending on the root structure of the polynomial $\pi(\rho)$ defined above). This density profile will be a parametric function of the current $q$ and the external field $E$, as well as the constants $K$ and $\lambda$, i.e. $\rho(x)=\rho(x;q,E,K,\lambda)$. These two latter constants must be fixed by imposing simultaneously the correct right boundary density $\rho_{\text{R}}$ and the total mass $m$, i.e.
\be
\rho(x=1;q,E,K,\lambda)=\rho_{\text{R}} \, , \qquad \qquad \int_0^1 \rho(x;q,E,K,\lambda) \, dx = m \, .
\ee
Although we find below explicit solutions for $\rho(x;q,E,K,\lambda)$, the simultaneous solution of the previous equations requires numerical methods to determine the values of $K$ and $\lambda$ associated to a joint fluctuation of the current $q$ and mass $m$ under external field $E$. Moreover, the lack of intuition about the possible values of the constants $K$ and $\lambda$ for a given set of parameters $(m,q,E)$ calls for an alternative codification of these two constants in terms of more physical quantities. In particular, defining $\rho'_{\text{L,R}}(m,q,E)\equiv \rho'(x=0,1)$ as the slope of the optimal density profile at the left (L) and right (R) boundary, respectively, which depend on the external parameters $(m,q,E)$, we can see from Eq. (\ref{ED2}) that
\be
\frac{1}{4}(\rho'_{\text{L,R}})^2 = q^2 + 2(K-\lambda\rho_{\text{L,R}})\rho_{\text{L,R}}(1-\rho_{\text{L,R}}) + E^2\rho_{\text{L,R}}^2(1-\rho_{\text{L,R}})^2 \, ,
\ee
which allows to write the constants $K$ and $\lambda$ in terms of the more intuitive boundary slopes $\rho'_{\text{L,R}}(m,q,E)$, i.e.
\be
K(m,q,E) = \frac{\Lambda_{\text{R}}(m,q,E)\rho_{\text{L}} - \Lambda_{\text{L}}(m,q,E)\rho_{\text{R}}}{\rho_{\text{L}}-\rho_{\text{R}}} \, , \qquad \lambda(m,q,E) = \frac{\Lambda_{\text{R}}(m,q,E) - \Lambda_{\text{L}}(m,q,E)}{\rho_{\text{L}}-\rho_{\text{R}}} \, ,
\label{slopes1}
\ee
where we have defined
\be
\Lambda_{\text{L,R}}(m,q,E) \equiv \frac{\frac{1}{4}(\rho'_{\text{L,R}})^2(m,q,E) - q^2 - E^2\rho_{\text{L,R}}^2(1-\rho_{\text{L,R}})^2}{2 \rho_{\text{L,R}}(1-\rho_{\text{L,R}})} \, .
\ee
Hence, for a given external field $E$ and fixed values of the current $q$ and the mass $m$, one has to find numerically the slopes $\rho'_{\text{L,R}}(m,q,E)$ such that
\be
\rho(x=1;q,E,\rho'_{\text{L}},\rho'_{\text{R}})=\rho_{\text{R}} \, , \qquad \qquad \int_0^1 \rho(x;q,\rho'_{\text{L}},\rho'_{\text{R}}) \, dx = m \, ,
\ee
where $\rho(x;q,E,\rho'_{\text{L}},\rho'_{\text{R}})$ is the optimal profile solution of our variational problem. Recalling now that $\rho_{\text{L}}\ge\rho_{\text{R}}$, it is interesting to note that
fixing the sign of the boundary slopes $\rho'_{\text{L,R}}(m,q,E)$ determines whether the resulting profiles is either monotonous ($\rho'_{\text{L}}<0, \rho'_{\text{R}}<0$) or exhibits a single maximum ($\rho'_{\text{L}}>0, \rho'_{\text{R}}<0$), a single minimum ($\rho'_{\text{L}}<0, \rho'_{\text{R}}>0$), or one maximum and one minimum ($\rho'_{\text{L}}>0, \rho'_{\text{R}}>0$; we discuss below the reason why the maximum comes before the minimum).

We turn now to the explicit solution of the ordinary differential equation (\ref{ED2}), which can be written as $\rho'(x) = \pm 2 |E| \sqrt{\pi(\rho)}$, 
where the sign depends on the section of the profile analyzed. Since $\rho_{\text{L}}\ge \rho_{\text{R}}$, monotonous profiles have $\rho'(x)\le 0$ 
$\forall x\in[0,1]$, and the differential equation can be integrated to yield
\be
2|E|x = \int_{\rho(x)}^{\rho_{\text{L}}} \frac{d\rho}{\sqrt{\pi(\rho)}} \qquad \qquad \text{(monotonous profile)}.
\label{prof1}
\ee
For optimal profiles containing a single maximum $\rho_+=\rho(x_+)$, such that $\pi(\rho_+)=0$, we have $\rho'(x)=+2|E|\sqrt{\pi(\rho)}$ $\forall x\in[0,x_+]$ and $\rho'(x)=-2|E|\sqrt{\pi(\rho)}$ $\forall x\in[x_+,1]$, and hence
\be
2|E|x = \left\{ 
\begin{array}{l l}
\displaystyle \int_{\rho_{\text{L}}}^{\rho(x)} \frac{d\rho}{\sqrt{\pi(\rho)}} & \quad 0\le x\le x_+ \\
\phantom{aaa} \\
\displaystyle 2|E| x_+ + \int_{\rho(x)}^{\rho_+} \frac{d\rho}{\sqrt{\pi(\rho)}} & \quad x_+<x\le 1  
\end{array} \right. \qquad \qquad \text{(single-maximum profile)},
\label{prof2}
\ee
where $\displaystyle 2|E| x_+=\int_{\rho_{\text{L}}}^{\rho_+} \frac{d\rho}{\sqrt{\pi(\rho)}}$ defines the position of the maximum. Next, for optimal profiles containing a single minimum $\rho_-=\rho(x_-)$, such that $\pi(\rho_-)=0$, one can show equivalently
\be
2|E|x = \left\{ 
\begin{array}{l l}
\displaystyle \int_{\rho(x)}^{\rho_{\text{L}}} \frac{d\rho}{\sqrt{\pi(\rho)}} & \quad 0\le x\le x_- \\
\phantom{aaa} \\
\displaystyle 2|E| x_- + \int_{\rho_-}^{\rho(x)} \frac{d\rho}{\sqrt{\pi	(\rho)}} & \quad x_-<x\le 1 
\end{array} \right. \qquad \qquad \text{(single-minimum profile)},
\label{prof3}
\ee
where now $\displaystyle 2|E| x_- = \int_{\rho_-}^{\rho_{\text{L}}} \frac{d\rho}{\sqrt{\pi(\rho)}}$ locates the minimum. Finally, for profiles with a maximum $\rho_+=\rho(x_+)$ and a minimum $\rho_-=\rho(x_-)$, with $\pi(\rho_+)=0=\pi(\rho-)$, it is easy to see that
\be
2|E|x = \left\{ 
\begin{array}{l l}
\displaystyle \int_{\rho_{\text{L}}}^{\rho(x)} \frac{d\rho}{\sqrt{\pi(\rho)}} & \quad 0\le x\le x_+ \, ,\\
\phantom{aaa} \\
\displaystyle 2|E| x_+ + \int_{\rho(x)}^{\rho_+} \frac{d\rho}{\sqrt{\pi(\rho)}} & \quad x_+<x\le x_- \, , \\
\phantom{aaa} \\
\displaystyle 2 |E| x_- + \int_{\rho_-}^{\rho(x)} \frac{d\rho}{\sqrt{\pi(\rho)}} & \quad x_-<x\le 1 \, . 
\end{array} \right. \qquad \qquad \text{(max-min profile)},
\label{prof4}
\ee
with
\ben
2 |E| x_+ &=& \displaystyle \int_{\rho_{\text{L}}}^{\rho_+} \frac{d\rho}{\sqrt{\pi(\rho)}} \, , \\
2 |E| x_- &=& \displaystyle 2 |E| x_+ + \int_{\rho_-}^{\rho_+} \frac{d\rho}{\sqrt{\pi(\rho)}} \, .
\een
Here we implicitly assume that $x_+<x_-$, i.e. the maximum comes before the minimum. This is a consequence of the choice $\rho_{\text{L}}\ge \rho_{\text{R}}$, which makes the cost of reversing the extrema non-optimal from a variational point of view, see Eq. (\ref{LDF1}). 

\begin{figure}
\includegraphics[width=18cm]{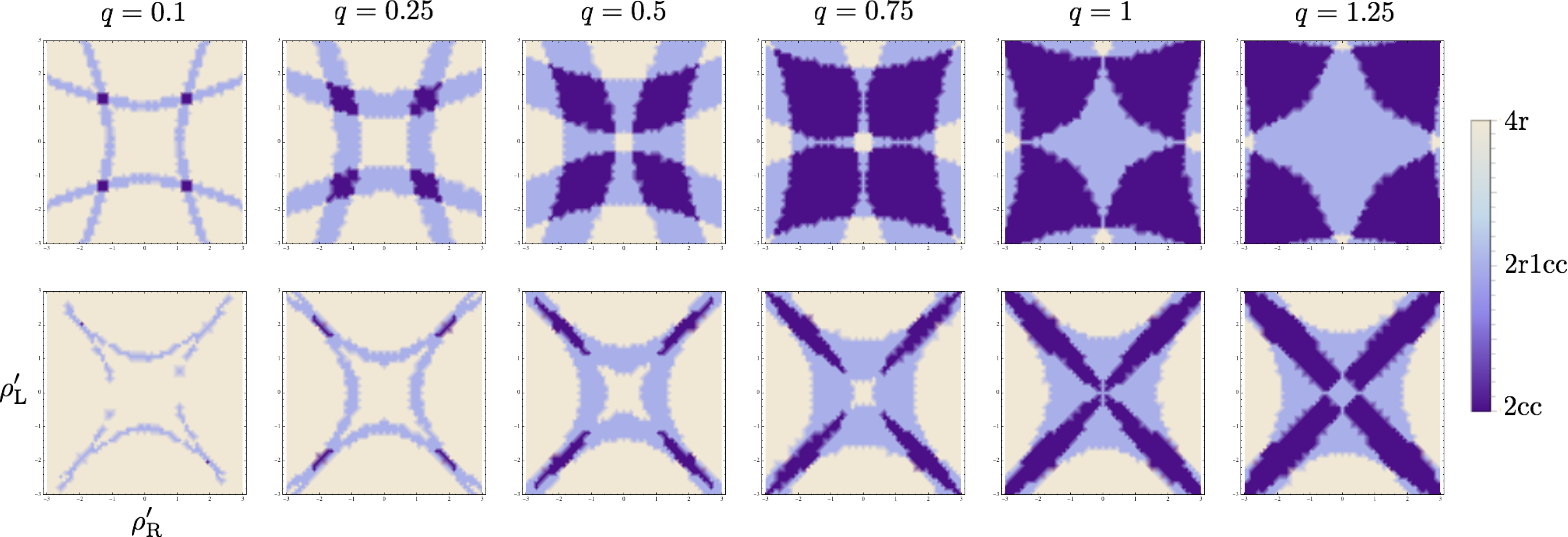}
\caption{Density plot of the structure of zeroes of the polynomial $\pi(\rho)$ as a function of the boundary slopes $\rho'_{\text{L,R}}(m,q,E)\in[-3,3]$ for external field $E=4$ and varying values of the current $q\in[0,1.25]$. Results for two density gradients are shown, namely $(\rho_{\text{L}}=0.8,\rho_{\text{R}}=0.2)$ (symmetric gradient, top row) and $(\rho_{\text{L}}=0.6,\rho_{\text{R}}=0.45)$  (asymmetric gradient, bottom row).}
\label{fig2SM}
\end{figure}

In all cases, the integrals appearing in Eqs. (\ref{prof1})-(\ref{prof4}) are elliptic integrals of the first kind, whose inverse solution can be written in terms of Jacobi elliptic functions \cite{byrd71a}, depending on the structure of zeroes of the 4$^{\text{th}}$-order polynomial $\pi(\rho)$. Since this polynomial is always real, its 4 roots can be either two pairs of complex conjugate numbers ($\rho_1,\rho_1^*,\rho_2,\rho_2^* \in \mathbb{C}$, denoted as case 2cc), two real roots accompanied by a single pair of complex conjugate roots ($\rho_1,\rho_2 \in \mathbb{R}, ~\rho_3,\rho_3^* \in \mathbb{C}$, denoted as case 2r1cc), or 4 different real roots ($\rho_1,\rho_2,\rho_3,\rho_4 \in \mathbb{R}$, denoted as case 4r). Note that all possible combinations do appear in the solution of this variational problem. As an example, Fig. \ref{fig2SM} shows density plots for the structure of zeroes of the polynomial $\pi(\rho)$ for a fixed external field $E=4$ (used below) as a function of the possible boundary slopes of the optimal density field, $\rho'_{\text{L,R}}(m,q,E)$, for two different density gradients. We now study each of the cases 
separately.

\subsection{Two pairs of complex conjugate roots}

In this case, due to the absence of real roots, the optimal density profile must be monotonous. This behavior will be dominant for small mass and current fluctuations, i.e. close to the average behavior. If we denote the complex roots as $\rho_1,\rho_1^*,\rho_2,\rho_2^* \in \mathbb{C}$, the polynomial can be written as $\pi(\rho)=  (\rho-\rho_1)(\rho-\rho_1^*)(\rho-\rho_2)(\rho-\rho_2^*)$.
Defining now $b_i\equiv \text{Re}(\rho_i)$ and $a_i\equiv |\text{Im}(\rho_i)|$, with $i=1,2$, and introducing the constants $A^2 \equiv (b_1-b_2)^2 + (a_1+a_2)^2$, $B^2 \equiv (b_1-b_2)^2 + (a_1-a_2)^2$ and $y_1\equiv b_1-a_1 g_1$, with 
\be
g_1^2\equiv\frac{4a_1^2-(A-B)^2}{(A+B)^2-4a_1^2} \, ,
\ee
we can solve \cite{byrd71a} the integral (\ref{prof1})
\be
2|E|x = \int_{y_1}^{\rho_{\text{L}}} \frac{d\rho}{\sqrt{\pi(\rho)}} - \int_{y_1}^{\rho(x)} \frac{d\rho}{\sqrt{\pi(\rho)}} = \frac{2}{A+B} \left[ F\left(\varphi(\rho_{\text{L}}),\frac{4AB}{(A+B)^2} \right) - F\left(\varphi(\rho(x)),\frac{4AB}{(A+B)^2} \right) \right] \, ,
\ee
with
\be
\varphi(z)\equiv \tan^{-1}\left( \frac{z-b_1+a_1 g_1}{a_1+g_1 b_1 - g_1 z} \right) \, ,
\ee
and where $F(\varphi(z),k^2)$ is the incomplete elliptic integral of the first kind of amplitude $\varphi(z)$ and modulus $k^2$ \cite{byrd71a}. As originally shown by Abel and Jacobi, this elliptic integral can be inverted \cite{byrd71a}. Indeed, if $u\equiv F(\varphi(z),k^2)$, then $\tan\varphi(z) = \text{tn}(u,k^2)$, where $\text{tn}(u,k^2)$ is the Jacobi tn elliptic function \cite{byrd71a}. Applying this inversion formula to
\be
F\Big(\varphi(\rho(x)),\kappa_\varphi^2 \Big) = F_\varphi^L - (A+B) |E| x \, ,
\ee
where we have defined for simplicity $\kappa_\varphi^2\equiv 4AB/(A+B)^2$ and $F_\varphi^L \equiv F\left(\varphi(\rho_{\text{L}}),\kappa_\varphi^2 \right)$, 
and solving for $\rho(x)$ we find for the case of two complex conjugate roots (2cc)
\be
\boxed{\rho_{\text{2cc}}(x)= \frac{\displaystyle (a_1+g_1 b_1) ~\text{tn}\Big[F_\varphi^L- (A+B) |E| x, \kappa_\varphi^2 \Big] + b_1-a_1 g_1 }{\displaystyle 1 + g_1 ~\text{tn}\Big[F_\varphi^L- (A+B) |E| x, \kappa_\varphi^2 \Big]}} \, .
\ee

\subsection{Two real roots, one pair of complex conjugate roots}

We denote the real roots as $\rho_1,\rho_2 \in \mathbb{R}$, while the pair of complex conjugate roots is $\rho_3,\rho_3^* \in \mathbb{C}$. We further assume without loss of generality that $\rho_1<\rho_2$.
Due to the presence of two real roots, the number of possibilities to study increases considerably. In particular, the two real roots can be either:

\begin{itemize}

\item[(i)] \underline{$\rho_1,\rho_2 \ge \rho_{\text{L}}$}. 

In this case the density profile can be monotonous (i1) or it may have a single maximum at $\rho_1$ (i2). 
The polynomial $\pi(\rho)$ can be now written in the region of interest as $\pi(\rho)=(\rho_1-\rho)(\rho_2-\rho)(\rho-\rho_3)(\rho-\rho_3^*)$. Defining now $b_3\equiv \text{Re}(\rho_3)$ and $a_3\equiv |\text{Im}(\rho_3)|$, and introducing the constants $A^2\equiv (\rho_1-b_3)^2 + a_3^2$ and $B^2\equiv (\rho_2-b_3)^2 + a_3^2$, we have for the case (i1) of monotonous profiles, see Eq. (\ref{prof1}), that
\be
2|E| x = \int_{\rho(x)}^{\rho_1} \frac{d\rho}{\sqrt{\pi(\rho)}} - \int_{\rho_{\text{L}}}^{\rho_1} \frac{d\rho}{\sqrt{\pi(\rho)}} = \frac{1}{\sqrt{AB}}\Bigg[ F\Big(\gamma(\rho(x)),\kappa_\gamma^2 \Big) - F_\gamma^L \Bigg] \, ,
\ee
where $F(\gamma(z),\kappa_\gamma^2)$ is the incomplete elliptic integral of the first kind of amplitude $\gamma(z)$ and modulus $\kappa_\gamma^2$ \cite{byrd71a}. 
We have further defined the amplitude function
\be
\gamma(z) \equiv \cos^{-1}\left(\frac{\displaystyle (A-B) z + \rho_1 B - \rho_2 A}{\displaystyle (A+B) z - \rho_1 B - \rho_2 A} \right)^{s_+} \, ,
\ee
as well as the modulus
\be
\kappa_\gamma^2 \equiv s_+ \frac{\displaystyle (A+s_+ B)^2 - (\rho_1-\rho_2)^2}{\displaystyle 4 A B} \, ,
\ee
and the constant $F_\gamma^L \equiv F\Big(\gamma(\rho_{\text{L}}),\kappa_\gamma^2 \Big)$, where we introduce for latter convenience the sign function $s_+\equiv (-1)^{n_+}$, with $n_+$ the number of real roots larger or equal than $\rho_L$ [note that for the current case (i) $s_+=+1$ as $n_+=2$].
As before, if $u\equiv F(\gamma(z),k^2)$, then $\cos\gamma(z) = \text{cn}(u,k^2)$, where $\text{cn}(u,k^2)$ is the Jacobi cosine elliptic function \cite{byrd71a}. Applying this inversion formula to
\be
F\Big(\gamma(\rho(x)),\kappa_\gamma^2 \Big) = F_\gamma^L + 2|E|\sqrt{AB} x 
\ee
and solving for $\rho(x)$ we obtain for the case of two real (2r) and one pair of complex conjugate roots (1cc) in the case (i1) of monotonous profiles
\be
\rho_{\text{2r1cc}}^{\text{(i1)}}(x) = \frac{\displaystyle (\rho_2 A - \rho_1 B) - (\rho_1 B + \rho_2 A)~\text{cn} \Big[F_\gamma^L + 2|E|\sqrt{AB} x, \kappa_\gamma^2 \Big]}{\displaystyle (A-B) - (A+B)~\text{cn} \Big[F_\gamma^L + 2|E|\sqrt{AB} x, \kappa_\gamma^2 \Big]} \, .
\label{casei1}
\ee
Next we consider a profile with a single maximum (i2). In this case, see Eq. (\ref{prof2}),
\be
2|E| x = \left\{ 
\begin{array}{l l}
\displaystyle \int_{\rho_{\text{L}}}^{\rho_1} \frac{d\rho}{\sqrt{\pi(\rho)}} - \int_{\rho(x)}^{\rho_1} \frac{d\rho}{\sqrt{\pi(\rho)}} = \frac{1}{\sqrt{AB}}\Bigg[ F_\gamma^L - F\Big(\gamma(\rho(x)),\kappa_\gamma^2 \Big) \Bigg] & \quad 0\le x\le x_+ \\
\phantom{aaa} \\
\displaystyle  \int_{\rho_{\text{L}}}^{\rho_1} \frac{d\rho}{\sqrt{\pi(\rho)}} + \int_{\rho(x)}^{\rho_1} \frac{d\rho}{\sqrt{\pi(\rho)}} = \frac{1}{\sqrt{AB}}\Bigg[ F_\gamma^L + F\Big(\gamma(\rho(x)),\kappa_\gamma^2 \Big) \Bigg]& \quad x_+<x\le 1  
\end{array} \right.
\ee
where $2|E|x_+ = F_\gamma^L/\sqrt{AB}$. We therefore have
\be
F\Big(\gamma(\rho(x)),\kappa_\gamma^2 \Big) 
= 2 |E| \sqrt{AB}  |x_+ - x| = |F_\gamma^L -  2 |E| \sqrt{AB}  x | \, ,
\ee
which can be inverted to obtain
\be
\rho_{\text{2r1cc}}^{\text{(i2)}}(x) = \frac{\displaystyle (\rho_2 A - \rho_1 B) - (\rho_1 B + \rho_2 A)~\text{cn} \Big[|F_\gamma^L -  2 |E| \sqrt{AB} x |, \kappa_\gamma^2 \Big]}{\displaystyle (A-B) - (A+B)~\text{cn} \Big[|F_\gamma^L -  2 |E| \sqrt{AB} x |, \kappa_\gamma^2 \Big]} \, .
\label{casei2}
\ee
The solution for both the monotonous (i1) and the single-maximum (i2) cases when $\rho_1,\rho_2 \ge \rho_{\text{L}}$ can be now unified by introducing the slope of the optimal profile at the left boundary and its sign. In particular, defining the boundary slopes $\rho'_{\text{L}}\equiv \rho'(0)$ and $\rho'_{\text{R}}\equiv \rho'(1)$, and introducing their sign $s_{\text{L,R}}\equiv \text{sign}(\rho'_{\text{L,R}})$, it's clear that the monotonous profile for $\rho_{\text{L}}\ge \rho_{\text{R}}$ corresponds to $s_{\text{L}}=-1$ while the single-maximum case corresponds to $s_{\text{L}}=+1$, and hence
\be
\rho_{\text{2r1cc}}^{\text{(i)}}(x) = \frac{\displaystyle (\rho_2 A - \rho_1 B) - (\rho_1 B + \rho_2 A)~\text{cn} \Big[|F_\gamma^L -  2 s_{\text{L}} |E| \sqrt{AB} x |, \kappa_\gamma^2 \Big]}{\displaystyle (A-B) - (A+B)~\text{cn} \Big[|F_\gamma^L -  2 s_{\text{L}} |E| \sqrt{AB} x |, \kappa_\gamma^2 \Big]} 
\label{casei}
\ee
represents both solutions for the case (i) $\rho_1,\rho_2 \ge \rho_{\text{L}}$.

\item[(ii)]  \underline{$\rho_1,\rho_2 \le \rho_{\text{R}}$}. 

In this case the density profile can be monotonous (ii1) or it may have a single minimum  (ii2) at $\rho_2$ (since in our notation $\rho_1<\rho_2$). Note that, as in case (i) above, the roots sign function is again $s_+=+1$ since $n_+=0$ here. We proceed now as above and write 
the polynomial $\pi(\rho)$ in the interesting regime as $\pi(\rho)=(\rho-\rho_1)(\rho-\rho_2)(\rho-\rho_3)(\rho-\rho_3^*)$. As before, for the case of monotnonous profiles we may write
\ben
2|E| x &=& \int_{\rho_2}^{\rho_{\text{L}}} \frac{d\rho}{\sqrt{\pi(\rho)}} -  \int_{\rho_2}^{\rho(x)} \frac{d\rho}{\sqrt{\pi(\rho)}} = \frac{1}{\sqrt{AB}}\Bigg[ F\Big(\pi - \gamma(\rho_{\text{L}}),\kappa_\gamma^2 \Big) -  F\Big(\pi - \gamma(\rho(x)),\kappa_\gamma^2 \Big) \Bigg] \nonumber \\
&=& \frac{1}{\sqrt{AB}}\Bigg[ F\Big(\gamma(\rho(x)),\kappa_\gamma^2 \Big) - F_\gamma^L \Bigg] \, ,
\een
where we have used that $\cos^{-1}(-z)=\pi-\cos^{-1}(z)$ and $F(\pi-\gamma,k^2)= 2 K(k^2) - F(\gamma,k^2)$, with $K(k^2)=F(\pi/2,k^2)$ the complete elliptic integral of the first kind. The previous equation once inverted in terms of Jacobi cosine elliptic functions and solved for $\rho(x)$ yields the same Eq. (\ref{casei1}) as in case (i1) above, i.e. $\rho_{\text{2r1cc}}^{\text{(ii1)}}(x) = \rho_{\text{2r1cc}}^{\text{(i1)}}(x)$.

In a similar way, when the profile has a single-minimum 
we have \cite{byrd71a}
\be
2|E| x = \left\{ 
\begin{array}{l l}
\displaystyle \int_{\rho_2}^{\rho_{\text{L}}} \frac{d\rho}{\sqrt{\pi(\rho)}} - \int_{\rho_2}^{\rho(x)} \frac{d\rho}{\sqrt{\pi(\rho)}} = \frac{1}{\sqrt{AB}}\Bigg[ F\Big(\pi-\gamma(\rho_{\text{L}}),\kappa_\gamma^2 \Big) - F\Big(\pi-\gamma(\rho(x)),\kappa_\gamma^2 \Big) \Bigg]  & \quad 0\le x\le x_- \\
\phantom{aaa} \\
\displaystyle  \int_{\rho_2}^{\rho_{\text{L}}} \frac{d\rho}{\sqrt{\pi(\rho)}} + \int_{\rho_2}^{\rho(x)} \frac{d\rho}{\sqrt{\pi(\rho)}} = \frac{1}{\sqrt{AB}}\Bigg[ F\Big(\pi-\gamma(\rho_{\text{L}}),\kappa_\gamma^2 \Big) + F\Big(\pi-\gamma(\rho(x)),\kappa_\gamma^2 \Big) \Bigg]  & \quad x_-<x\le 1  
\end{array} \right.
\ee
or equivalently
\be
2|E| x = \left\{ 
\begin{array}{l l}
\displaystyle \frac{1}{\sqrt{AB}}\Bigg[ F\Big(\gamma(\rho(x)),\kappa_\gamma^2 \Big) - F_\gamma^L \Bigg] & \quad 0\le x\le x_- \\
\phantom{aaa} \\
\displaystyle  \frac{1}{\sqrt{AB}}\Bigg[ 4K(k^2) - F\Big(\gamma(\rho(x)),\kappa_\gamma^2 \Big) - F_\gamma^L \Bigg] & \quad x_-<x\le 1  
\end{array} \right.
\ee
Solving for $F\Big(\gamma(\rho(x)),\kappa_\gamma^2 \Big)$ in the previous piece-wise equation, applying the inversion formula and noting that $\text{cn}(u,k^2)$ is even in $u$ and periodic with period $4K(k^2)$, i.e. $\text{cn}(u+4 K(k^2),k^2) = \text{cn}(u) = \text{cn}(-u)$, see Ref. \cite{byrd71a}, we thus find after solving for the density profile
\be
\rho_{\text{2r1cc}}^{\text{(ii2)}}(x) = \frac{\displaystyle (\rho_2 A - \rho_1 B) - (\rho_1 B + \rho_2 A)~\text{cn} \Big[F_\gamma^L + 2|E|\sqrt{AB} x, \kappa_\gamma^2 \Big]}{\displaystyle (A-B) - (A+B)~\text{cn} \Big[F_\gamma^L + 2|E|\sqrt{AB} x, \kappa_\gamma^2 \Big]} = \rho_{\text{2r1cc}}^{\text{(ii1)}}(x) = \rho_{\text{2r1cc}}^{\text{(i1)}}(x)\, ,
\ee
so the general formula (\ref{casei}) for case (i) is also valid for case (ii) [note that in the latter case the sign of the profile slope at the left boundary is $s_{\text{L}}=-1$].

\item[(iii)]  \underline{$\rho_1\le \rho_{\text{R}},~\rho_2 \ge \rho_{\text{L}}$}. 

In this case the density profile can be monotonous (iii1) or it may a single maximum (iii2), a single minimum (iii3), or a maximum and a minimum (iii4). In all cases the roots sign function is now $s_+=-1$ since $n_+=1$. The polynomial $\pi(\rho)$ can be decomposed as $\pi(\rho)=(\rho-\rho_1)(\rho_2-\rho)(\rho-\rho_3)(\rho-\rho_3^*)$, and for the case (iii1) of monotonous profiles --see Eq. (\ref{prof1})-- we find
\be
2|E| x = \int_{\rho_1}^{\rho_{\text{L}}} \frac{d\rho}{\sqrt{\pi(\rho)}} - \int_{\rho_1}^{\rho(x)} \frac{d\rho}{\sqrt{\pi(\rho)}} = \frac{1}{\sqrt{AB}}\Bigg[ F_\gamma^L - F\Big(\gamma(\rho(x)),\kappa_\gamma^2 \Big) \Bigg] \, ,
\ee
and therefore
\be
\rho_{\text{2r1cc}}^{\text{(iii1)}}(x) = \frac{\displaystyle (\rho_2 A - \rho_1 B) - (\rho_1 B + \rho_2 A)~\left(\text{cn} \Big[F_\gamma^L - 2|E|\sqrt{AB} x, \kappa_\gamma^2 \Big]\right)^{-1}}{\displaystyle (A-B) - (A+B)~\left(\text{cn} \Big[F_\gamma^L - 2|E|\sqrt{AB} x, \kappa_\gamma^2 \Big]\right)^{-1}} \, .
\label{caseiii1}
\ee
When a single maximum is presents, case (iii2), we have
\be
2|E| x = \left\{ 
\begin{array}{l l}
\displaystyle \int_{\rho_1}^{\rho(x)} \frac{d\rho}{\sqrt{\pi(\rho)}} - \int_{\rho_1}^{\rho_{\text{L}}} \frac{d\rho}{\sqrt{\pi(\rho)}}  = \frac{1}{\sqrt{AB}}\Bigg[ F\Big(\gamma(\rho(x)),\kappa_\gamma^2 \Big) - F_\gamma^L \Bigg] & \quad 0\le x\le x_+ \\
\phantom{aaa} \\
\displaystyle  2 \int_{\rho_1}^{\rho_2} \frac{d\rho}{\sqrt{\pi(\rho)}} - \int_{\rho_1}^{\rho_{\text{L}}} \frac{d\rho}{\sqrt{\pi(\rho)}} - \int_{\rho_1}^{\rho(x)} \frac{d\rho}{\sqrt{\pi(\rho)}} = \frac{1}{\sqrt{AB}}\Bigg[ 4 K(\kappa_\gamma^2) - F_\gamma^L - F\Big(\gamma(\rho(x)),\kappa_\gamma^2 \Big) \Bigg]& \quad x_+<x\le 1  
\end{array} \right.
\ee
where the maximum location is given now by $2|E|\sqrt{AB} x_+ = 4 K(\kappa_\gamma^2) - F_\gamma^L/\sqrt{AB}$. Solving for $F\Big(\gamma(\rho(x)),\kappa_\gamma^2 \Big)$, applying the inversion formula and recalling that $\text{cn}(u+4 K(k^2),k^2) = \text{cn}(u) = \text{cn}(-u)$, we thus find after solving for the density profile
\be
\rho_{\text{2r1cc}}^{\text{(iii2)}}(x) = \frac{\displaystyle (\rho_2 A - \rho_1 B) - (\rho_1 B + \rho_2 A)~\left(\text{cn} \Big[F_\gamma^L + 2|E|\sqrt{AB} x, \kappa_\gamma^2 \Big]\right)^{-1}}{\displaystyle (A-B) - (A+B)~\left(\text{cn} \Big[F_\gamma^L + 2|E|\sqrt{AB} x, \kappa_\gamma^2 \Big]\right)^{-1}} \, .
\ee
For the single-minimum case (iii3) we have
\be
2|E| x = \left\{ 
\begin{array}{l l}
\displaystyle \int_{\rho_1}^{\rho_{\text{L}}} \frac{d\rho}{\sqrt{\pi(\rho)}} - \int_{\rho_1}^{\rho(x)} \frac{d\rho}{\sqrt{\pi(\rho)}}  = \frac{1}{\sqrt{AB}}\Bigg[ F_\gamma^L - F\Big(\gamma(\rho(x)),\kappa_\gamma^2 \Big)  \Bigg] & \quad 0\le x\le x_- \\
\phantom{aaa} \\
\displaystyle  \int_{\rho_1}^{\rho_{\text{L}}} \frac{d\rho}{\sqrt{\pi(\rho)}} + \int_{\rho_1}^{\rho(x)} \frac{d\rho}{\sqrt{\pi(\rho)}} = \frac{1}{\sqrt{AB}}\Bigg[ F_\gamma^L + F\Big(\gamma(\rho(x)),\kappa_\gamma^2 \Big) \Bigg]& \quad x_-<x\le 1  
\end{array} \right.
\ee
with $x_- = F_\gamma^L/(2|E|\sqrt{AB})$,
and therefore
\be
\rho_{\text{2r1cc}}^{\text{(iii3)}}(x) = \frac{\displaystyle (\rho_2 A - \rho_1 B) - (\rho_1 B + \rho_2 A)~\left(\text{cn} \Big[|F_\gamma^L - 2|E|\sqrt{AB} x|, \kappa_\gamma^2 \Big]\right)^{-1}}{\displaystyle (A-B) - (A+B)~\left(\text{cn} \Big[|F_\gamma^L - 2|E|\sqrt{AB} x|, \kappa_\gamma^2 \Big]\right)^{-1}} \, .
\ee
Finally, for the case (iii4) with a maximum and a minimum, we can write
\be
2|E| x = \left\{ 
\begin{array}{l l}
\displaystyle \int_{\rho_1}^{\rho(x)} \frac{d\rho}{\sqrt{\pi(\rho)}} - \int_{\rho_1}^{\rho_{\text{L}}} \frac{d\rho}{\sqrt{\pi(\rho)}}  = \frac{1}{\sqrt{AB}}\Bigg[ F\Big(\gamma(\rho(x)),\kappa_\gamma^2 \Big) - F_\gamma^L \Bigg] & \quad 0\le x\le x_+ \\
\phantom{aaa} \\
\displaystyle  2 \int_{\rho_1}^{\rho_2} \frac{d\rho}{\sqrt{\pi(\rho)}} - \int_{\rho_1}^{\rho_{\text{L}}} \frac{d\rho}{\sqrt{\pi(\rho)}} - \int_{\rho_1}^{\rho(x)} \frac{d\rho}{\sqrt{\pi(\rho)}} = \frac{1}{\sqrt{AB}}\Bigg[ 4 K(\kappa_\gamma^2) - F_\gamma^L - F\Big(\gamma(\rho(x)),\kappa_\gamma^2 \Big) \Bigg]& \quad x_+<x\le x_-  \\
\phantom{aaa} \\
\displaystyle  2 \int_{\rho_1}^{\rho_2} \frac{d\rho}{\sqrt{\pi(\rho)}} - \int_{\rho_1}^{\rho_{\text{L}}} \frac{d\rho}{\sqrt{\pi(\rho)}} + \int_{\rho_1}^{\rho(x)} \frac{d\rho}{\sqrt{\pi(\rho)}} = \frac{1}{\sqrt{AB}}\Bigg[ 4 K(\kappa_\gamma^2) - F_\gamma^L + F\Big(\gamma(\rho(x)),\kappa_\gamma^2 \Big) \Bigg]& \quad x_-<x\le 1  
\end{array} \right.
\ee
or equivalently
\be
F\Big(\gamma(\rho(x)),\kappa_\gamma^2 \Big) = \left\{ 
\begin{array}{l l}
\displaystyle F_\gamma^L + 2|E| \sqrt{AB} x & \quad 0\le x\le x_+ \\
\phantom{aaa} \\
\displaystyle  4 K(\kappa_\gamma^2) - (F_\gamma^L + 2|E| \sqrt{AB} x) & \quad x_+<x\le x_-  \\
\phantom{aaa} \\
\displaystyle  (F_\gamma^L + 2|E| \sqrt{AB} x) - 4 K(\kappa_\gamma^2) & \quad x_-<x\le 1  
\end{array} \right.
\ee
where $x_+=(2 K(\kappa_\gamma^2) - F_\gamma^L)/(2|E|\sqrt{AB})$ and $x_-=(4 K(\kappa_\gamma^2) - F_\gamma^L)/(2|E|\sqrt{AB})$. Inverting the previous piecewise equation, taking into account the periodicity of the Jacobi cosine elliptic function $\text{cn}(u,k^2)$, and solving for the density we thus find
\be
\rho_{\text{2r1cc}}^{\text{(iii4)}}(x) = \frac{\displaystyle (\rho_2 A - \rho_1 B) - (\rho_1 B + \rho_2 A)~\left(\text{cn} \Big[F_\gamma^L + 2|E|\sqrt{AB} x, \kappa_\gamma^2 \Big]\right)^{-1}}{\displaystyle (A-B) - (A+B)~\left(\text{cn} \Big[F_\gamma^L + 2|E|\sqrt{AB} x, \kappa_\gamma^2 \Big]\right)^{-1}} \, .
\ee
It is now clear that the four different options for case (iii) with $\rho_1\le \rho_{\text{R}},~\rho_2 \ge \rho_{\text{L}}$ can be unified into a single expression using the sign of the left boundary slope $s_{\text{L}}$, i.e. with the argument of the $\text{cn}$ function written as $|F_\gamma^L + 2s_{\text{L}}|E|\sqrt{AB} x|$. Moreover, using also the roots sign function $s_+$ defined above, we may write the general solution for the case of two real roots and one pair of complex conjugate roots for $\pi(\rho)$ in a compact form
\be
\boxed{\rho_{\text{2r1cc}}(x) = \frac{\displaystyle (\rho_2 A - \rho_1 B) - (\rho_1 B + \rho_2 A)~\left(\text{cn} \Big[|F_\gamma^L - 2 s_{\text{L}} s_+|E|\sqrt{AB} x|, \kappa_\gamma^2 \Big]\right)^{s_+}}{\displaystyle (A-B) - (A+B)~\left(\text{cn} \Big[|F_\gamma^L - 2 s_{\text{L}} s_+|E|\sqrt{AB} x|, \kappa_\gamma^2 \Big]\right)^{s_+}}} \, .
\ee

\end{itemize}

\subsection{Four real roots}

We denote the real roots as $\rho_1<\rho_2<\rho_3<\rho_4 \in \mathbb{R}$, where the label ordering is arbitrary. As in Section A.3.2 above, we should now explore all possible orderings of these 4 real roots with respect to the boundary densities $\rho_{\text{L}}\ge \rho_{\text{R}}$. However, one can check numerically that the only ordering appearing in all cases of interest is that of two real roots above $\rho_{\text{L}}$ and two real roots below $\rho_{\text{R}}$, i.e. $\rho_1<\rho_2< \rho_{\text{R}} \le \rho_{\text{L}} < \rho_3<\rho_4$, in which case the polynomial can be written in the regime of interest as $\pi(\rho)=(\rho_1-\rho)(\rho_2-\rho)(\rho-\rho_3)(\rho-\rho_4)$. Due to the presence of two real roots bracketing the boundary densities, the resulting density profile can be monotonous (iv1), or it may have a single maximum (iv2), a single minimum (iv3), or a maximum and a minimum (iv4). Defining now the constant $g_\phi\equiv \sqrt{(\rho_4-\rho_2)(\rho_3-\rho_1)}$ and 
the amplitude function
\be
\phi(z) \equiv \sin^{-1} \sqrt{\frac{(\rho_4-\rho_2)(\rho_3-z)}{(\rho_3-\rho_2)(\rho_4-z)}} \, ,
\label{iv1}
\ee
together with the modulus
\be
\kappa_\phi^2 \equiv \frac{(\rho_3-\rho_2)(\rho_4-\rho_1)}{(\rho_4-\rho_2)(\rho_3-\rho_1)} \, ,
\label{iv2}
\ee
we find for the monotonous case (iv1) that
\be
2|E| x = \int_{\rho(x)}^{\rho_3} \frac{d\rho}{\sqrt{\pi(\rho)}} - \int_{\rho_{\text{L}}}^{\rho_3} \frac{d\rho}{\sqrt{\pi(\rho)}} = \frac{2}{g_\phi} \Bigg[F\Big(\phi(\rho(x)),\kappa_\phi^2 \Big) - F_\phi^L \Bigg] \, 
\ee
where $F(\phi(z),\kappa_\phi^2)$ is the incomplete elliptic integral of the first kind with amplitude $\phi(z)$ and modulus $\kappa_\phi^2$, see Eqs. (\ref{iv1})  and (\ref{iv2}), and $F_\phi^L\equiv F(\phi(\rho_{\text{L}}),\kappa_\phi^2)$. By noting that if $u\equiv F(\phi(z),k^2)$, then $\sin\gamma(z) = \text{sn}(u,k^2)$, where $\text{sn}(u,k^2)$ is the Jacobi sine elliptic function \cite{byrd71a}, we thus find
\be
\frac{(\rho_4-\rho_2)(\rho_3-\rho(x))}{(\rho_3-\rho_2)(\rho_4-\rho(x))} = \text{sn}^2\Big(g_\phi |E| x + F_\phi^L,\kappa_\phi^2\Big) \, ,
\ee
which can be solved for $\rho(x)$ to yield
\be
\rho_{\text{4r}}^{\text{(iv1)}}(x) = \rho_4 \frac{\displaystyle A_\phi ~\text{sn}^2\Big(g_\phi |E| x + F_\phi^L,\kappa_\phi^2\Big) - \rho_3/\rho_4}{\displaystyle A_\phi ~\text{sn}^2\Big(g_\phi |E| x + F_\phi^L,\kappa_\phi^2\Big) -  1}  \, ,
\label{soliv1}
\ee
where $A_\phi\equiv (\rho_3-\rho_2)/(\rho_4-\rho_2)$ is another constant. For the case (iv2) of profiles exhibiting a single maximum, proceeding as in previous examples one simply obtains
\be
F\Big(\phi(\rho(x)),\kappa_\phi^2 \Big) = \left\{ 
\begin{array}{l l}
\displaystyle F_\phi^L - g_\phi |E| x  & \quad 0\le x\le x_+ \\
\phantom{aaa} \\
\displaystyle  -(F_\phi^L - g_\phi |E| x) & \quad x_+<x\le 1  
\end{array} \right.
\ee
where the maximum location is defined by $g_\phi |E| x_+ = F_\phi^L$. Inverting the previous equation and solving for the density field we hence find
\be
\rho_{\text{4r}}^{\text{(iv2)}}(x) = \rho_4 \frac{\displaystyle A_\phi ~\text{sn}^2\Big(\big|F_\phi^L - g_\phi |E| x\big|,\kappa_\phi^2\Big) - \rho_3/\rho_4}{\displaystyle A_\phi ~\text{sn}^2\Big(\big|F_\phi^L - g_\phi |E| x\big|,\kappa_\phi^2\Big) -  1}  \, .
\label{soliv2}
\ee
For the single minimum case (iv3), we have
\be
2|E| x = \left\{ 
\begin{array}{l l}
\displaystyle \int_{\rho(x)}^{\rho_3} \frac{d\rho}{\sqrt{\pi(\rho)}} - \int_{\rho_{\text{L}}}^{\rho_3} \frac{d\rho}{\sqrt{\pi(\rho)}}  = \frac{2}{g_\phi}\Bigg[ F\Big(\phi(\rho(x)),\kappa_\phi^2 \Big) - F_\phi^L \Bigg] & \quad 0\le x\le x_- \\
\phantom{aaa} \\
\displaystyle  2 \int_{\rho_2}^{\rho_3} \frac{d\rho}{\sqrt{\pi(\rho)}} - \int_{\rho_{\text{L}}}^{\rho_3} \frac{d\rho}{\sqrt{\pi(\rho)}} -  \int_{\rho(x)}^{\rho_3} \frac{d\rho}{\sqrt{\pi(\rho)}}   = \frac{2}{g_\phi}\Bigg[ 2 K(\kappa_\phi^2) - F_\phi^L - F\Big(\phi(\rho(x)),\kappa_\phi^2 \Big) \Bigg] & \quad x_-<x\le 1  
\end{array} \right.
\ee
or equivalently
\be
F\Big(\phi(\rho(x)),\kappa_\phi^2 \Big) = \left\{ 
\begin{array}{l l}
\displaystyle F_\phi^L + g_\phi |E| x  & \quad 0\le x\le x_- \\
\phantom{aaa} \\
\displaystyle  2 K(\kappa_\phi^2) - (F_\phi^L + g_\phi |E| x)  & \quad x_-<x\le 1  
\end{array} \right.
\ee
This expression can be easily inverted by noting \cite{byrd71a} that $\text{sn}(u+2K(k^2),k^2) = -\text{sn}(u,k^2) = \text{sn}(-u,k^2)$, and solving for the density profile we thus obtain $\rho_{\text{4r}}^{\text{(iv3)}}(x)=\rho_{\text{4r}}^{\text{(iv1)}}(x)$, i.e. the same expression as in case (iv1) above, see Eq. (\ref{soliv1}).
Finally, for the case (iv4) of a profile with a maximum and a minimum, we have
\be
2|E| x = \left\{ 
\begin{array}{l l}
\displaystyle \int_{\rho_{\text{L}}}^{\rho_3} \frac{d\rho}{\sqrt{\pi(\rho)}} - \int_{\rho(x)}^{\rho_3} \frac{d\rho}{\sqrt{\pi(\rho)}} = \frac{2}{g_\phi}\Bigg[ F_\phi^L - F\Big(\phi(\rho(x)),\kappa_\phi^2 \Big) \Bigg] & \quad 0\le x\le x_+ \\
\phantom{aaa} \\
\displaystyle \int_{\rho_{\text{L}}}^{\rho_3} \frac{d\rho}{\sqrt{\pi(\rho)}} + \int_{\rho(x)}^{\rho_3} \frac{d\rho}{\sqrt{\pi(\rho)}} = \frac{2}{g_\phi}\Bigg[ F_\phi^L + F\Big(\phi(\rho(x)),\kappa_\phi^2 \Big) \Bigg] & \quad x_+<x\le x_-  \\
\phantom{aaa} \\
\displaystyle 2 \int_{\rho_2}^{\rho_3} \frac{d\rho}{\sqrt{\pi(\rho)}} + \int_{\rho_{\text{L}}}^{\rho_3} \frac{d\rho}{\sqrt{\pi(\rho)}} - \int_{\rho(x)}^{\rho_3} \frac{d\rho}{\sqrt{\pi(\rho)}} = \frac{2}{g_\phi}\Bigg[ 2 K(\kappa_\phi^2) + F_\phi^L - F\Big(\phi(\rho(x)),\kappa_\phi^2 \Big) \Bigg] & \quad x_-<x\le 1  
\end{array} \right.
\ee
or equivalently
\be
F\Big(\phi(\rho(x)),\kappa_\phi^2 \Big) = \left\{ 
\begin{array}{l l}
\displaystyle F_\phi^L - g_\phi |E| x & \quad 0\le x\le x_+ \\
\phantom{aaa} \\
\displaystyle  -(F_\phi^L - g_\phi |E| x) & \quad x_+<x\le x_-  \\
\phantom{aaa} \\
\displaystyle  2 K(\kappa_\phi^2) + F_\phi^L - g_\phi |E| x & \quad x_-<x\le 1  
\end{array} \right.
\ee
with $g_\phi |E| x_+ = F_\phi^L$ and $g_\phi |E| x_- = F_\phi^L + K(\kappa_\phi^2)$. Using again the periodicity of the Jacobi elliptic sn function, and solving for the density profile, it is easy to find that $\rho_{\text{4r}}^{\text{(iv4)}}(x)=\rho_{\text{4r}}^{\text{(iv2)}}(x)$, i.e. the same expression as in case (iv2) above, see Eq. (\ref{soliv2}). Moreover, all expressions for cases (iv1)--(iv4) (when $\pi(\rho)$ has four real roots) can be unified into a single formula by making use again of the left boundary slope sign function $s_{\text{L}}$, i.e. the sign of the slope of the density field $\rho(x)$ at $x=0$. The result is
\be
\boxed{\rho_{\text{4r}}(x) = \rho_4 \frac{\displaystyle A_\phi ~\text{sn}^2\Big(\big|F_\phi^L - s_{\text{L}} g_\phi |E| x\big|,\kappa_\phi^2\Big) - \rho_3/\rho_4}{\displaystyle A_\phi ~\text{sn}^2\Big(\big|F_\phi^L - s_{\text{L}} g_\phi |E| x\big|,\kappa_\phi^2\Big) -  1}} \, .
\ee
In summary, the general solution for the optimal density field associated to a joint mass and current fluctuation in the $1d$ weakly assymmetric simple exclusion process in contact with boundary reservoirs at densities $\rho_{\text{L}}\ge \rho_{\text{R}}$ and subject to an external driving field $E$ can be written as
\be
\boxed{
\rho(x) = \left\{ 
\begin{array}{l l}
\displaystyle  \frac{\displaystyle (a_1+g_1 b_1) ~\text{tn}\Big[F_\varphi^L- (A+B) |E| x, \kappa_\varphi^2 \Big] + b_1-a_1 g_1 }{\displaystyle 1 + g_1 ~\text{tn}\Big[F_\varphi^L- (A+B) |E| x, \kappa_\varphi^2 \Big]}  & \quad \text{(2cc)}  \\
\phantom{aaa} \\
\displaystyle  \frac{\displaystyle (\rho_2 A - \rho_1 B) - (\rho_1 B + \rho_2 A)~\left(\text{cn} \Big[|F_\gamma^L - 2 s_{\text{L}} s_+|E|\sqrt{AB} x|, \kappa_\gamma^2 \Big]\right)^{s_+}}{\displaystyle (A-B) - (A+B)~\left(\text{cn} \Big[|F_\gamma^L - 2 s_{\text{L}} s_+|E|\sqrt{AB} x|, \kappa_\gamma^2 \Big]\right)^{s_+}}   & \quad \text{(2r1cc)} \\
\phantom{aaa} \\
\displaystyle \rho_4 \frac{\displaystyle A_\phi ~\text{sn}^2\Big(\big|F_\phi^L - s_{\text{L}} g_\phi |E| x\big|,\kappa_\phi^2\Big) - \rho_3/\rho_4}{\displaystyle A_\phi ~\text{sn}^2\Big(\big|F_\phi^L - s_{\text{L}} g_\phi |E| x\big|,\kappa_\phi^2\Big) -  1} & \quad \text{(4r)} 
\end{array} \right.
}
\ee
where the relevant constants in each case are defined above.

\begin{figure}
\includegraphics[width=18cm]{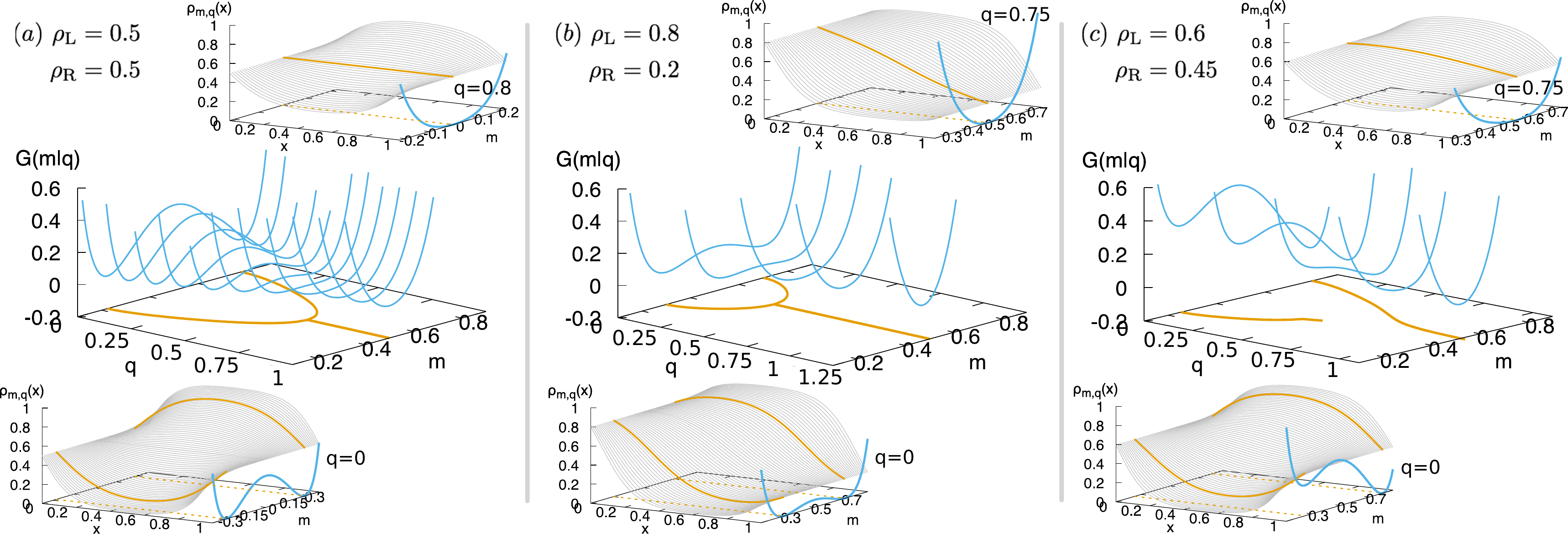}
\caption{Middle row: Conditional LDF $G(m|q)=G(m,q)-G(q)$ as a function of the mass $m$ for different currents $q$ for three different boundary drivings, namely (a) $\rho_{\text{L}}=0.5, \rho_{\text{R}}=0.5$ (symmetric driving), (b) $\rho_{\text{L}}=0.8, \rho_{\text{R}}=0.2$ (symmetric driving), and (c) $\rho_{\text{L}}=0.6, \rho_{\text{R}}=0.45$ (asymmetric driving). The lines projected in the $m-q$ plane correspond to the local minima of the LDF $G(m|q)$, which define the mass $m_q$ associated to a current fluctuation $q$. In the symmetry-broken regime this defines the low- and high-mass branches  $m_q^\pm$. Botton row: optimal density profiles $\rho_{m,q}(x)$ obtained for $q=0$ and the three different boundary drivings. The thick lines are the optimal profiles associated to the local minima $m_q^\pm$ of $G(m|q)$. For completeness the associated $G(m|q)$ is also shown. Top row: optimal density profiles in each caso, for a current in the PH-symmetric region, $|q|>q_c$.}
\label{fig1SM}
\end{figure}

Using this result, it is now possible to study analytically the dynamical phase transition described in the main text for arbitrary boundary gradient (symmetric or asymmetric), well beyond the perturbative nonequilibrium linear regime. In particular, for PH-symmetric boundaries ($\rho_{\text{R}}=1-\rho_{\text{L}}$), the conditional mass-current LDF $G(m|q)\equiv G(m,q)-G(q)$ exhibits a peculiar change of behavior at a critical current $|q_c|$, see Figs. \ref{fig1SM}.a-b: while for $|q|>|q_c|$ the LDF $G(m|q)$ displays a single minimum at $m_q=1/2$, with an associated PH-symmetric optimal profile (top insets in Figs. \ref{fig1SM}.a-b), for $|q|<|q_c|$ two equivalent minima $m_q^\pm$ appear in $G(m|q)$, each one associated with a PH-symmetry-broken optimal profile $\rho_q^\pm(x)$, see bottom insets in Figs. \ref{fig1SM}.a-b, such that $\rho_q^\pm(x) \to 1-\rho_q^\mp(1-x)$. The emergence of this non-convex regime in $G(m|q)$ signals a $2^{\text{nd}}$-order DPT to a PH-symmetry-broken dynamical phase. Note that this happens both for equal boundary densities ($\rho_{\text{R}}=0.5=\rho_{\text{L}}$, Fig. \ref{fig1SM}.a) and for large but symmetric boundary gradients ($\rho_{\text{L}}=0.8, \rho_{\text{R}}=0.2$, Fig. \ref{fig1SM}.b). On the other hand, for PH-asymmetric boundaries ($\rho_{\text{R}}\neq 1-\rho_{\text{L}}$, as e.g. $\rho_{\text{L}}=0.6, \rho_{\text{R}}=0.45$, see Fig. \ref{fig1SM}.c), the governing action (\ref{Ipath}) is no longer PH-symmetric: the asymmetry favors one of the mass branches and the associated $G(m|q)$ displays a single \emph{global} minimum $\forall q$, see Fig. \ref{fig1SM}.c, and an unique optimal profile. Still, $G(m|q)$ becomes non-convex for low enough currents, and for weak gradient asymmetry, as is the case for $\rho_{\text{L}}=0.6, \rho_{\text{R}}=0.45$ shown in Fig. \ref{fig1SM}.c, metastable-like local minima in $G(m|q)$ may appear.

The mass $m_q$ where the minima of $G(m|q)$ appear for a fixed $q$ is evaluated by demanding $\frac{d G(m|q)}{dm} = \frac{d G(m,q)}{dm}=0$. The $m$-slope of the LDF $G(m,q)$ at a given $(m,q)$-point is simply given by the Lagrange multiplier $\lambda(m,q)$ used to impose the mass constraint, so
\be
\frac{d G(m,q)}{dm}\bigg\vert_{m_q} = \lambda(m_q,q) = 0 \quad \Rightarrow \quad \Lambda_{\text{L}}(m_q,q,E) = \Lambda_{\text{R}}(m,q,E) \, ,
\ee
where we have used the formula which relates the Lagrange multiplier $\lambda(m,q)$ with the boundary slopes $\rho'_{\text{L,R}}(m,q,E)$ of the optimal density profile, see Eq. (\ref{slopes1}) in \S\ref{app3} above, with the definition
\be
\Lambda_{\text{L,R}}(m,q,E) \equiv \frac{\frac{1}{4}(\rho'_{\text{L,R}})^2(m,q,E) - q^2 - E^2\rho_{\text{L,R}}^2(1-\rho_{\text{L,R}})^2}{2 \rho_{\text{L,R}}(1-\rho_{\text{L,R}})} \, .
\ee
In this way, defining $\sigma_{\text{L,R}}\equiv \rho_{\text{L,R}}(1-\rho_{\text{L,R}})$, the equation for the mass minima $m_q$ for a fixed $q$ is
\be
\frac{1}{4\sigma_{\text{L}}}(\rho'_{\text{L}})^2(m_q,q,E) - \frac{1}{4\sigma_{\text{R}}}(\rho'_{\text{R}})^2(m_q,q,E) = q^2\left(\frac{1}{\sigma_{\text{L}}} - \frac{1}{\sigma_{\text{R}}} \right) + E^2 (\sigma_{\text{L}}-\sigma_{\text{R}}) \, .
\ee
The critical current $q_c$ can be evaluated as well by demanding that
\be
\frac{d G(m,q)}{dm}\bigg\vert_{m_{q_c},q_c} = 0 = \frac{d^2 G(m,q)}{dm^2}\bigg\vert_{m_{q_c},q_c} \, ,
\ee
which leads to the following pair of equations
\ben
\frac{1}{4\sigma_{\text{L}}}(\rho'_{\text{L}})^2(m_{q_c},q_c,E) - \frac{1}{4\sigma_{\text{R}}}(\rho'_{\text{R}})^2(m_{q_c},q_c,E) &=&  q_c^2\left(\frac{1}{\sigma_{\text{L}}} - \frac{1}{\sigma_{\text{R}}} \right) + E^2 (\sigma_{\text{L}}-\sigma_{\text{R}}) \, , \\
\frac{\rho'_{\text{L}}(m,q,E)}{\sigma_{\text{L}}} \frac{d \rho'_{\text{L}}(m,q,E)}{dm} \bigg\vert_{m_{q_c},q_c} & = & \frac{\rho'_{\text{R}}(m,q,E)}{\sigma_{\text{R}}} \frac{d \rho'_{\text{R}}(m,q,E)}{dm} \bigg\vert_{m_{q_c},q_c} \, .
\een
Note that these equations for $m_q$ and for $q_c$ must be solved numerically due to the nonlinear character of the problem.

\newpage
\section{Instanton solution, Maxwell-like construction and violation of additivity principle}
\label{app4}

In this section we build a time-dependent, instanton-like solution for the optimal density and current fields responsible of a joint fluctuation of the empirical current and mass. We further show that this solution improves the additivity principle prediction (i.e. yields a better minimizer of the MFT action) in the regime where the joint current-mass LDF becomes non-convex. 
This result demonstrates that time-dependent solutions of the MFT problem in open systems exist and dominate fluctuation behavior in dynamical coexistence regimes emerging at DPTs. 

We start from the general expression derived above for the joint mass-current LDF, see Eq. (\ref{LDF0}),
\be
G(m,q)=\lim_{\tau \rightarrow \infty} \frac{1}{\tau}\min_{\{\rho,j\}_0^{\tau}}\int_0^{\tau}dt\int_0^1dx \frac{[j+D(\rho)\partial_x\rho-E \sigma(\rho)]^2}{2\sigma(\rho)} \, ,
\label{LDFA0}
\ee
with the fields $\rho(x,t)$ and $j(x,t)$ coupled at every point of space and time via the continuity equation, $\partial_t\rho + \partial_x j = 0$. Moreover, the density and current fields are further constrained to yield empirical values
\ben
q &= & \frac{1}{\tau} \int_0^\tau dt\int_0^1 dx \, j(x,t) \, , \label{Acurrent0}\\
m &= & \frac{1}{\tau} \int_0^\tau dt\int_0^1 dx \, \rho(x,t) \label{Amass0} \, ,
\een
and boundary conditions for the density field are such that $\rho(0,t)=\rho_{\text{L}}$ and $\rho(1,t)=\rho_{\text{R}}$ $\forall t$. We have seen in previous sections of the SM that, under the additivity conjecture \cite{bodineau04a}, the joint mass-current LDF is simplified to
\be
G_{\text{ad}}(m,q) = \min_{\rho(x)} \int_0^1 dx \frac{\displaystyle \left[q+D(\rho)\rho'(x)-\sigma(\rho)E \right]^2}{\displaystyle 2\sigma(\rho)} \, ,
\label{LDFA1}
\ee
with a reduced set of constraints (i.e. boundary densities, and total mass). We denote in this section as $\rho_{m,q}^{\text{ad}}(x)$ the optimal density profile responsible of a joint mass and current fluctuation under the additivity hypothesis. 
To search for violations of the additivity principle, we focus our attention in current fluctuations $|q|\le q_c$ below the critical point in systems driven by a \emph{symmetric} density gradient ($\rho_{\text{R}}=1-\rho_{\text{L}}$). In this regime we conjecture a solution for the
optimal \emph{trajectory} responsible of a given mass-current fluctuation, which is time-dependent for masses where $G(m,q)$ is non-convex.
In particular, our ansatz in this regime is
\be
\rho_{m,q}(x,t) = \left \{ 
\begin{array}{ll}
\rho_{m,q}^{\text{ad}}(x) & \text{if } m<m_q^- \text{ or } m>m_q^+ \\
\phantom{aaa} & \phantom{bbb} \\
\rho_{m_q^-,q}^{\text{ad}}(x) \left[1-\phi(t-t_{m,q}) \right] + \rho_{m_q^+,q}^{\text{ad}}(x)~\phi(t-t_{m,q}) & \text{if } m_q^-\le m\le m_q^+ 
\end{array}
\right.
\label{inst1}
\ee 
where $m_q^\pm$ are the masses of the optimal density profiles associated to a current fluctuation $|q|\le q_c$ in the PH symmetry broken regime along the high-mass ($+$) and low-mass ($-$) branches. 
The time-dependent function $\phi(t)$ is a sufficiently smooth localized crossover function such that $\phi(t)=0$ $\forall t< -\frac{\delta t}{2}$ and $\phi(t)=1$ $\forall t> \frac{\delta t}{2}$,
with $\delta t$ a fixed timescale. 
The crossover time $t_{m,q}$ in Eq (\ref{inst1}) can be determined now by imposing the constraint on the empirical mass, Eq. (\ref{Amass0}). In particular
\ben
\hspace{-2cm} m &=&  \frac{1}{\tau} \int_0^\tau dt\int_0^1 dx \, \rho_{m,q}(x,t) = \left(\frac{t_{m,q}-\frac{\delta t}{2}}{\tau}\right) m_q^- +  \left(\frac{\tau-(t_{m,q}+\frac{\delta t}{2})}{\tau}\right) m_q^+ + \frac{1}{\tau} \int_{t_{m,q}-\frac{\delta t}{2}}^{t_{m,q}+\frac{\delta t}{2}} dt \int_0^1 dx \rho_{m,q}(x,t) \nonumber \\
&=&  \frac{t_{m,q}}{\tau} m_q^- + \left(1-\frac{t_{m,q}}{\tau}\right) m_q^+ + \frac{1}{\tau}\left[ -\delta t + \int_{t_{m,q}-\frac{\delta t}{2}}^{t_{m,q}+\frac{\delta t}{2}} dt \int_0^1 dx \rho_{m,q}(x,t)\right] \, .
\een
The third term in the rhs of the last equation is $\sim{\cal O}(\delta t/\tau)$, so in the long-time limit ($\tau \to \infty$) and for a fixed crosscover time $\delta t$ this term tends to zero, and hence we find $t_{m,q} = p~\tau$ with the definition
\be
p = \frac{m_q^+-m}{m_q^+-m_q^-} \, .
\ee

As mentioned above, the time-dependent optimal density field $\rho_{m,q}(x,t)$ must obey at all points of space and time a continuity equation $\partial_t \rho_{m,q}(x,t) + \partial_x j_{m,q}(x,t)=0$. To obtain the optimal time-dependent current field $j_{m,q}(x,t)$ for $m_q^-\le m\le m_q^+$, we first note that in this case
\be
\partial_t \rho_{m,q}(x,t) = \left \{ 
\begin{array}{ll}
0 & \text{if } t\notin [t_{m,q}-\frac{\delta t}{2},t_{m,q}+\frac{\delta t}{2}] \\
\phantom{aaa} & \phantom{bbb} \\
\left[ \rho_{m_q^+,q}^{\text{ad}}(x)-\rho_{m_q^-,q}^{\text{ad}}(x) \right] ~\phi'(t-t_{m,q}) & \text{if } t\in [t_{m,q}-\frac{\delta t}{2},t_{m,q}+\frac{\delta t}{2}] 
\end{array}
\right.
\label{inst2}
\ee 
Therefore the continuity constraint in the mass regime $m_q^-\le m\le m_q^+$ leads to the following optimal current trajectory
\be
j_{m,q}(x,t) = \left \{ 
\begin{array}{ll}
q & \text{if } t\notin [t_{m,q}-\frac{\delta t}{2},t_{m,q}+\frac{\delta t}{2}]  \\
\phantom{aaa} & \phantom{bbb} \\
\chi(x)~\phi'(t-t_{m,q}) & \text{if } t\in [t_{m,q}-\frac{\delta t}{2},t_{m,q}+\frac{\delta t}{2}] 
\end{array}
\right.
\label{inst3}
\ee 
where we have already taken into account the constraint on the empirical current $q$, see Eq. (\ref{Acurrent0}). The function $\chi(x)$ is such that $\chi'(x) = \rho_{m_q^+,q}^{\text{ad}}(x)-\rho_{m_q^-,q}^{\text{ad}}(x)$, and we note that the transient regime where $j_{m,q}(x,t)$ is different from $q$ does not contribute to the final value of the empirical current, Eq. (\ref{Acurrent0}), as this transient is negligible against the long-time limit for $\tau$.

Using this ansatz for the optimal trajectory responsible of a mass and current fluctuation in Eq. (\ref{LDFA0}), we obtain for the associated joint LDF 
\ben
G(m,q)&=&\lim_{\tau \rightarrow \infty} \frac{1}{\tau} \int_0^{\tau}dt\int_0^1dx \frac{[j_{m,q}(x,t)+D(\rho_{m,q})\partial_x\rho_{m,q}(x,t)-E \sigma(\rho_{m,q})]^2}{2\sigma(\rho_{m,q})} \nonumber \\
&=& \lim_{\tau \rightarrow \infty} \left[ \left(\frac{t_{m,q}-\frac{\delta t}{2}}{\tau}\right) G_{\text{ad}}(m_q^-,q) + \left(\frac{\tau-(t_{m,q}+\frac{\delta t}{2})}{\tau}\right) G_{\text{ad}}(m_q^+,q) + \frac{1}{\tau} {\cal I}
\right] \, ,
\label{LDFA2}
\een
with the definition
\be
{\cal I} \equiv \int_{t_{m,q}-\frac{\delta t}{2}}^{t_{m,q}+\frac{\delta t}{2}} dt \int_0^1 dx \frac{[j_{m,q}(x,t)+D(\rho_{m,q})\partial_x\rho_{m,q}(x,t)-E \sigma(\rho_{m,q})]^2}{2\sigma(\rho_{m,q})} \, .
\ee
Noting that ${\cal I}\sim {\cal O}(\delta t)$ and using the same arguments as above, we find in the long-time limit $\tau\to\infty$ that
\be
G(m,q) = p~G_{\text{ad}}(m_q^-,q) + (1-p) G_{\text{ad}}(m_q^+,q) \, ,
\ee
which corresponds to the Maxwell construction obtained from $G_{\text{ad}}(m,q)$ in the mass regime $m_q^-\le m\le m_q^+$ where this joint LDF is non-convex (for $|q|\le q_c$), as described above and in the main text. Note that an equivalent argument can be developed for the conditional mass-current LDF $G(m|q)=G(m,q)-G(q)$. This instanton solution corresponds to the dynamical coexistence of the different symmetry-broken phases which appear for $|q|\le q_c$, a behavior typical of $1^{\text{st}}$-order DPTs. Note also that one can generalize the previous solution to PH-asymmetric boundaries in regimes where $G(m,q)$ is non-convex. Finally, we would like to mention that some subtleties of the instanton solution appear for $|q|\approx q_c$ related to the order of the $L\to\infty$ and $\tau\to\infty$ limits, see Ref. \cite{baek18a} for a discussion of this issue.

\newpage
\section{Spectral analysis of the dynamical generator and metastable manifold}
\label{app5}

In this section we perform a spectral analysis of the microscopic dynamics of the $1d$ WASEP in order to better understand the DPT demonstrated above from a microscopic point of view. 
In particular, we will focus on the quasi-degenerate (metastable) states $\ket{P_{MS}^{c_1}}$ and $\ket{P_{MS}^{c_2}}$ introduced in the main text, which contain the information about the optimal trajectories in the symmetry-broken phase. 

At the microscopic level, a configuration of the $1d$ WASEP is given by $C=\{n_k\}_{k=1,\ldots,L}$, where $n_k=0,1$ is the occupation number of the $k^{\text{th}}$-site of the lattice. Within the quantum Hamiltonian formalism for the master equation \cite{schutz01a}, each configuration is then represented as a vector in a Hilbert space
\be
\ket{C}=\bigotimes_{k=1}^L \begin{pmatrix} n_k\\1-n_k \end{pmatrix} \, ,
\ee
and the complete information about the system is contained in a vector $\ket{P}=(P(C_1),P(C_2),...)^T=\sum_i P(C_i)\ket{C_i}$, with $^T$ denoting transposition, such that $P(C_i)$ represents the probability of configuration $C_i$. This probability vector is normalized such that $\langle-|P\rangle=1$ where $\bra{-}=\sum_i\bra{C_i}$ is the vector representing the sum over all possible configurations and $\langle C_i| C_j \rangle=\delta_{ij}$. The probability vector $\ket{P}$ evolves in time according to the master equation 
\be
\partial_t\ket{P}={\mathbb W} \ket{P} \, ,
\ee
where ${\mathbb W}$ defines the Markov generator of the dynamics. Such generator can be \emph{tilted} ${\mathbb W}^{\mu,\lambda}$\cite{touchette09a,hurtado14a} to bias the original stochastic dynamics in order to favor large (low) mass for  $\mu<0$ ($\mu>0$) and large (low) currents for $\lambda > 0$ ($\lambda < 0$), with $\mu$ and $\lambda$ the conjugate parameters to the microscopic mass and current observables, respectively. In particular, the tilted dynamical generator for the $1d$ open WASEP is
\begin{eqnarray}
\nonumber
{\mathbb W}^{\mu,\lambda}&=&\sum_{k=1}^{L-1}[\frac{1}{2}e^{(\lambda+E)/(L-1)} \sigma_{k+1}^+\sigma_k^- + \frac{1}{2}e^{-(\lambda+E)/(L-1)} \sigma_{k}^+\sigma_{k+1}^- \\ \nonumber
&-&\frac{1}{2}e^{E/(L-1)}{\hat n}_k ({\mathbb 1}-{\hat n}_{k+1} )-\frac{1}{2}e^{-E/(L-1)}{\hat n}_{k+1} ({\mathbb 1}-{\hat n}_k )]\\ \nonumber
&+&\alpha [\sigma_1^+ - ({\mathbb 1}-{\hat n}_1)] + \gamma [\sigma_1^- - {\hat n}_1 ]\\ 
&+&\delta [\sigma_L^+ - ({\mathbb 1}-{\hat n}_L)] + \beta [\sigma_L^- - {\hat n}_L ] - \frac{\mu}{L} \sum_{k=1}^L {\hat n}_k\, ,
\label{biasedgenSM}
\end{eqnarray}
and we recall (see main text)
that $\alpha$ and $\gamma$ ($\delta$ and $\beta$) are the injection and extraction rates at the leftmost (rightmost) site, respectively.
In the previous expression ${\mathbb 1}$ is the identity matrix and ${\hat n}_k=\sigma_k^+ \sigma_k^-$ is the number operator at site $k\in[1,L]$, where $\sigma_k^+$ and $\sigma_k^-$ are the creation and annihilation operators given by $\sigma_k^{\pm}=(\sigma_k^x \pm i \sigma_k^y)/2$ respectively, with $\sigma_k^{x,y}$ the standard $x,y$-Pauli matrices acting on site $k$. The connection between the biased dynamics and the large deviation properties of the $1d$ WASEP is established through the largest eigenvalue of ${\mathbb W}^{\mu,\lambda}$ \cite{touchette09a,garrahan18a}. Such eigenvalue, denoted by $\theta_0(\mu,\lambda)$, is nothing but the cumulant generating function of the observables $m$ and $q$, related to the LDF $G(m,q)$ via a Legendre transform,
\be
\theta_0(\mu,\lambda)=L^{-1}\max_{m,q}[\lambda q - \mu L m - G(m,q)] \, .
\ee

The average of an observable $b$ at a final time $t$ in the unbiased ($\lambda=\mu=0$) dynamics can be written in operator notation as $\la b(t) \ra\equiv\la -|\hat{b}e^{t\!~ {\mathbb W}^{0,0}}| P_0 \ra$. We can write the time evolution operator for long times as $e^{t\!~ {\mathbb W}^{0,0}}\sim \ket{P_{ss}}\bra{-}$, with $\ket{P_{ss}}$ being the stationary state probability vector. Thus, as $\langle - | P_{0} \rangle=1$ the average of $b$ is $\la b(t) \ra\equiv\la -|\hat{b}\ket{P_{ss}}$. Since we are in the unbiased dynamics this average is the same at both the final time $t$ and the intermediate times $0 \ll \tau \ll t$, so that $\la b(t) \ra=\la b(\tau) \ra$ \cite{garrahan09a}. However, for a biased dynamics such as ${\mathbb W}^{0,\lambda}$, we are interested in computing the average of observables at intermediate times, since the rare event sustained by ${\mathbb W}^{0,\lambda}$ presents time-boundary effects which make the average at final and at intermediate times no longer equivalent \cite{garrahan09a}. Hence, in order to make these averages equivalent in the biased dynamics, we transform the non-stochastic generator ${\mathbb W}^{0,\lambda}$ (note that it does not conserve probability $\bra{-}{\mathbb W}^{0,\lambda}\neq 0$) into a physical stochastic generator via the Doob transform \cite{chetrite15a,carollo2018making}:
\be
{\mathbb W}^{0,\lambda}_{Doob}=\hat{L}_0{\mathbb W}^{0,\lambda}\hat{L}_0^{-1}-\theta_0(\lambda)\, ,
\label{DoobgenSM}
\ee
which is a proper stochastic generator (now $\bra{-}{\mathbb W}^{0,\lambda}_{Doob}=0$), with largest eigenvalue equal to zero, generating the same trajectories as ${\mathbb W}^{0,\lambda}$. Here $\hat{L}_0$ is a diagonal matrix whose elements $(\hat{L}_0)_{ii}$ are the $i$-th entries of the left eigenvector $\bra{L_0}$ of the biased generator ${\mathbb W}^{0,\lambda}$ associated with its largest eigenvalue $\theta_0(\lambda)$. Thus, with this new generator ${\mathbb W}^{0,\lambda}_{Doob}$ we can compute the average of any observable $b$ at intermediate times as 
\be
\la b(\tau) \ra_{\lambda}=\la b(t) \ra_{\lambda}\equiv \frac{\la -|\hat{b}e^{t\!~ {\mathbb W}^{0,\lambda}_{Doob}}| P_0 \ra}{\la -|e^{t\!~ {\mathbb W}^{0,\lambda}_{Doob}}| P_0 \ra}\, .
\label{DoobaveSM}
\ee
In what follows we show how the previous average takes different forms depending on whether or not the largest eigenvalue of the biased generator ${\mathbb W}^{0,\lambda}$ is degenerate.

\subsection{Non-degenerate largest eigenvalue (PH symmetric phase)}
If $\theta_0(\lambda)$ is non-degenerate, the time evolution operator for long times is $e^{t\!~ {\mathbb W}^{0,\lambda}}\sim e^{t\theta_0(\lambda) }\ket{R_0}\bra{L_0}$. Then by using \eqref{DoobgenSM} the asymptotic Doob time evolution operator reads 
$$
e^{t\!~ {\mathbb W}^{0,\lambda}_{Doob}}\sim \hat{L}_0\ket{R_0}\bra{L_0}\hat{L}_0^{-1}=\hat{L}_0\ket{R_0}\bra{-}\, ,
$$
with $\ket{R_0}$ being the right eigenvector of ${\mathbb W}^{0,\lambda}$ associated with its largest eigenvalue $\theta_0(\lambda)$. Additionally we can normalize eigenvectors so that 
$$
\la L_i|R_j\ra=\delta_{ij}~~~~~\text{and}~~~~~\la -|R_0\ra=1\,.
$$
Thus the time-evolved initial probability vector is 
\be
e^{t\!~ {\mathbb W}^{0,\lambda}_{Doob}}|P_0\ra\sim \hat{L}_0\ket{R_0} \,.
\label{singleSSM}
\ee
As a consequence the average \eqref{DoobaveSM} is given by
$$
\la b(\tau) \ra_{\lambda}=\frac{\la -|\hat{b}\hat{L}_0\ket{R_0}}{\la -|\hat{L}_0\ket{R_0}}=\frac{\la -|\hat{b}\hat{L}_0\ket{R_0}}{\la L_0|R_0\ra}=\la -|\hat{b}\hat{L}_0\ket{R_0}
$$
where in the last equality we have used the fact that eigenvectors are normalized. This is how we calculate, from the microscopic dynamics, the optimal density profiles associated with current fluctuations ($\lambda\neq 0$) in the particle-hole (PH) symmetric phase. The optimal particle density in the large size limit at $x=k/L$, with $L$ being the total number of sites, is thus given by
$$
\rho(x)=\la \hat{n}_k (\tau)\ra_{\lambda}=\la -|\hat{n}_k \hat{L}_0\ket{R_0}\, .
$$

\subsection{Degenerate largest eigenvalue (PH symmetry-broken phase)}
As we have seen in the main text, for $\lambda_c^- \le \lambda \le \lambda_c^+$ (or equivalently $|q|\le q_c$), the largest eigenvalue of ${\mathbb W}^{0,\lambda}$ becomes degenerate in the large size limit, $L\to\infty$. This is reflected in the diffusively-scaled spectral gap, $L^2[\theta_0(0,\lambda) - \theta_1(0,\lambda)]$, with $\theta_1(0,\lambda)$ the next-to-leading eigenvalue of ${\mathbb W}^{0,\lambda}$, which tends to zero as $L$ increases in this $\lambda$-region. In this case, defining as $\ket{R_1}$ and $\bra{L_1}$ the right and left eigenvectors associated to $\theta_1(0,\lambda)$,
we have that the time evolution operator can be written for long times as $e^{t\!~ {\mathbb W}^{0,\lambda}}\sim e^{t\theta_0(\lambda)}(\ket{R_0}\bra{L_0}+\ket{R_1}\bra{L_1})$. Hence, by using \eqref{DoobgenSM} the asymptotic Doob time evolution operator reads 
$$
e^{t\!~ {\mathbb W}^{0,\lambda}_{Doob}}\sim \hat{L}_0\ket{R_0}\bra{L_0}L_0^{-1}+\hat{L}_0\ket{R_1}\bra{L_1}\hat{L}_0^{-1}=\hat{L}_0\ket{R_0}\bra{-}+\hat{L}_0\ket{R_1}\bra{L_1}\hat{L}_0^{-1}\, .
$$
Thus the time-evolved initial vector probability is 
\be
\boxed{
e^{t\!~ {\mathbb W}^{0,\lambda}_{Doob}}|P_0\ra\sim \hat{L}_0\ket{R_0} + c \hat{L}_0\ket{R_1} \,
\label{MSSM}
}\, ,
\ee
with $c=\bra{L_1}\hat{L}_0^{-1}|P_0 \ra$. Note that, since $\la - |P_0\ra = 1$ then $c\in[c_1,c_2]$ with $c_1=\min\left(\bra{L_1}\hat{L}_0^{-1}\right)$ and $c_2=\max\left(\bra{L_1}\hat{L}_0^{-1}\right)$, where $\min$ and $\max$ correspond to the minimum and maximum entries of the vector $\bra{L_1}\hat{L}_0^{-1}$. Thus, Eq. \eqref{MSSM} defines the set of metastable states $|P_{MS}^c\ra$ of the main text, whose extremes are given by $|P_{MS}^{c_1}\ra$ and $|P_{MS}^{c_2}\ra$. As a consequence the average \eqref{DoobaveSM} is given by
$$
\la b(\tau) \ra_{\lambda}=\frac{\la -|\hat{b}\hat{L}_0\ket{R_0} + c \la -|\hat{b}\hat{L}_0\ket{R_1}}{\la -|\hat{L}_0\ket{R_0} + c \la -|\hat{L}_0\ket{R_1} }=\la -|\hat{b}\hat{L}_0\ket{R_0} + c \la -|\hat{b}\hat{L}_0\ket{R_1}
$$
where in the last equality we have used the fact that eigenvectors are normalized. This is how we calculate, from the microscopic dynamics, the optimal density profiles associated with current fluctuations ($\lambda \ne 0$) in the symmetry-broken phase. The optimal particle densities in the large size limit at $x=k/L$, are thus given by
$$
\rho_1(x)=\la \hat{n}_k (\tau)\ra_{\lambda}=\la -|\hat{n}_k \hat{L}_0\ket{R_0} + c_1 \la -|\hat{n}_k \hat{L}_0\ket{R_1}\, 
$$
and 
$$
\rho_2(x)=\la \hat{n}_k (\tau)\ra_{\lambda}=\la -|\hat{n}_k \hat{L}_0\ket{R_0} + c_2 \la -|\hat{n}_k \hat{L}_0\ket{R_1}\, ,
$$
which correspond to the metastable density profiles for $L=10$ and $L=20$ of Fig.~4 in the main text.


\end{document}